\documentclass[11pt]{article}

\usepackage[final]{acl}

\usepackage{times}
\usepackage{latexsym}
\usepackage[T1]{fontenc}
\usepackage[utf8]{inputenc}
\usepackage{microtype}
\usepackage{inconsolata}
\usepackage{graphicx}
\usepackage{caption} 
\usepackage[inline]{enumitem}
\usepackage{amsmath}

\usepackage{booktabs}
\usepackage[table]{xcolor} 
\usepackage{multirow}
\usepackage{siunitx}
\usepackage{xcolor}
\usepackage{enumitem}
\usepackage{amssymb}
\usepackage{subcaption}
\usepackage{placeins}
\usepackage{cuted}
\usepackage{array}
\usepackage{tikz}
\usetikzlibrary{tikzmark,fit}
\usepackage{pgfplots}
\pgfplotsset{compat=1.18}
\usepgfplotslibrary{colormaps}

\usepackage[dvipsnames]{xcolor}
\usepackage{booktabs}

\usepackage{booktabs}
\usepackage{colortbl}
\usepackage{xcolor}
\usepackage{pgf}
\usepackage{multirow}
\usepackage{makecell}
\usepackage{adjustbox} 
\usepackage{diagbox}

\usepackage{nicematrix, tikz}

\usepackage{xparse}
\usepackage{tikz}
\usetikzlibrary{calc}

\definecolor{SoftRed}{HTML}{FF9999}
\definecolor{SoftBlue}{HTML}{99CCFF}

\newcommand{\DropWhite}{0.95}
\newcommand{\DropRed}{0.75}

\usepackage{pdfrender}

\newcommand{\gradcell}[4]{%
    \pgfmathsetmacro{\Ratio}{#3/#2}%
    \pgfmathparse{\Ratio >= \DropWhite ? 1 : 0}%
    \ifnum\pgfmathresult=1%
        \pgfmathsetmacro{\BlueIntensity}{((\Ratio - \DropWhite) / (1 - \DropWhite)) * 100}%
        \edef\mycolor{\noexpand\cellcolor{SoftBlue!\BlueIntensity!white}}%
        \mycolor%
    \else%
        \pgfmathparse{\Ratio >= \DropRed ? 1 : 0}%
        \ifnum\pgfmathresult=1%
            \pgfmathsetmacro{\RedIntensity}{((\DropWhite - \Ratio) / (\DropWhite - \DropRed)) * 100}%
            \edef\mycolor{\noexpand\cellcolor{SoftRed!\RedIntensity!white}}%
            \mycolor%
        \else%
            \cellcolor{SoftRed}%
        \fi%
    \fi%
    \pgfmathsetmacro{\MeanPct}{100*(#3)}%
    \ifnum#1=1%
        \textpdfrender{TextRenderingMode=FillStroke, LineWidth=0.2pt}{%
            \textbf{\num[round-mode=places, round-precision=1]{\MeanPct}}%
        }%
    \else%
        \num[round-mode=places, round-precision=1]{\MeanPct}%
    \fi%
}

\newcommand{\DropWhiteStructural}{0.70}
\newcommand{\DropRedStructural}{0.30}

\newcommand{\gradcellStructural}[4]{%
    \pgfmathsetmacro{\Ratio}{#3/#2}%
    \pgfmathparse{\Ratio >= \DropWhiteStructural ? 1 : 0}%
    \ifnum\pgfmathresult=1%
        \pgfmathsetmacro{\BlueIntensity}{((\Ratio - \DropWhiteStructural) / (1 - \DropWhiteStructural)) * 100}%
        \edef\temp{\noexpand\cellcolor{SoftBlue!\BlueIntensity!white}}%
        \temp%
    \else%
        \pgfmathparse{\Ratio >= \DropRedStructural ? 1 : 0}%
        \ifnum\pgfmathresult=1%
            \pgfmathsetmacro{\RedIntensity}{((\DropWhiteStructural - \Ratio) / (\DropWhiteStructural - \DropRedStructural)) * 100}%
            \edef\temp{\noexpand\cellcolor{SoftRed!\RedIntensity!white}}%
            \temp%
        \else%
            \cellcolor{SoftRed}%
        \fi%
    \fi%
    \pgfmathsetmacro{\MeanPct}{#3}%
    \ifnum#1=1%
        \textpdfrender{TextRenderingMode=FillStroke, LineWidth=0.2pt}{%
            \textbf{\num[round-mode=places, round-precision=2]{\MeanPct}}%
        }%
    \else%
        \num[round-mode=places, round-precision=2]{\MeanPct}%
    \fi%
}

\newcommand{\DropWhiteStructuralConnection}{0.0}
\newcommand{\DropRedStructuralConnection}{-0.3}

\newcommand{\gradcellStructuralConnection}[4]{%
    \pgfmathsetmacro{\Ratio}{#3/#2}%
    \pgfmathparse{\Ratio >= \DropWhiteStructuralConnection ? 1 : 0}%
    \ifnum\pgfmathresult=1%
        \pgfmathsetmacro{\BlueIntensity}{((\Ratio - \DropWhiteStructuralConnection) / (1 - \DropWhiteStructuralConnection)) * 100}%
        \edef\temp{\noexpand\cellcolor{SoftBlue!\BlueIntensity!white}}%
        \temp%
    \else%
        \pgfmathparse{\Ratio >= \DropRedStructuralConnection ? 1 : 0}%
        \ifnum\pgfmathresult=1%
            \pgfmathsetmacro{\RedIntensity}{((\DropWhiteStructuralConnection - \Ratio) / (\DropWhiteStructuralConnection - \DropRedStructuralConnection)) * 100}%
            \edef\temp{\noexpand\cellcolor{SoftRed!\RedIntensity!white}}%
            \temp%
        \else%
            \cellcolor{SoftRed}%
        \fi%
    \fi%
    \pgfmathsetmacro{\MeanPct}{#3}%
    \ifnum#1=1%
        \textpdfrender{TextRenderingMode=FillStroke, LineWidth=0.2pt}{%
            \textbf{\num[round-mode=places, round-precision=2]{\MeanPct}}%
        }%
    \else%
        \num[round-mode=places, round-precision=2]{\MeanPct}%
    \fi%
}

\graphicspath{{images/figures/}}


\title{Superficial Success vs. Internal Breakdown: An Empirical Study of Generalization in Adaptive Multi-Agent Systems}

\author{First Author \\
  Affiliation / Address line 1 \\
  Affiliation / Address line 2 \\
  Affiliation / Address line 3 \\
  \texttt{email@domain} \\\And
  Second Author \\
  Affiliation / Address line 1 \\
  Affiliation / Address line 2 \\
  Affiliation / Address line 3 \\
  \texttt{email@domain} \\}
\author{
    Namyoung So$^\dag$\quad 
    Seokgyu Jang$^\dag$\quad 
    Taeuk Kim$^*$ \\
  Department of Computer Science, Hanyang University, Seoul, Republic of Korea \\
  \texttt{\{thskadud, diamondgyu, kimtaeuk\}@hanyang.ac.kr}
  }

\begin{document}
\maketitle

\begin{abstract}
Adaptive multi-agent systems (MAS) are increasingly adopted to tackle complex problems.
However, the narrow task coverage of their optimization raises the question of whether they can function as general-purpose systems.
To address this gap, we conduct an extensive empirical study of adaptive MAS, revealing two key findings: (1) \textbf{topological overfitting}---they fail to generalize across different domains; and (2) \textbf{illusory coordination}---they achieve reasonable surface-level accuracy while the underlying agent interactions diverge from ideal MAS behavior, raising concerns about their practical utility.
These findings highlight the pressing need to prioritize generalization in MAS development and motivate evaluation protocols that extend beyond simple final-answer correctness.
We release our code at \href{https://github.com/HYU-NLP/Superficial-Internal.git}{this https URL}.
\end{abstract}

\renewcommand{\thefootnote}{} 
\footnotetext{\textsuperscript{\dag}Equal contribution. \textsuperscript{*}Corresponding author.}
\renewcommand{\thefootnote}{\arabic{footnote}} 

\begin{table*}[!t]
    \centering
    \scriptsize
    \renewcommand{\arraystretch}{1.2}
    \setlength{\tabcolsep}{2.7pt}

    \begin{minipage}{0.48\textwidth}
        \centering
        \newcommand{\cOneMaxAF}{0.635} \newcommand{\cTwoMaxAF}{0.490}
        \newcommand{\cThreeMaxAF}{0.584} \newcommand{\cFourMaxAF}{0.418}
        \newcommand{\cFiveMaxAF}{0.655} \newcommand{\cSixMaxAF}{0.753}

        \begin{subtable}{\textwidth}
            \centering
            \begin{tabular}{l c c c c c c}
                \toprule
                \textbf{Training / Test (Domain)} & \textbf{L} & \textbf{D} & \textbf{MH} & \textbf{S} & \textbf{MA} & \textbf{CS} \\
                \midrule
                CaseHOLD (\textbf{L}egal)
                & \gradcell{1}{\cOneMaxAF}{0.635}{0}
                & \gradcell{0}{\cTwoMaxAF}{0.442}{0}
                & \gradcell{0}{\cThreeMaxAF}{0.574}{0}
                & \gradcell{0}{\cFourMaxAF}{0.418}{0}
                & \gradcell{0}{\cFiveMaxAF}{0.655}{0}
                & \gradcell{0}{\cSixMaxAF}{0.700}{0} \\
                COM$^2$ (\textbf{D}etective)
                & \gradcell{0}{\cOneMaxAF}{0.534}{0}
                & \gradcell{1}{\cTwoMaxAF}{0.479}{0}
                & \gradcell{0}{\cThreeMaxAF}{0.538}{0}
                & \gradcell{0}{\cFourMaxAF}{0.358}{0}
                & \gradcell{0}{\cFiveMaxAF}{0.544}{0}
                & \gradcell{0}{\cSixMaxAF}{0.195}{0} \\
                MuSiQue (\textbf{M}ulti-\textbf{H}op)
                & \gradcell{0}{\cOneMaxAF}{0.632}{0}
                & \gradcell{0}{\cTwoMaxAF}{0.490}{0}
                & \gradcell{1}{\cThreeMaxAF}{0.584}{0}
                & \gradcell{0}{\cFourMaxAF}{0.401}{0}
                & \gradcell{0}{\cFiveMaxAF}{0.654}{0}
                & \gradcell{0}{\cSixMaxAF}{0.738}{0} \\
                SciBench (\textbf{S}cience)
                & \gradcell{0}{\cOneMaxAF}{0.618}{0}
                & \gradcell{0}{\cTwoMaxAF}{0.342}{0}
                & \gradcell{0}{\cThreeMaxAF}{0.549}{0}
                & \gradcell{1}{\cFourMaxAF}{0.389}{0}
                & \gradcell{0}{\cFiveMaxAF}{0.628}{0}
                & \gradcell{0}{\cSixMaxAF}{0.475}{0} \\
                TheoremQA (\textbf{Ma}th)
                & \gradcell{0}{\cOneMaxAF}{0.622}{0}
                & \gradcell{0}{\cTwoMaxAF}{0.472}{0}
                & \gradcell{0}{\cThreeMaxAF}{0.575}{0}
                & \gradcell{0}{\cFourMaxAF}{0.369}{0}
                & \gradcell{1}{\cFiveMaxAF}{0.638}{0}
                & \gradcell{0}{\cSixMaxAF}{0.751}{0} \\
                StrategyQA (\textbf{C}ommon\textbf{s}ense)
                & \gradcell{0}{\cOneMaxAF}{0.006}{0}
                & \gradcell{0}{\cTwoMaxAF}{0.005}{0}
                & \gradcell{0}{\cThreeMaxAF}{0.415}{0}
                & \gradcell{0}{\cFourMaxAF}{0.001}{0}
                & \gradcell{0}{\cFiveMaxAF}{0.157}{0}
                & \gradcell{1}{\cSixMaxAF}{0.725}{0} \\
                \midrule
                Multi-Domain Training
                & \gradcell{0}{\cOneMaxAF}{0.602}{0}
                & \gradcell{0}{\cTwoMaxAF}{0.467}{0}
                & \gradcell{0}{\cThreeMaxAF}{0.529}{0}
                & \gradcell{0}{\cFourMaxAF}{0.411}{0}
                & \gradcell{0}{\cFiveMaxAF}{0.644}{0}
                & \gradcell{0}{\cSixMaxAF}{0.753}{0} \\
                \bottomrule
            \end{tabular}
            \caption{AgentDropout}
        \end{subtable}
    \end{minipage}
    \hfill
    \begin{minipage}{0.48\textwidth}
        \centering
        \newcommand{\cOneMaxAF}{0.633} \newcommand{\cTwoMaxAF}{0.414}
        \newcommand{\cThreeMaxAF}{0.521} \newcommand{\cFourMaxAF}{0.408}
        \newcommand{\cFiveMaxAF}{0.664} \newcommand{\cSixMaxAF}{0.738}

        \begin{subtable}{\textwidth}
            \centering
            \begin{tabular}{l c c c c c c}
                \toprule
                \textbf{Training / Test (Domain)} & \textbf{L} & \textbf{D} & \textbf{MH} & \textbf{S} & \textbf{MA} & \textbf{CS} \\
                \midrule
                CaseHOLD (\textbf{L}egal)
                & \gradcell{1}{\cOneMaxAF}{0.617}{0}
                & \gradcell{0}{\cTwoMaxAF}{0.412}{0}
                & \gradcell{0}{\cThreeMaxAF}{0.476}{0}
                & \gradcell{0}{\cFourMaxAF}{0.408}{0}
                & \gradcell{0}{\cFiveMaxAF}{0.664}{0}
                & \gradcell{0}{\cSixMaxAF}{0.689}{0} \\
                COM$^2$ (\textbf{D}etective)
                & \gradcell{0}{\cOneMaxAF}{0.633}{0}
                & \gradcell{1}{\cTwoMaxAF}{0.408}{0}
                & \gradcell{0}{\cThreeMaxAF}{0.496}{0}
                & \gradcell{0}{\cFourMaxAF}{0.389}{0}
                & \gradcell{0}{\cFiveMaxAF}{0.623}{0}
                & \gradcell{0}{\cSixMaxAF}{0.552}{0} \\
                MuSiQue (\textbf{M}ulti-\textbf{H}op)
                & \gradcell{0}{\cOneMaxAF}{0.517}{0}
                & \gradcell{0}{\cTwoMaxAF}{0.414}{0}
                & \gradcell{1}{\cThreeMaxAF}{0.481}{0}
                & \gradcell{0}{\cFourMaxAF}{0.393}{0}
                & \gradcell{0}{\cFiveMaxAF}{0.620}{0}
                & \gradcell{0}{\cSixMaxAF}{0.489}{0} \\
                SciBench (\textbf{S}cience)
                & \gradcell{0}{\cOneMaxAF}{0.560}{0}
                & \gradcell{0}{\cTwoMaxAF}{0.392}{0}
                & \gradcell{0}{\cThreeMaxAF}{0.476}{0}
                & \gradcell{1}{\cFourMaxAF}{0.330}{0}
                & \gradcell{0}{\cFiveMaxAF}{0.572}{0}
                & \gradcell{0}{\cSixMaxAF}{0.715}{0} \\
                TheoremQA (\textbf{Ma}th)
                & \gradcell{0}{\cOneMaxAF}{0.485}{0}
                & \gradcell{0}{\cTwoMaxAF}{0.378}{0}
                & \gradcell{0}{\cThreeMaxAF}{0.468}{0}
                & \gradcell{0}{\cFourMaxAF}{0.359}{0}
                & \gradcell{1}{\cFiveMaxAF}{0.598}{0}
                & \gradcell{0}{\cSixMaxAF}{0.581}{0} \\
                StrategyQA (\textbf{C}ommon\textbf{s}ense)
                & \gradcell{0}{\cOneMaxAF}{0.382}{0}
                & \gradcell{0}{\cTwoMaxAF}{0.344}{0}
                & \gradcell{0}{\cThreeMaxAF}{0.338}{0}
                & \gradcell{0}{\cFourMaxAF}{0.342}{0}
                & \gradcell{0}{\cFiveMaxAF}{0.560}{0}
                & \gradcell{1}{\cSixMaxAF}{0.738}{0} \\
                \midrule
                Multi-Domain Training
                & \gradcell{0}{\cOneMaxAF}{0.622}{0}
                & \gradcell{0}{\cTwoMaxAF}{0.406}{0}
                & \gradcell{0}{\cThreeMaxAF}{0.521}{0}
                & \gradcell{0}{\cFourMaxAF}{0.390}{0}
                & \gradcell{0}{\cFiveMaxAF}{0.659}{0}
                & \gradcell{0}{\cSixMaxAF}{0.735}{0} \\
                \bottomrule
            \end{tabular}
            \caption{AFlow}
        \end{subtable}
    \end{minipage}

    \vspace{-2mm}
    \caption{
    In-domain and OOD performance of AgentDropout and AFlow on \texttt{GPT-oss-20B}.
    Cell colors are normalized per column by the column-wise maximum: values $\ge 95\%$ max are shaded in \textcolor{blue}{blue} (i.e., successful transfer), and values $< 70\%$ max are in \textcolor{red}{red} (i.e., failed transfer).
    While domain transfer is often reasonable, it fails in many others.
    Multi-domain training shows promising results compared to \textbf{in-domain} (bolded, diagonal) baselines.
    See Appendix \ref{sec:performance_measure_appendix} for results on \texttt{Qwen3-30B-A3B} and full three-run details.
    }
    \label{tab:main_table}
\end{table*}

\section{Introduction}

Alongside recent advances in large language models (LLMs) \cite{gemini3flash_modelcard,openai_gpt5p1_systemcard}, multi-agent systems (MAS) \cite{qian-etal-2024-chatdev, khan2024debating, chen2024agentverse, yu2024fincon}---which treat each model as an agent and coordinate multiple agents to accomplish challenging tasks---have attracted growing attention.
In this paradigm, agents collaborate as a unified system, iteratively exchanging feedback to refine their conclusions. 

Among various lines of work, \textbf{adaptive MAS} \cite{zhuge2024gptswarm,zhang2025multi,li2025assemble} have emerged as a prominent direction, where agent roles and communication topologies are adapted to a given task and objective, analogous to optimization in standard supervised learning.\footnote{We use the term \textit{topology} to denote the configurations of both agent roles (nodes) and their connections (edges).}

Yet this trend is paradoxical: although adaptive MAS are built from general-purpose LLMs, they are often heavily tailored to a narrow set of tasks.
Consequently, despite their potential for broad generalization, much of the literature focuses on optimizing MAS for specific domains, leaving it unclear whether such systems generalize beyond their training scope.
This is not merely a theoretical concern: constructing a MAS entails substantial costs from orchestrating LLMs, making it impractical to deploy a separate MAS for each task of interest.

\begin{figure}[t]
    \centering
    \includegraphics[width=0.48\textwidth,trim=0 0 0 0,clip]{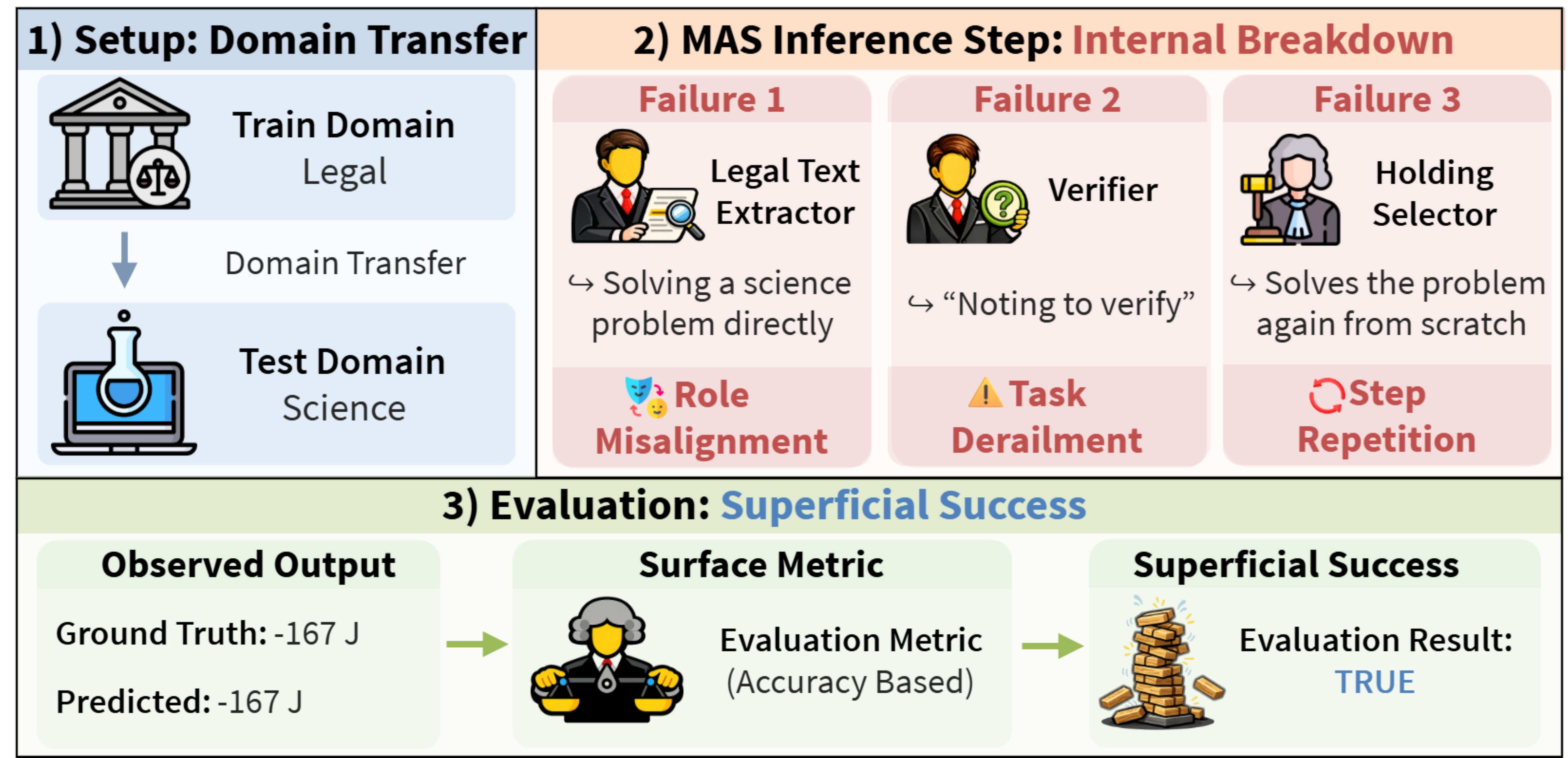}
    \caption{
        Example of superficial success vs. internal breakdown under domain transfer.
        When an adaptive MAS trained on the legal domain is applied to science, the agents make multiple errors during collaboration, yet the final answer remains correct due to the strength of individual LLMs---illustrating illusory coordination.
        }
\label{fig:figure1}
\end{figure}
In this work, we investigate the generalization capabilities of adaptive MAS approaches and identify two failure modes. 
First, we show that \textbf{topological overfitting} (\S\ref{sec:performance_measure})---performance degradation when an adaptive MAS is evaluated on out-of-distribution (OOD) tasks---is prevalent, revealing a clear failure of domain and task transfer. 

Meanwhile, even MAS that appear to generalize across domains often do so for the wrong reasons.
As illustrated in Figure \ref{fig:figure1}, closer inspection of their execution traces reveals that collaboration mechanisms frequently break down, and the system instead relies on the brute-force reasoning of individual LLMs rather than collective intelligence---a phenomenon we term \textbf{illusory coordination} (\S\ref{sec:collab_diagnosis}).

To better understand this issue, we conduct complementary qualitative and quantitative analyses.
Qualitatively, we apply the Multi-Agent System Failure Taxonomy (MAST) of \citet{cemri2025multi} and find that approximately 60\% of failures in adaptive MAS stem from role non-adherence and miscommunication between agents.

Quantitatively, we propose \textbf{Role Alignment} ($\mathcal{R}$) and \textbf{Connection Significance} ($\mathcal{O}$)---two new metrics that respectively evaluate role preservation and inter-agent information flow---and use them to formalize illusory coordination: a regime where accuracy remains high despite $\mathcal{R}$ and/or $\mathcal{O}$ being low.
Overall, topologies learned by adaptive MAS are often excessively tailored to specific tasks, giving rise to various forms of internal collapse.

In summary, our findings suggest that the community should look beyond final-answer accuracy and adopt more rigorous evaluation schemes that examine the internal dynamics of collaboration.

\section{Background}

\paragraph{Formulation of Adaptive MAS}
An \textit{adaptive} MAS is an agent collaboration framework defined by a tuple $(\mathcal{A}, \mathcal{C})$, where $\mathcal{A}$ denotes the set of agents and $\mathcal{C}$ their connections.\footnote{Here, an agent is an LLM instantiated with a specific role and instructions; a single LLM may serve as multiple agents.}
Notably, both $\mathcal{A}$ and $\mathcal{C}$ are learned from data: given a task-specific training set $\mathcal{D}_{\mathrm{train}}$ and a fixed base LLM, an optimization method $\mathbb{M}$ searches for an optimal topology:
\begin{equation*}
    (\mathcal{A}^{*}, \mathcal{C}^{*}) = \mathbb{M}\!\left(\mathcal{D}_{\mathrm{train}},\, \mathrm{LLM}\right),
\end{equation*}
where $\mathcal{A}^{*}$ and $\mathcal{C}^{*}$ denote the optimized agent roles and connections, respectively.

As the base LLM, we employ \texttt{GPT-oss-20B} \cite{agarwal2025gpt}. 
Additional results with \texttt{Qwen3-30B-A3B} \cite{qwen3technicalreport}, which show similar trends, are reported in Appendix \ref{sec:performance_measure_appendix}.
\paragraph{Adaptive MAS Algorithms}
We evaluate two representative algorithms: \textbf{AFlow} \cite{zhang2024aflow}, which incrementally constructs communication paths (bottom-up), and \textbf{AgentDropout} \cite{wang2025agentdropout}, which prunes redundant links from a fully connected graph (top-down).
AFlow jointly optimizes $\mathcal{A}$ and $\mathcal{C}$ during search, whereas AgentDropout optimizes only $\mathcal{C}$ after $\mathcal{A}$ has been determined. 
For AgentDropout, we use AgentInit \cite{tian2025agentinit} to optimize $\mathcal{A}$ beforehand.

For both ID and OOD evaluation, we use the same learned topology without further adaptation, varying only the test domain.
Further procedural details are provided in Appendix \ref{sec:algorithms_explained}.
To support reproducibility, we include the judging prompts (Appendix \ref{sec:prompts_appendix}), three-run results (Appendix \ref{sec:performance_measure_appendix}), and the full evaluation configurations (Appendix \ref{sec:exp_details}).

\begin{table}[t]
    \centering
    \scriptsize
    \renewcommand{\arraystretch}{1.08}
    \setlength{\tabcolsep}{3.2pt}
    \begin{tabular}{l l p{0.44\columnwidth}}
        \toprule
        \textbf{Dataset} & \textbf{Domain} & \textbf{Specification} \\
        \midrule
        \makecell[l]{CaseHOLD \\ \cite{chalkidis2022lexglue}} & \textbf{L}egal & Judicial decision (holding) identification from case contexts. \\
        \makecell[l]{COM$^2$ \\ \cite{xiong20252}} & \textbf{D}etective & Perpetrator identification under multi-logical constraints. \\
        \makecell[l]{MuSiQue \\ \cite{trivedi2022musique}} & \textbf{M}ulti-\textbf{H}op & QA requiring integration across multiple long-context documents. \\
        \makecell[l]{SciBench \\ \cite{wang2023scibench}} & \textbf{S}cientific & University-level scientific reasoning and calculations. \\
        \makecell[l]{TheoremQA \\ \cite{chen2023theoremqa}} & \textbf{Ma}th & Theorem-based STEM problem solving. \\
        \makecell[l]{StrategyQA \\ \cite{geva2021did}} & \textbf{C}ommon\textbf{s}ense & Multi-step fact synthesis for boolean QA. \\
        \bottomrule
    \end{tabular}
    \caption{Overview of the six datasets used in our study, selected to cover diverse domains and reasoning types.}
    \label{tab:benchmark_overview}
\end{table}

\paragraph{Datasets}
\begin{table*}
    \centering
    \scriptsize
    \begin{tabular}{l p{11cm}} 
        \toprule
        \textbf{Failure Category} & \textbf{Example} \\ \midrule
        \makecell[tl]{\textbullet\ Disobey role \\ \hspace*{0.3em} specification} & 
        \textbf{Question:} Calculate the Carnot efficiency on given condition... \newline
        \hspace*{0.25em} $\hookrightarrow$ \textcolor{red}{\textbf{Legal Text Extractor}}: 1. Convert the temperatures: ..., 2. Compute the efficiency: ... \textcolor{red}{Final solution: 0.107}  \\ \midrule
        
        \makecell[tl]{\textbullet\ Step repetition \\ \textbullet\ No verification \\ \textbullet\ Ignores input} & 
        \textbf{Question:} Two sets of points are linearly separable when... true or false? \newline
        \hspace*{0.25em} $\hookrightarrow$ \textbf{Clue Fact Extractor:} Two sets are linearly separable when... If the hulls are disjoint, ...
        \textcolor{blue}{The answer is: True} \newline
        \hspace*{1.8em} $\hookrightarrow$ \textbf{Validator:} If they are linearly separable, then... Consequently, ... \textcolor{blue}{The answer is: True} \\ \midrule
        
        \makecell[tl]{\textbullet\ Disobey task \\ \hspace*{0.3em} specification \\ \textbullet\ Task derailment} & 
        \textbf{Question:} What actions could have been taken to prevent crime when... \textcolor{purple}{(A):..., (B):..., ...} \newline
        \hspace*{0.4em}$\hookrightarrow$ \textbf{Context Analyzer:} ... (B) and (D) are good for purposes, while other candidates are... \newline
        \hspace*{1.7em} $\hookrightarrow$ \textbf{Answer Synthesizer:} \textcolor{purple}{True} \\ \bottomrule
    \end{tabular}
    \caption{Case studies of internal breakdown.
    Reasoning traces flow downward (e.g., Clue Fact Extractor $\rightarrow$ Validator). Domain shifts induce multiple failure modes: \textcolor{red}{role misalignment} (the Legal Text Extractor attempts physics calculations), \textcolor{blue}{input neglect} (the Validator disregards prior inputs and re-solves the problem), and \textcolor{purple}{task violation} (the Answer Synthesizer responds ``True'' to a multiple-choice question), all leading to undesired outcomes.}
    \label{tab:case_study_table}
\end{table*}


We employ six datasets that span diverse domains and reasoning challenges (see Table~\ref{tab:benchmark_overview}).
Adaptive MAS have primarily been developed with an emphasis on data-efficient optimization under limited training data \cite{zhang2024cut, wang2025agentdropout, zhang2024aflow, zhang2024g}.
To remain faithful to these original experimental settings, we cap the number of training samples at 100 for AFlow and 60 for AgentDropout.\footnote{We compare the performance of AgentDropout with \texttt{GPT-oss-20B} using 60 vs. 200 training examples. The difference is negligible (Appendix~\ref{sec:training_set_size_impact}), supporting our decision to limit the instance count following prior conventions.}
Dataset statistics are reported in Appendix \ref{sec:dataset_stats}.
\section{Finding 1: Topological Overfitting}\label{sec:performance_measure}
We first investigate the capacity of adaptive MAS topologies trained on specialized domains. 
All experiments are conducted with three separate runs.

\subsection{Evaluation Criteria}

\paragraph{In-Domain (ID) Performance}
Accuracy is measured on the test split of the source dataset used for optimization, following standard practice in the literature \cite{zhang2024aflow, wang2025agentdropout}.

\paragraph{Out-of-Distribution (OOD) Performance}
To probe cross-domain robustness, we evaluate MAS topologies learned in one domain on datasets from unseen domains. 
Under our setup, OOD evaluation spans the five domains excluded from training.


\paragraph{Multi-Domain Performance}
As a simple recipe to mitigate overfitting, we test multi-domain training by mixing data from six domains while keeping the total number of training instances consistent.

\subsection{Experimental Results}\label{sec:result_analysis}

Table \ref{tab:main_table} shows that single-domain MAS optimization often leads to over-specialization that fails under distribution shifts. 
They exhibit a marked decline in OOD performance—e.g., on the legal domain, AgentDropout drops from 63.5\% in-domain to 55.78\% OOD (avg.).
In addition, MAS optimized for commonsense reasoning often fail to transfer to numerical or multiple-choice problems. 
In such cases, accuracy degrades severely, as topologies optimized for binary (true/false) questions are unable to produce valid solutions in richer answer formats.

In contrast, a few cases show promising robustness—e.g., AgentDropout trained on MuSiQue and CaseHOLD—and the multi-domain variants also perform reasonably, suggesting potential for generalization.
However, it remains unclear whether this apparent generalization reflects effective collaboration or simply the strength of individual agents.



\section{Finding 2: Illusory Coordination}\label{sec:collab_diagnosis}
Motivated by the question posed in \S\ref{sec:result_analysis}, we conduct an in-depth analysis of the inner workings of adaptive MAS from both qualitative and quantitative perspectives, focusing on AgentDropout.
\subsection{Qualitative Analysis}
\begin{figure}[t]
    \centering
    \includegraphics[width=0.46\textwidth]{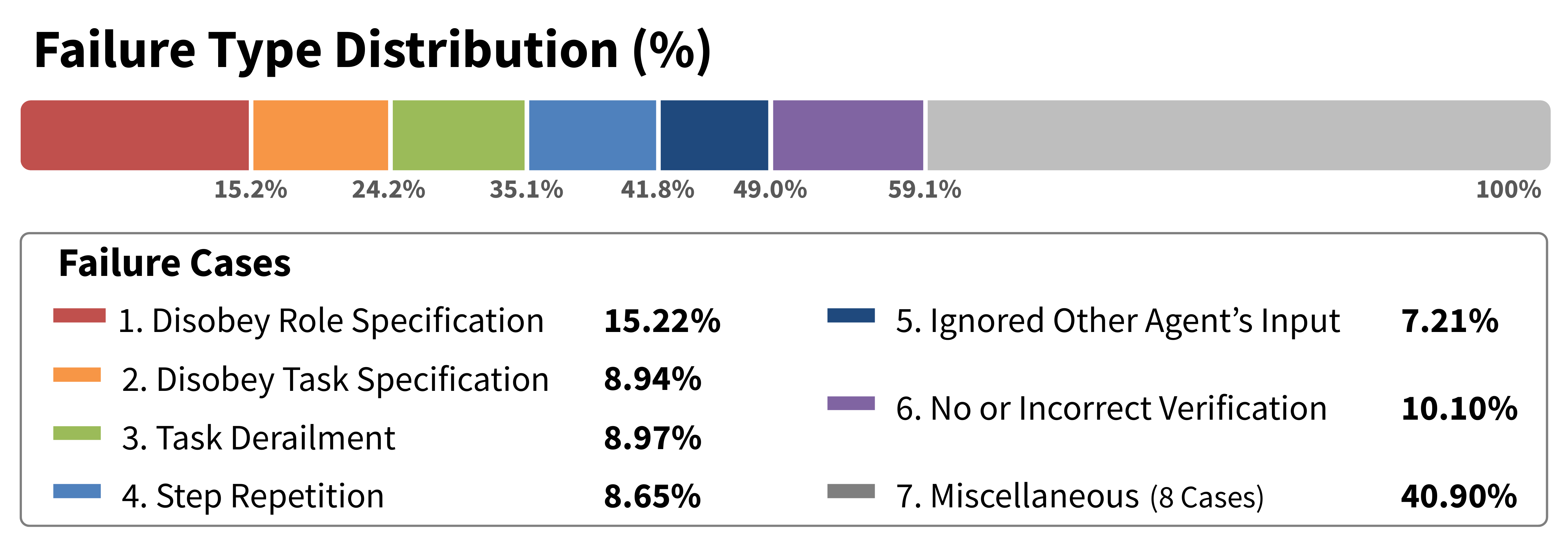}
    \vspace{-2mm}
    \caption{Failure types under domain transfer, grouped by MAST. Role-related and connection-related failures (Cases 1–6) account for around 60\%, indicating they are primary sources of collaboration breakdown.}
    \label{fig:case_study_stats}
\end{figure}
\begin{table*}[!t]
    \centering
    \scriptsize
    \renewcommand{\arraystretch}{1.15}
    {

\begin{minipage}{0.48\textwidth}
    \centering
    \newcommand{\cOneMaxAF}{1} 
    \newcommand{\cTwoMaxAF}{1} 
    \newcommand{\cThreeMaxAF}{1}
    \newcommand{\cFourMaxAF}{1} 
    \newcommand{\cFiveMaxAF}{1} 
    \newcommand{\cSixMaxAF}{1}
    \newcommand{\cSevenMaxAF}{1}
    \begin{subtable}{\textwidth}
            \centering
            \setlength{\tabcolsep}{2pt}
            \begin{tabular}{l c c c c c c}
                \toprule
                \textbf{Training / Test (Domain) } & \textbf{L} & \textbf{D} & \textbf{MH} & \textbf{S} & \textbf{MA} & \textbf{CS}\\
                \midrule
                CaseHOLD(\textbf{L}egal)   & 
                \gradcellStructural{1}{\cOneMaxAF}{1}{0} &
                \gradcellStructural{0}{\cOneMaxAF}{0.561}{0} & 
                \gradcellStructural{0}{\cOneMaxAF}{0.042}{0} &
                \gradcellStructural{0}{\cOneMaxAF}{0.216}{0} & 
                \gradcellStructural{0}{\cOneMaxAF}{0.252}{0} & 
                \gradcellStructural{0}{\cOneMaxAF}{0.5377}{0}\\
                COM$^2$(\textbf{D}etective)    & 
                \gradcellStructural{0}{\cTwoMaxAF}{0.786}{0} & 
                \gradcellStructural{1}{\cTwoMaxAF}{1}{0} & 
                \gradcellStructural{0}{\cTwoMaxAF}{0.0445}{0} &
                \gradcellStructural{0}{\cTwoMaxAF}{0.43}{0} & 
                \gradcellStructural{0}{\cTwoMaxAF}{0.47}{0} &
                \gradcellStructural{0}{\cTwoMaxAF}{0.8225}{0} \\
                MuSiQue(\textbf{M}ulti-\textbf{H}op)    &
                \gradcellStructural{0}{\cThreeMaxAF}{0.6855}{0} & 
                \gradcellStructural{0}{\cThreeMaxAF}{1}{0} & 
                \gradcellStructural{1}{\cThreeMaxAF}{0.3823}{0} & 
                \gradcellStructural{0}{\cThreeMaxAF}{0.502}{0} & 
                \gradcellStructural{0}{\cThreeMaxAF}{0.582}{0} & 
                \gradcellStructural{0}{\cThreeMaxAF}{0.580}{0}\\
                SciBench(\textbf{S}cience)   & 
                \gradcellStructural{0}{\cFourMaxAF}{0.443}{0} & 
                \gradcellStructural{0}{\cFourMaxAF}{0.487}{0} &
                \gradcellStructural{0}{\cFourMaxAF}{0.043}{0} & 
                \gradcellStructural{1}{\cFourMaxAF}{1}{0} & 
                \gradcellStructural{0}{\cFourMaxAF}{0.6199}{0} & 
                \gradcellStructural{0}{\cFourMaxAF}{0.459}{0} \\
                TheoremQA(\textbf{Ma}th)  & 
                \gradcellStructural{0}{\cFiveMaxAF}{0.379}{0} &
                \gradcellStructural{0}{\cFiveMaxAF}{0.357}{0} &
                \gradcellStructural{0}{\cFiveMaxAF}{0.041}{0} &
                \gradcellStructural{0}{\cFiveMaxAF}{0.600}{0} &
                \gradcellStructural{1}{\cFiveMaxAF}{1}{0} &
                \gradcellStructural{0}{\cFiveMaxAF}{0.3181}{0} \\
                StrategyQA(\textbf{C}ommon\textbf{s}ense) & 
                \gradcellStructural{0}{\cSixMaxAF}{1}{0} &
                \gradcellStructural{0}{\cSixMaxAF}{0.963}{0} & 
                \gradcellStructural{0}{\cSixMaxAF}{0.07}{0} &
                \gradcellStructural{0}{\cSixMaxAF}{0.318}{0} &
                \gradcellStructural{0}{\cSixMaxAF}{0.45}{0} & 
                \gradcellStructural{1}{\cSixMaxAF}{0.95}{0} \\
                \midrule
                Multi-Domain Training    & 
                \gradcellStructural{0}{\cSevenMaxAF}{0.89}{0} &
                \gradcellStructural{0}{\cSevenMaxAF}{1}{0} & 
                \gradcellStructural{0}{\cSevenMaxAF}{0.31}{0} & 
                \gradcellStructural{0}{\cSevenMaxAF}{0.58}{0} &
                \gradcellStructural{0}{\cSevenMaxAF}{0.619}{0} &
                \gradcellStructural{0}{\cSevenMaxAF}{0.976}{0} \\
                \bottomrule
            \end{tabular}
            \caption{Role Alignment ($\mathcal{R}$) Heatmap}
        \end{subtable}
    \end{minipage}
\hfill
}
    {\begin{minipage}{0.48\textwidth}
\centering
\newcommand{\cOneMaxAF}{1} 
\newcommand{\cTwoMaxAF}{1} 
\newcommand{\cThreeMaxAF}{1}
\newcommand{\cFourMaxAF}{1} 
\newcommand{\cFiveMaxAF}{1} 
\newcommand{\cSixMaxAF}{1}
\newcommand{\cSevenMaxAF}{1}
\begin{subtable}{\textwidth}
        \centering
        {\setlength{\tabcolsep}{1.2pt}
        \begin{tabular}{l c c c c c c}
            \toprule
            \textbf{Training / Test (Domain)} & \textbf{L} & \textbf{D} & \textbf{MH} & \textbf{S} & \textbf{MA} & \textbf{CS}\\
            \midrule
            CaseHOLD(\textbf{L}egal)   & 
            \gradcellStructural{1}{\cOneMaxAF}{1}{0} & 
            \gradcellStructural{0}{\cOneMaxAF}{0.065975417}{0} & 
            \gradcellStructural{0}{\cOneMaxAF}{-1.792234797}{0} & 
            \gradcellStructural{0}{\cOneMaxAF}{-2.069561746}{0} & 
            \gradcellStructural{0}{\cOneMaxAF}{-1.894698701}{0} & 
            \gradcellStructural{0}{\cOneMaxAF}{-1.560472056}{0}\\
            COM$^2$(\textbf{D}etective)    & 
            \gradcellStructural{0}{\cTwoMaxAF}{0.900871611}{0} & 
            \gradcellStructural{1}{\cTwoMaxAF}{1}{0} & 
            \gradcellStructural{0}{\cTwoMaxAF}{0.172767277}{0} & 
            \gradcellStructural{0}{\cTwoMaxAF}{0.645121668}{0} &
            \gradcellStructural{0}{\cTwoMaxAF}{0.579350006}{0} & 
            \gradcellStructural{0}{\cTwoMaxAF}{0.458512116}{0}\\
            MuSiQue(\textbf{M}ulti-\textbf{H}op)    & 
            \gradcellStructural{0}{\cThreeMaxAF}{1}{0} &
            \gradcellStructural{0}{\cThreeMaxAF}{0.963715604}{0} & 
            \gradcellStructural{1}{\cThreeMaxAF}{0.151295037}{0} & 
            \gradcellStructural{0}{\cThreeMaxAF}{0.949550395}{0} & 
            \gradcellStructural{0}{\cThreeMaxAF}{0.795387389}{0} & 
            \gradcellStructural{0}{\cThreeMaxAF}{0.860299933}{0} \\
            SciBench(\textbf{S}cience)   & 
            \gradcellStructural{0}{\cFourMaxAF}{-0.753037489}{0} &
            \gradcellStructural{0}{\cFourMaxAF}{-0.500635673}{0} &
            \gradcellStructural{0}{\cFourMaxAF}{-0.56529475}{0} &
            \gradcellStructural{1}{\cFourMaxAF}{1}{0} &
            \gradcellStructural{0}{\cFourMaxAF}{0.766412615}{0} &
            \gradcellStructural{0}{\cFourMaxAF}{-0.069934519}{0} \\
            TheoremQA(\textbf{Ma}th)  &
            \gradcellStructural{0}{\cSixMaxAF}{-0.115869431}{0} &
            \gradcellStructural{0}{\cSixMaxAF}{0.211042946}{0} &
            \gradcellStructural{0}{\cSixMaxAF}{-0.071023375}{0} &
            \gradcellStructural{0}{\cSixMaxAF}{1}{0} &
            \gradcellStructural{1}{\cSixMaxAF}{0.883151411}{0} &
            \gradcellStructural{0}{\cSixMaxAF}{0.475497823}{0}\\
            StrategyQA(\textbf{C}ommon\textbf{s}ense) & 
            \gradcellStructural{0}{\cFiveMaxAF}{0.927250397}{0} & 
            \gradcellStructural{0}{\cFiveMaxAF}{0.947135932}{0} & 
            \gradcellStructural{0}{\cFiveMaxAF}{0.174206893}{0} & 
            \gradcellStructural{0}{\cFiveMaxAF}{1}{0} & 
            \gradcellStructural{0}{\cFiveMaxAF}{0.898038563}{0} & 
            \gradcellStructural{1}{\cFiveMaxAF}{0.813220025}{0} \\
            \midrule
            Multi-Domain Training    & 
            \gradcellStructural{0}{\cSevenMaxAF}{0.691655183}{0} &
            \gradcellStructural{0}{\cSevenMaxAF}{0.992485165}{0} &
            \gradcellStructural{0}{\cSevenMaxAF}{-0.234043067}{0} &
            \gradcellStructural{0}{\cSevenMaxAF}{0.877032275}{0} &
            \gradcellStructural{0}{\cSevenMaxAF}{0.845137686}{0} &
            \gradcellStructural{0}{\cSevenMaxAF}{1}{0} \\
            \bottomrule
        \end{tabular}
        }
        \caption{Connection Significance ($\mathcal{O}$) Heatmap}
    \end{subtable}
\end{minipage}
}
    \caption{
    Illusory coordination of AgentDropout detected by $\mathcal{R}$ and $\mathcal{O}$. 
    All entries are row-wise normalized by the maximum value in each row (i.e., each cell reports value / max(row)). 
    Cell colors follow the normalized ratios: values $\ge 0.70$ are shaded \textcolor{blue}{blue} (i.e., successful transfer), values $< 0.70$ are shaded \textcolor{red}{red} (i.e., failed transfer). 
    In-domain results are in bold. 
    Results provide new insights into MAS dynamics. 
    Further details are in Appendix~\ref{sec:appendix_internal_failure}.
    }
    \label{tab:structural_failure}
\end{table*}

As shown in Figure \ref{fig:figure1} and Appendix \ref{sec:appendix_case_study}, we discover cases where collaborations collapse under domain transfer.
We organize these patterns using MAST~\citep{cemri2025multi}, which defines 14 MAS failure categories,\footnote{Detailed definitions are provided in Appendix~\ref{sec:mast_taxonomy}.} assigning 100 execution logs to one or more categories via an LLM judge (\texttt{GPT-oss-120B}).

In Figure \ref{fig:case_study_stats}, we focus on 6 topology-related failure types, including role violations and step repetition.
They account for a majority (59.1\%) of errors, suggesting that domain shift mainly induces systemic failures tied to role adherence and information flow.
Table \ref{tab:case_study_table} further presents case studies of internal collapse in MAS.


\subsection{Quantitative Analysis}
\label{sec:role_ridigity}

Based on the qualitative analysis, we devise two novel metrics for additional quantitative analysis. 


\paragraph{Role Alignment ($\mathcal{R}$)} 

Our previous observations suggest that failures often stem from breakdowns in role adherence: agents in an MAS should preserve role diversity rather than collapse into identical behaviors.
To quantify adherence to this principle, we introduce \textbf{Role Alignment} ($\mathcal{R}$).



Formally, let $\mathcal{A}$ denote the set of agents.
For each agent $i \in \mathcal{A}$, let $p_i$ represent its role prompt and $y_i$ for its output.
We utilize an encoder $\mathbf{e}(\cdot)$ (i.e., \texttt{all-MiniLM-L6-v2}) to map these texts into a shared embedding space.
We then compute $S_{1,i}$, the cosine similarity between the agent's role definition and its output: 
\begin{equation*}
\scalebox{0.9}{$
S_{1,i} = \cos\!\big(\mathbf{e}(p_i), \mathbf{e}(y_i)\big)$.
}
\end{equation*}
A higher $S_{1,i}$ suggests stronger semantic alignment with the assigned role.
Second, we compute $S_{2,i}$, which represents the average similarity between agent $i$'s output and the outputs of all other agents $j \in \mathcal{A} \setminus \{i\}$ for the same instance:
\begin{equation*}
\scalebox{0.9}{$
S_{2,i} = \frac{1}{|\mathcal{A}| - 1} \sum_{j \neq i} \cos\!\big(\mathbf{e}(y_i), \mathbf{e}(y_j)\big).
$}
\end{equation*}
A large $S_{2,i}$ indicates that agents produce generic, similar responses.
Finally, Role Alignment for agent $i$ is defined as $\mathcal{R}_i = S_{1,i} \times (1 - S_{2,i})$.
Thus, a large $\mathcal{R}_i$ signifies a robust topology where agents contribute unique, role-specific information.

\paragraph{Connection Significance ($\mathcal{O})$}
\label{sec:connection_Significance}

\begin{table}[t]
    \centering
    \scriptsize
    \setlength{\tabcolsep}{6pt}
    \renewcommand{\arraystretch}{1.18}
    \begin{tabular}{c c c p{0.48\columnwidth}}
        \toprule
        \textbf{\(\mathcal{O}_i\)} & \textbf{\(\alpha_{i,\ell}\)} & \textbf{\(s_{i,\ell}\)} & \textbf{Interpretation} \\
        \midrule
        \(\approx 0\) & small or mixed & either & Negligible net message impact. \\
        \(< 0\)       & large          & \(-1\) & Influential but unhelpful messages. \\
        \(> 0\)       & large          & \(+1\) & Influential and useful messages. \\
        \bottomrule
    \end{tabular}
    \caption{Interpretation of $\mathcal{O}_i$ in terms of message influence $\alpha_{i,\ell}$ and usefulness $s_{i,\ell}$.}
    \label{tab:o_interpretation}
\end{table}

To diagnose information flow, such as identifying disregarded messages, we define \textbf{Connection Significance} ($\mathcal{O}$), which quantifies how incoming messages from other agents influence an agent's output $y_i$ beyond the contribution of the task query $q$ and the agent's role $p_i$.
Here, a \textit{message} refers to the output of a predecessor agent that serves as input to agent $i$.

Let $\mathcal{X}_i=\{x_1,\dots,x_m\}$ denote the set of all incoming messages to agent $i$.
We then compute influence weights $\alpha_{i,\ell}$ by comparing message utility against the static priors $p_i$ and $q$:
\begin{equation*}
\scalebox{0.9}{$
\alpha_{i,\ell} = \frac{\exp\!\big(sim(x_\ell, y_i)\big)}{\sum_{\hat{z} \in \mathcal{X}_i \cup \{p_i,q\}}\exp\!\big(sim(\hat{z}, y_i)\big)},
$}
\end{equation*}
where $\hat{z}$ ranges over all prompt components in $\mathcal{X}_i \cup \{p_i,q\}$, and $sim(a,b) = \cos\!\big(\mathbf{e}(a),\mathbf{e}(b)\big)$.
That is, $\alpha_{i,\ell}$ is obtained by applying a softmax over both the incoming messages and the static priors $(p_i, q)$.
As a result, a message receives a smaller weight when $y_i$ is better explained by the agent's role $p_i$ or the task query $q$ than by the incoming messages.

Next, we determine the message usefulness \(s_{i,\ell} = \mathrm{LaaJ}(q, p_i, x_\ell) \in \{+1, -1\}\) using an LLM-as-a-judge (\texttt{GPT-oss-20B}),\footnote{The \(\mathrm{LaaJ}\) prompt is provided in Appendix~\ref{sec:prompts_appendix}.} and aggregate the weighted scores over all input messages:
\begin{equation*}
\scalebox{0.9}{$
    \mathcal{O}_i = \sum_{x_\ell \in \mathcal{X}_i} \alpha_{i,\ell}\, s_{i,\ell}.
$}
\end{equation*}
Thus, \(\mathcal{O}_i\) captures the aggregate effect of all inputs to agent \(i\), jointly reflecting their influence \((\alpha)\) and usefulness \((s)\): values near zero indicate negligible net impact, negative values indicate influential but unhelpful messages, and positive values reflect influential and useful ones (see Table \ref{tab:o_interpretation}).



\paragraph{Formal Definition of Illusory Coordination}\label{sec:structural_results_analysis}

Finally, for each (train, test) pair, we report in Table~\ref{tab:structural_failure} a single score for each metric, obtained by averaging $\mathcal{R}_i$ and $\mathcal{O}_i$ over all agents $i \in \mathcal{A}$ and then over test instances $\mathcal{D}_{\mathrm{test}}$:
\begin{equation*}
\scalebox{0.9}{$
(\mathcal{R}, \mathcal{O})
=
\frac{1}{|\mathcal{D}_{\mathrm{test}}|}
\sum_{q \in \mathcal{D}_{\mathrm{test}}}
\left(
\frac{1}{|\mathcal{A}|}
\sum_{i \in \mathcal{A}}
(\mathcal{R}_i(q), \mathcal{O}_i(q))
\right).
$}
\end{equation*}

Building on these values, we define \textbf{illusory coordination} as settings where task accuracy is high while internal collaboration quality is poor---i.e., low $\mathcal{R}$ and/or low (possibly negative) $\mathcal{O}$.
This definition is diagnostic: rather than claiming collaboration is always illusory, our goal is to identify cases where surface-level success is not supported by the intended collaborative mechanism.




\paragraph{Results with $\mathcal{R}$}\label{results_with_r}



While Table~\ref{tab:main_table} suggests that CaseHOLD~(\textbf{L}) transfers well to SciBench~(\textbf{S}), Table~\ref{tab:structural_failure} reveals a different picture: $\mathcal{R}$ drops to only 22\% of its in-domain value, indicating that this apparent success is not supported by role adherence.
Conversely, when transferring from COM$^2$~(\textbf{D}) to StrategyQA~(\textbf{CS}), $\mathcal{R}$ remains high (82\% of the in-domain value) despite the poor accuracy reported in Table~\ref{tab:main_table}.
These observations point to a \textit{dissociation} between accuracy and role alignment.
Table \ref{tab:rebuttal_analysis} further supports this interpretation: the Pearson correlation between accuracy and $\mathcal{R}$ (\textbf{Acc--$\mathcal{R}$}) is near zero in most cases, suggesting that final-answer correctness does not necessarily reflect whether the intended role structure is preserved.




\paragraph{Results with $\mathcal{O}$}\label{results_with_o}


Table \ref{tab:structural_failure} reveals degradation in some OOD scenarios, where agents fail to integrate incoming messages and instead rely on independent reasoning---a hallmark of illusory coordination.
Three trends emerge:
(1) StrategyQA(\textbf{CS})- and MuSiQue(\textbf{MH})-trained topologies maintain positive $\mathcal{O}$ across domains, indicating robust message utilization;
(2) CaseHOLD(\textbf{L})-trained topologies show strongly negative $\mathcal{O}$ in most transfers despite high accuracy; and
(3) multi-domain training mitigates fluctuations, yielding more stable connections.
Refer to Appendix \ref{sec:appendix_internal_failure} for further discussion.


\section{Correlation and Ablation Studies} 

Table \ref{tab:rebuttal_analysis} reports our correlation and ablation results.
In the correlation analysis, we confirm that accuracy is only weakly correlated with $\mathcal{R}$ and $\mathcal{O}$, again highlighting the risk of surface-level evaluation.

In the ablation study, we isolate the contribution of each topological component by fixing one and replacing the other with its OOD counterpart: \textit{Connection-OOD} retains the in-domain roles while replacing the connections with OOD ones; \textit{Role-OOD} does the opposite.
Both ablations cause performance drops across all benchmarks except COM$^2$~(\textbf{D}), indicating that roles and connections are each susceptible to topological overfitting (\S\ref{sec:performance_measure}).
Moreover, \textit{Role-OOD} causes a substantially larger average drop than \textit{Connection-OOD} (-13.00 vs.\ -1.24 percentage points), implying that roles are on average more task-specific than connections.

MuSiQue~(\textbf{MH}), however, is a notable exception: under \textit{Connection-OOD}, performance drops by 5.36 points, compared with an average of only 0.418 points across the other five datasets.
This suggests that, for multi-hop reasoning tasks, valid connections are particularly important.
\begin{table}[t]
    \centering
    \scriptsize
    \renewcommand{\arraystretch}{1.15}
    \setlength{\tabcolsep}{2.8pt}

    \begin{adjustbox}{max width=\columnwidth}
        \begin{tabular}{l c c c c c}
            \toprule
            \multirow{2}{*}{\textbf{Benchmark}} & \multicolumn{2}{c}{\textbf{Pearson Correlation}} & \multicolumn{3}{c}{\cellcolor{black!10}\textbf{Ablation Accuracy}} \\
            \cmidrule(lr){2-3} \cmidrule(lr){4-6}
            & \textbf{Acc--$\mathcal{R}$} & \textbf{Acc--$\mathcal{O}$}
            & \cellcolor{black!5}\textbf{In-Domain}
            & \cellcolor{blue!8}\textbf{Connection-OOD}
            & \cellcolor{red!10}\textbf{Role-OOD} \\
            \midrule
            CaseHOLD & -0.007 & 0.0002 & 63.50 & 62.88 (\textcolor[HTML]{BF4E4E}{-0.62}) & 48.26 (\textcolor[HTML]{BF4E4E}{-15.24}) \\
            COM$^2$ & -0.035$^{**}$ & 0.045$^{***}$ & 47.90 & 50.68 (\textcolor[HTML]{A0BF5E}{+2.78}) & 34.50 (\textcolor[HTML]{BF4E4E}{-13.40}) \\
            MuSiQue & 0.003 & 0.123$^{***}$ & 58.40 & 53.04 (\textcolor[HTML]{BF4E4E}{-5.36}) & 48.44 (\textcolor[HTML]{BF4E4E}{-9.96}) \\
            SciBench & 0.084$^{***}$ & -0.039$^{*}$ & 38.90 & 38.69 (\textcolor[HTML]{BF4E4E}{-0.21}) & 30.29 (\textcolor[HTML]{BF4E4E}{-8.61}) \\
            TheoremQA & 0.113$^{***}$ & -0.081$^{***}$ & 63.80 & 61.26 (\textcolor[HTML]{BF4E4E}{-2.54}) & 51.64 (\textcolor[HTML]{BF4E4E}{-12.16}) \\
            StrategyQA & -0.096$^{***}$ & 0.067$^{**}$ & 72.50 & 71.00 (\textcolor[HTML]{BF4E4E}{-1.50}) & 53.89 (\textcolor[HTML]{BF4E4E}{-18.61}) \\
            \bottomrule
        \end{tabular}
    \end{adjustbox}

    \caption{
    Correlation and ablation results for AgentDropout across 6 datasets.
    \textbf{Acc--$\mathcal{R}$} and \textbf{Acc--$\mathcal{O}$} columns report Pearson correlations ($^*p<0.05$, $^{**}p<0.01$, $^{***}p<0.001$). 
    The right part shows ablation results; parenthesized values indicate accuracy changes (pp).
    See Appendix~\ref{sec:results_interchange} for details.
    }
    \label{tab:rebuttal_analysis}
\end{table}

\FloatBarrier

\section{Conclusion}
This work examines generalization failures in adaptive MAS. 
We show that adaptive MAS exhibit mixed generalization, succeeding in some settings but failing in others (\textbf{topological overfitting}). 
Going beyond surface-level accuracy, we provide analysis using two new metrics and reveal \textbf{illusory coordination}, where strong accuracy masks flawed internal collaboration. 
These results call for designing adaptive MAS with generalizability in mind and evaluating them through rigorous internal analysis rather than accuracy in isolation.
\section*{Limitations}
\paragraph{Focused Evaluation}
To present a generalizable analysis of adaptive MAS, we focused on two representative frameworks: AFlow, which exemplifies constructive topology search, and AgentDropout, which implements dynamic pruning. 
These methods reflect contrasting design philosophies and our analysis sheds light on topological overfitting and illusory coordination under domain transfer.
However, they do not cover the full range of adaptive MAS structures.
Other frameworks—potentially using alternative optimization strategies or architectural assumptions—may exhibit different performance profiles.
In particular, our scope is centered on adaptive topology-learning MAS in data-scarce settings and does not directly include tool-using or web-surfing agents.
Exploring a wider range of adaptive MAS frameworks remains an important direction for future work.

\paragraph{Extension of Multi-Domain Learning}
Our results show that multi-domain training yields notable performance gains compared to topologies optimized for a single dataset.
However, our analysis covered only simple, initial attempts within this paradigm; discovering robust methods thus remains an important and promising direction. Such an extension may be essential for shifting the system from task-specific coordination toward more generalized and robust collaborative intelligence.

\section*{Acknowledgments}

This work was supported by Institute of Information \& Communications Technology Planning \& Evaluation (IITP) grant funded by the Korea government (MSIT) (No.\ RS-2020-II201373, Artificial Intelligence Graduate School Program (Hanyang University)). This work was supported by Institute of Information \& Communications Technology Planning \& Evaluation (IITP) under the artificial intelligence semiconductor support program to nurture the best talents (IITP-(2026)-RS-2023-00253914) grant funded by the Korea government (MSIT). This work was supported by the National Research Foundation of Korea (NRF) grant funded by the Korea government (MSIT) (RS-2025-00558151).

\bibliography{anthology} 

\newpage
\appendix
\twocolumn
\appendix
\part*{Appendix}


\section{Dataset Statistics}\label{sec:dataset_stats}
Please refer to Table \ref{tab:dataset_stats}. For the training set, we used 60 for AgentDropout and 100 for AFlow.
\begin{table}[ht]
    \centering
    \small 
    \setlength{\tabcolsep}{3pt}
    \begin{tabular}{lccc}
    \toprule
    \makecell[b]{\textbf{Dataset Name} \\ \textbf{(Domain)}} & 
    \makecell[b]{\textbf{Total} \\ \textbf{Train Split}} & 
    \makecell[b]{\textbf{Training} \\ \textbf{Set Size}} & 
    \makecell[b]{\textbf{Test} \\ \textbf{Set Size}} \\ \midrule
    CaseHOLD (Legal)            & 45,000 & 60 / 100 & 1,000 \\
    COM$^2$ (Detective)       & 251    & 60 / 100 & 1,004 \\
    MuSiQue (Multi-hop)       & 1,174  & 60 / 100 & 405   \\
    SciBench (Science)        & 138    & 60 / 100 & 552   \\
    TheoremQA (Math)          & 149    & 60 / 100 & 555   \\
    StrategyQA (CS) & 2,061  & 60 / 100 & 229   \\ \bottomrule
    \end{tabular}
    
    \caption{Dataset statistics across diverse domains. CS denotes Commonsense.}
    \label{tab:dataset_stats}
\end{table}

\section{Adaptive MAS Algorithms}\label{sec:algorithms_explained}
We summarize the adaptive MAS algorithms used in our experiments and describe their operating mechanisms.

\paragraph{AFlow} 
is a topology optimization framework that focuses on automatically discovering the best way for agents to collaborate.
Instead of using a predefined communication structure (like a debate or a pipeline), AFlow learns the optimal communication graph from scratch.

The central idea is that the best ``workflow'' or communication pattern for agents is not known beforehand and should be learned. 
AFlow takes a constructive approach, starting with nothing and building up the communication graph. The detailed process is described below:
\begin{enumerate}
    \item Start with Isolated Agents: The process begins with a blank process with a simple `solve it' prompt.
    \item Propose and Evaluate Extensions: The framework iteratively proposes adding new workflows by MCTS.
    \item Score and Select: It evaluates how much each potential new workflow improves the system's performance on a given task.
    \item Build the Graph: The workflows that provide the most significant performance boost are permanently added to the graph.
    \item Iterate: This process repeats, gradually building a complex and effective topology (workflow) optimized for the specific task.
\end{enumerate}

\paragraph{AgentInit}
is an automated MAS \emph{initialization} method that focuses on forming a strong agent team (roles) \emph{before} running the inference framework.
Instead of optimizing a communication graph directly, AgentInit first generates a pool of candidate agents, standardizes their role specifications, and then selects a compact, complementary team by jointly balancing \emph{task relevance} and \emph{intra-team diversity}.

The overall procedure is as follows:
\begin{enumerate}
    \item Multi-round Candidate Generation: A Planner decomposes the user query into sub-tasks and drafts candidate agent roles, while an Observer reviews the decomposition and role assignments and provides feedback. This refinement repeats for multiple rounds.
    \item NL-to-Format Standardization: A Formatter converts each candidate agent role from free-form natural language into a standardized representation (e.g., JSON) to ensure consistency for downstream comparison.
    \item Construct Candidate Teams: From the candidate agent pool, enumerate possible teams whose size lies within predefined bounds.
    \item Score Teams by Relevance and Diversity: Compute a relevance score between each agent (and team) and the query using embedding-based cosine similarity, and measure intra-team diversity using an embedding-similarity matrix (e.g., via Vendi-style diversity).
    \item Pareto-based Selection: Identify the Pareto-optimal set of teams that are non-dominated with respect to relevance and diversity, and use a Selector (LLM-powered) to choose the final team for deployment.
\end{enumerate}

\paragraph{AgentDropout} 
is another topology optimization framework, but it takes the opposite approach to AFlow. 
It aims to create a communication structure that is not only high-performing but also token-efficient. 
Inspired by the "dropout" technique in neural networks, AgentDropout assumes that in any given multi-agent team, some agents or communication links might be redundant or even harmful. 
It improves performance and efficiency by dynamically pruning or eliminating these non-essential components. The detailed process is described below:
\begin{enumerate}
    \item Start with a Dense Graph: The process typically begins with a highly-connected graph where most agents can communicate with each other.
    \item Identify Redundancy: During different rounds of communication, the framework uses an optimization method to score the importance of each agent and each communication link.
    \item Dynamically "Drop" Agents: Agents or links with low importance scores are temporarily dropped out for that round. This forces the system to solve the problem without relying on every single voice, making it more robust.
    \item Optimize for Efficiency and Performance: By removing unnecessary communication, the method significantly reduces the number of tokens required, lowering computational costs while often improving the final answer by reducing noise.
\end{enumerate}

\section{Additional Experimental Results}
\label{sec:performance_measure_appendix}

\subsection{Cross-Model Comparison on Qwen3-30B-A3B}
For completeness, Table~\ref{tab:qwen_main_table} reports the corresponding cross-domain generalization results on \texttt{Qwen3-30B-A3B}, using the same layout as Table~\ref{tab:main_table}.
\begin{table*}[!t]
    \centering
    \scriptsize
    \renewcommand{\arraystretch}{1.2}
    \setlength{\tabcolsep}{2.7pt}

    \begin{minipage}{0.48\textwidth}
        \centering
        \newcommand{\cOneMaxAF}{0.700} \newcommand{\cTwoMaxAF}{0.402}
        \newcommand{\cThreeMaxAF}{0.368} \newcommand{\cFourMaxAF}{0.478}
        \newcommand{\cFiveMaxAF}{0.661} \newcommand{\cSixMaxAF}{0.635}

        \begin{subtable}{\textwidth}
            \centering
            \begin{tabular}{l c c c c c c}
                \toprule
                \textbf{Training / Test (Domain)} & \textbf{L} & \textbf{D} & \textbf{MH} & \textbf{S} & \textbf{MA} & \textbf{CS} \\
                \midrule
                CaseHOLD (\textbf{L}egal)
                & \gradcell{1}{\cOneMaxAF}{0.698}{0}
                & \gradcell{0}{\cTwoMaxAF}{0.339}{0}
                & \gradcell{0}{\cThreeMaxAF}{0.368}{0}
                & \gradcell{0}{\cFourMaxAF}{0.478}{0}
                & \gradcell{0}{\cFiveMaxAF}{0.661}{0}
                & \gradcell{0}{\cSixMaxAF}{0.635}{0} \\
                COM$^2$ (\textbf{D}etective)
                & \gradcell{0}{\cOneMaxAF}{0.693}{0}
                & \gradcell{1}{\cTwoMaxAF}{0.402}{0}
                & \gradcell{0}{\cThreeMaxAF}{0.243}{0}
                & \gradcell{0}{\cFourMaxAF}{0.135}{0}
                & \gradcell{0}{\cFiveMaxAF}{0.294}{0}
                & \gradcell{0}{\cSixMaxAF}{0.144}{0} \\
                MuSiQue (\textbf{M}ulti-\textbf{H}op)
                & \gradcell{0}{\cOneMaxAF}{0.678}{0}
                & \gradcell{0}{\cTwoMaxAF}{0.318}{0}
                & \gradcell{1}{\cThreeMaxAF}{0.280}{0}
                & \gradcell{0}{\cFourMaxAF}{0.372}{0}
                & \gradcell{0}{\cFiveMaxAF}{0.605}{0}
                & \gradcell{0}{\cSixMaxAF}{0.132}{0} \\
                SciBench (\textbf{S}cience)
                & \gradcell{0}{\cOneMaxAF}{0.692}{0}
                & \gradcell{0}{\cTwoMaxAF}{0.296}{0}
                & \gradcell{0}{\cThreeMaxAF}{0.197}{0}
                & \gradcell{1}{\cFourMaxAF}{0.341}{0}
                & \gradcell{0}{\cFiveMaxAF}{0.616}{0}
                & \gradcell{0}{\cSixMaxAF}{0.342}{0} \\
                TheoremQA (\textbf{Ma}th)
                & \gradcell{0}{\cOneMaxAF}{0.695}{0}
                & \gradcell{0}{\cTwoMaxAF}{0.337}{0}
                & \gradcell{0}{\cThreeMaxAF}{0.265}{0}
                & \gradcell{0}{\cFourMaxAF}{0.395}{0}
                & \gradcell{1}{\cFiveMaxAF}{0.607}{0}
                & \gradcell{0}{\cSixMaxAF}{0.530}{0} \\
                StrategyQA (\textbf{C}ommon\textbf{s}ense)
                & \gradcell{0}{\cOneMaxAF}{0.090}{0}
                & \gradcell{0}{\cTwoMaxAF}{0.093}{0}
                & \gradcell{0}{\cThreeMaxAF}{0.019}{0}
                & \gradcell{0}{\cFourMaxAF}{0.001}{0}
                & \gradcell{0}{\cFiveMaxAF}{0.156}{0}
                & \gradcell{1}{\cSixMaxAF}{0.630}{0} \\
                \midrule
                Multi-Domain Training
                & \gradcell{0}{\cOneMaxAF}{0.700}{0}
                & \gradcell{0}{\cTwoMaxAF}{0.274}{0}
                & \gradcell{0}{\cThreeMaxAF}{0.091}{0}
                & \gradcell{0}{\cFourMaxAF}{0.056}{0}
                & \gradcell{0}{\cFiveMaxAF}{0.070}{0}
                & \gradcell{0}{\cSixMaxAF}{0.127}{0} \\
                \bottomrule
            \end{tabular}
            \caption{AgentDropout}
        \end{subtable}
    \end{minipage}
    \hfill
    \begin{minipage}{0.48\textwidth}
        \centering
        \newcommand{\cOneMaxAF}{0.697} \newcommand{\cTwoMaxAF}{0.388}
        \newcommand{\cThreeMaxAF}{0.491} \newcommand{\cFourMaxAF}{0.427}
        \newcommand{\cFiveMaxAF}{0.636} \newcommand{\cSixMaxAF}{0.793}

        \begin{subtable}{\textwidth}
            \centering
            \begin{tabular}{l c c c c c c}
                \toprule
                \textbf{Training / Test (Domain)} & \textbf{L} & \textbf{D} & \textbf{MH} & \textbf{S} & \textbf{MA} & \textbf{CS} \\
                \midrule
                CaseHOLD (\textbf{L}egal)
                & \gradcell{1}{\cOneMaxAF}{0.697}{0}
                & \gradcell{0}{\cTwoMaxAF}{0.370}{0}
                & \gradcell{0}{\cThreeMaxAF}{0.454}{0}
                & \gradcell{0}{\cFourMaxAF}{0.360}{0}
                & \gradcell{0}{\cFiveMaxAF}{0.548}{0}
                & \gradcell{0}{\cSixMaxAF}{0.689}{0} \\
                COM$^2$ (\textbf{D}etective)
                & \gradcell{0}{\cOneMaxAF}{0.667}{0}
                & \gradcell{1}{\cTwoMaxAF}{0.388}{0}
                & \gradcell{0}{\cThreeMaxAF}{0.491}{0}
                & \gradcell{0}{\cFourMaxAF}{0.427}{0}
                & \gradcell{0}{\cFiveMaxAF}{0.636}{0}
                & \gradcell{0}{\cSixMaxAF}{0.793}{0} \\
                MuSiQue (\textbf{M}ulti-\textbf{H}op)
                & \gradcell{0}{\cOneMaxAF}{0.686}{0}
                & \gradcell{0}{\cTwoMaxAF}{0.368}{0}
                & \gradcell{1}{\cThreeMaxAF}{0.436}{0}
                & \gradcell{0}{\cFourMaxAF}{0.393}{0}
                & \gradcell{0}{\cFiveMaxAF}{0.613}{0}
                & \gradcell{0}{\cSixMaxAF}{0.694}{0} \\
                SciBench (\textbf{S}cience)
                & \gradcell{0}{\cOneMaxAF}{0.697}{0}
                & \gradcell{0}{\cTwoMaxAF}{0.306}{0}
                & \gradcell{0}{\cThreeMaxAF}{0.382}{0}
                & \gradcell{1}{\cFourMaxAF}{0.391}{0}
                & \gradcell{0}{\cFiveMaxAF}{0.592}{0}
                & \gradcell{0}{\cSixMaxAF}{0.610}{0} \\
                TheoremQA (\textbf{Ma}th)
                & \gradcell{0}{\cOneMaxAF}{0.678}{0}
                & \gradcell{0}{\cTwoMaxAF}{0.345}{0}
                & \gradcell{0}{\cThreeMaxAF}{0.337}{0}
                & \gradcell{0}{\cFourMaxAF}{0.297}{0}
                & \gradcell{1}{\cFiveMaxAF}{0.509}{0}
                & \gradcell{0}{\cSixMaxAF}{0.700}{0} \\
                StrategyQA (\textbf{C}ommon\textbf{s}ense)
                & \gradcell{0}{\cOneMaxAF}{0.692}{0}
                & \gradcell{0}{\cTwoMaxAF}{0.376}{0}
                & \gradcell{0}{\cThreeMaxAF}{0.394}{0}
                & \gradcell{0}{\cFourMaxAF}{0.303}{0}
                & \gradcell{0}{\cFiveMaxAF}{0.493}{0}
                & \gradcell{1}{\cSixMaxAF}{0.693}{0} \\
                \midrule
                Multi-Domain Training
                & \gradcell{0}{\cOneMaxAF}{0.614}{0}
                & \gradcell{0}{\cTwoMaxAF}{0.367}{0}
                & \gradcell{0}{\cThreeMaxAF}{0.385}{0}
                & \gradcell{0}{\cFourMaxAF}{0.307}{0}
                & \gradcell{0}{\cFiveMaxAF}{0.537}{0}
                & \gradcell{0}{\cSixMaxAF}{0.555}{0} \\
                \bottomrule
            \end{tabular}
            \caption{AFlow}
        \end{subtable}
    \end{minipage}

    \vspace{-2mm}
    \caption{
    In-domain and OOD performance of AgentDropout and AFlow on \texttt{Qwen3-30B-A3B}, formatted identically to Table~\ref{tab:main_table}.
    Cell colors are normalized per column by the column-wise maximum: values $\ge 95\%$ max are shaded in \textcolor{blue}{blue}, and values $< 70\%$ max are in \textcolor{red}{red}.
    This table is provided for cross-model comparison with the main-paper \texttt{GPT-oss-20B} results.
    }
    \label{tab:qwen_main_table}
\end{table*}
\begin{table*}[!t]
    \centering
    \scriptsize
    \renewcommand{\arraystretch}{1.15}
    {\begin{minipage}{0.48\textwidth}
    \centering
    \newcommand{\cOneMaxAF}{1}
    \newcommand{\cTwoMaxAF}{1}
    \newcommand{\cThreeMaxAF}{1}
    \newcommand{\cFourMaxAF}{1}
    \newcommand{\cFiveMaxAF}{1}
    \newcommand{\cSixMaxAF}{1}
    \newcommand{\cSevenMaxAF}{1}
    \begin{subtable}{\textwidth}
        \centering
        \setlength{\tabcolsep}{2pt}
        \begin{tabular}{l c c c c c c}
            \toprule
            \textbf{Training / Test (Domain)} & \textbf{L} & \textbf{D} & \textbf{MH} & \textbf{S} & \textbf{MA} & \textbf{CS} \\
            \midrule
            CaseHOLD(\textbf{L}egal) &
            \gradcellStructural{1}{\cOneMaxAF}{1.0000}{0} &
            \gradcellStructural{0}{\cOneMaxAF}{0.3329}{0} &
            \gradcellStructural{0}{\cOneMaxAF}{0.1184}{0} &
            \gradcellStructural{0}{\cOneMaxAF}{0.1908}{0} &
            \gradcellStructural{0}{\cOneMaxAF}{0.1421}{0} &
            \gradcellStructural{0}{\cOneMaxAF}{0.4289}{0} \\
            COM$^2$(\textbf{D}etective) &
            \gradcellStructural{0}{\cTwoMaxAF}{0.2209}{0} &
            \gradcellStructural{1}{\cTwoMaxAF}{1.0000}{0} &
            \gradcellStructural{0}{\cTwoMaxAF}{0.2282}{0} &
            \gradcellStructural{0}{\cTwoMaxAF}{0.2398}{0} &
            \gradcellStructural{0}{\cTwoMaxAF}{0.2820}{0} &
            \gradcellStructural{0}{\cTwoMaxAF}{0.3706}{0} \\
            MuSiQue(\textbf{M}ulti-\textbf{H}op) &
            \gradcellStructural{0}{\cThreeMaxAF}{0.7882}{0} &
            \gradcellStructural{0}{\cThreeMaxAF}{0.9603}{0} &
            \gradcellStructural{1}{\cThreeMaxAF}{0.8725}{0} &
            \gradcellStructural{0}{\cThreeMaxAF}{1.0000}{0} &
            \gradcellStructural{0}{\cThreeMaxAF}{0.9632}{0} &
            \gradcellStructural{0}{\cThreeMaxAF}{0.8245}{0} \\
            SciBench(\textbf{S}cience) &
            \gradcellStructural{0}{\cFourMaxAF}{0.8156}{0} &
            \gradcellStructural{0}{\cFourMaxAF}{0.8332}{0} &
            \gradcellStructural{0}{\cFourMaxAF}{0.8002}{0} &
            \gradcellStructural{1}{\cFourMaxAF}{0.9851}{0} &
            \gradcellStructural{0}{\cFourMaxAF}{1.0000}{0} &
            \gradcellStructural{0}{\cFourMaxAF}{0.9323}{0} \\
            TheoremQA(\textbf{Ma}th) &
            \gradcellStructural{0}{\cFiveMaxAF}{0.7381}{0} &
            \gradcellStructural{0}{\cFiveMaxAF}{0.7442}{0} &
            \gradcellStructural{0}{\cFiveMaxAF}{0.5518}{0} &
            \gradcellStructural{0}{\cFiveMaxAF}{0.7479}{0} &
            \gradcellStructural{1}{\cFiveMaxAF}{1.0000}{0} &
            \gradcellStructural{0}{\cFiveMaxAF}{0.9050}{0} \\
            StrategyQA(\textbf{C}ommon\textbf{s}ense) &
            \gradcellStructural{0}{\cSixMaxAF}{0.8040}{0} &
            \gradcellStructural{0}{\cSixMaxAF}{0.8436}{0} &
            \gradcellStructural{0}{\cSixMaxAF}{0.5446}{0} &
            \gradcellStructural{0}{\cSixMaxAF}{0.8022}{0} &
            \gradcellStructural{0}{\cSixMaxAF}{0.8620}{0} &
            \gradcellStructural{1}{\cSixMaxAF}{1.0000}{0} \\
            \midrule
            Multi-Domain Training &
            \gradcellStructural{0}{\cSevenMaxAF}{0.9386}{0} &
            \gradcellStructural{0}{\cSevenMaxAF}{0.9358}{0} &
            \gradcellStructural{0}{\cSevenMaxAF}{0.5562}{0} &
            \gradcellStructural{0}{\cSevenMaxAF}{0.6110}{0} &
            \gradcellStructural{0}{\cSevenMaxAF}{0.6733}{0} &
            \gradcellStructural{0}{\cSevenMaxAF}{1.0000}{0} \\
            \bottomrule
        \end{tabular}
        \label{tab:qwen3_role_alignment}
    \end{subtable}
\end{minipage}}
    {\begin{minipage}{0.48\textwidth}
    \centering
    \newcommand{\cOneMaxAF}{1}
    \newcommand{\cTwoMaxAF}{1}
    \newcommand{\cThreeMaxAF}{1}
    \newcommand{\cFourMaxAF}{1}
    \newcommand{\cFiveMaxAF}{1}
    \newcommand{\cSixMaxAF}{1}
    \newcommand{\cSevenMaxAF}{1}
    \begin{subtable}{\textwidth}
        \centering
        \setlength{\tabcolsep}{2pt}
        \begin{tabular}{l c c c c c c}
            \toprule
            \textbf{Training / Test (Domain)} & \textbf{L} & \textbf{D} & \textbf{MH} & \textbf{S} & \textbf{MA} & \textbf{CS} \\
            \midrule
            CaseHOLD(\textbf{L}egal) &
            \gradcellStructural{1}{\cOneMaxAF}{1.0000}{0} &
            \gradcellStructural{0}{\cOneMaxAF}{-0.4845}{0} &
            \gradcellStructural{0}{\cOneMaxAF}{-0.9332}{0} &
            \gradcellStructural{0}{\cOneMaxAF}{-0.7088}{0} &
            \gradcellStructural{0}{\cOneMaxAF}{-0.8115}{0} &
            \gradcellStructural{0}{\cOneMaxAF}{-0.6516}{0} \\
            COM$^2$(\textbf{D}etective) &
            \gradcellStructural{0}{\cTwoMaxAF}{-0.0072}{0} &
            \gradcellStructural{1}{\cTwoMaxAF}{1.0000}{0} &
            \gradcellStructural{0}{\cTwoMaxAF}{-1.2093}{0} &
            \gradcellStructural{0}{\cTwoMaxAF}{-0.6085}{0} &
            \gradcellStructural{0}{\cTwoMaxAF}{-0.6008}{0} &
            \gradcellStructural{0}{\cTwoMaxAF}{-0.3527}{0} \\
            MuSiQue(\textbf{M}ulti-\textbf{H}op) &
            \gradcellStructural{0}{\cThreeMaxAF}{-0.8179}{0} &
            \gradcellStructural{0}{\cThreeMaxAF}{-1.0000}{0} &
            \gradcellStructural{1}{\cThreeMaxAF}{-0.7134}{0} &
            \gradcellStructural{0}{\cThreeMaxAF}{-0.9045}{0} &
            \gradcellStructural{0}{\cThreeMaxAF}{-0.8478}{0} &
            \gradcellStructural{0}{\cThreeMaxAF}{-0.9224}{0} \\
            SciBench(\textbf{S}cience) &
            \gradcellStructural{0}{\cFourMaxAF}{-0.1993}{0} &
            \gradcellStructural{0}{\cFourMaxAF}{0.3188}{0} &
            \gradcellStructural{0}{\cFourMaxAF}{-0.0036}{0} &
            \gradcellStructural{1}{\cFourMaxAF}{1.0000}{0} &
            \gradcellStructural{0}{\cFourMaxAF}{0.9420}{0} &
            \gradcellStructural{0}{\cFourMaxAF}{0.3768}{0} \\
            TheoremQA(\textbf{Ma}th) &
            \gradcellStructural{0}{\cFiveMaxAF}{-0.1351}{0} &
            \gradcellStructural{0}{\cFiveMaxAF}{0.5086}{0} &
            \gradcellStructural{0}{\cFiveMaxAF}{0.2701}{0} &
            \gradcellStructural{0}{\cFiveMaxAF}{1.0000}{0} &
            \gradcellStructural{1}{\cFiveMaxAF}{0.9943}{0} &
            \gradcellStructural{0}{\cFiveMaxAF}{0.5431}{0} \\
            StrategyQA(\textbf{C}ommon\textbf{s}ense) &
            \gradcellStructural{0}{\cSixMaxAF}{0.3603}{0} &
            \gradcellStructural{0}{\cSixMaxAF}{0.7500}{0} &
            \gradcellStructural{0}{\cSixMaxAF}{0.9596}{0} &
            \gradcellStructural{0}{\cSixMaxAF}{-0.0110}{0} &
            \gradcellStructural{0}{\cSixMaxAF}{-0.1324}{0} &
            \gradcellStructural{1}{\cSixMaxAF}{1.0000}{0} \\
            \midrule
            Multi-Domain Training &
            \gradcellStructural{0}{\cSevenMaxAF}{1.0000}{0} &
            \gradcellStructural{0}{\cSevenMaxAF}{0.9946}{0} &
            \gradcellStructural{0}{\cSevenMaxAF}{-0.8098}{0} &
            \gradcellStructural{0}{\cSevenMaxAF}{0.8967}{0} &
            \gradcellStructural{0}{\cSevenMaxAF}{0.9348}{0} &
            \gradcellStructural{0}{\cSevenMaxAF}{0.4457}{0} \\
            \bottomrule
        \end{tabular}
        \label{tab:qwen3_connection_significance}
    \end{subtable}
\end{minipage}}
    \caption{
    Illusory coordination of AgentDropout on \texttt{Qwen3-30B-A3B} detected by $\mathcal{R}$ and $\mathcal{O}$. 
    All entries are row-wise normalized by the maximum value in each row (i.e., each cell reports value / max(row)); for the MuSiQue row in the connection metric $\mathcal{O}$, where all entries are negative, row-wise absolute maximum normalization is used instead. 
    Cell colors follow the normalized ratios: values $\ge 0.70$ are shaded \textcolor{blue}{blue} (i.e., successful transfer), values $< 0.70$ are shaded \textcolor{red}{red} (i.e., failed transfer). 
    In-domain results are in bold. 
    Results provide new insights into MAS dynamics. 
    }
    \label{tab:new_matric_qwen}
\end{table*}

\subsection{Analysis of Qwen3-30B-A3B Results}
\label{sec:qwen_results}

\paragraph{Generality of Topological Overfitting}
The results presented in Table~\ref{tab:qwen_main_table} demonstrate that the limitations of single-domain optimization observed in the GPT-oss-20B environment also persist in the Qwen3-30B-A3B setting. Specifically, when the AgentDropout algorithm is optimized for the Legal domain, it achieves a solid in-domain performance of 69.7\%. However, when this same configuration is transferred to out-of-distribution (OOD) domains such as Detective or Science, the accuracy plummets to 30.6\% and 39.1\%, respectively. In certain configurations, this performance degradation is even more pronounced than what was observed with GPT-oss-20B. These findings empirically validate that `topological overfitting' is not a phenomenon restricted to a specific base LLM but is a pervasive challenge across adaptive multi-agent systems (MAS), regardless of the underlying model or the training domain.

\paragraph{Quantitative Evidence of Illusory Coordination}
While some domain transfer settings might appear successful based on final accuracy, a quantitative analysis of internal system metrics reveals a breakdown in coordination, a phenomenon we term `illusory coordination'. This is clearly supported by the `Connection Significance ($\mathcal{O}$)' analysis in Table~\ref{tab:new_matric_qwen}. For instance, when a topology optimized on the Multi-Hop (MuSiQue) domain is transferred, the utility of information exchange between agents—measured by connection significance—recorded negative values across all tested domains, such as -1.00 in the Detective domain and -0.92 in the Commonsense domain. These negative scores indicate that the messages exchanged between agents do not meaningfully contribute to the final answer. This highlights that seemingly high performance in OOD tasks is not a result of generalized MAS coordination, but rather a reliance on the strong standalone reasoning capabilities of the underlying Qwen3 model, masking the failure of the intended multi-agent collaboration.
\subsection{Training Set Size Impact Analysis}\label{sec:training_set_size_impact}
\begin{figure}[t]
    \centering
    \includegraphics[width=\linewidth]{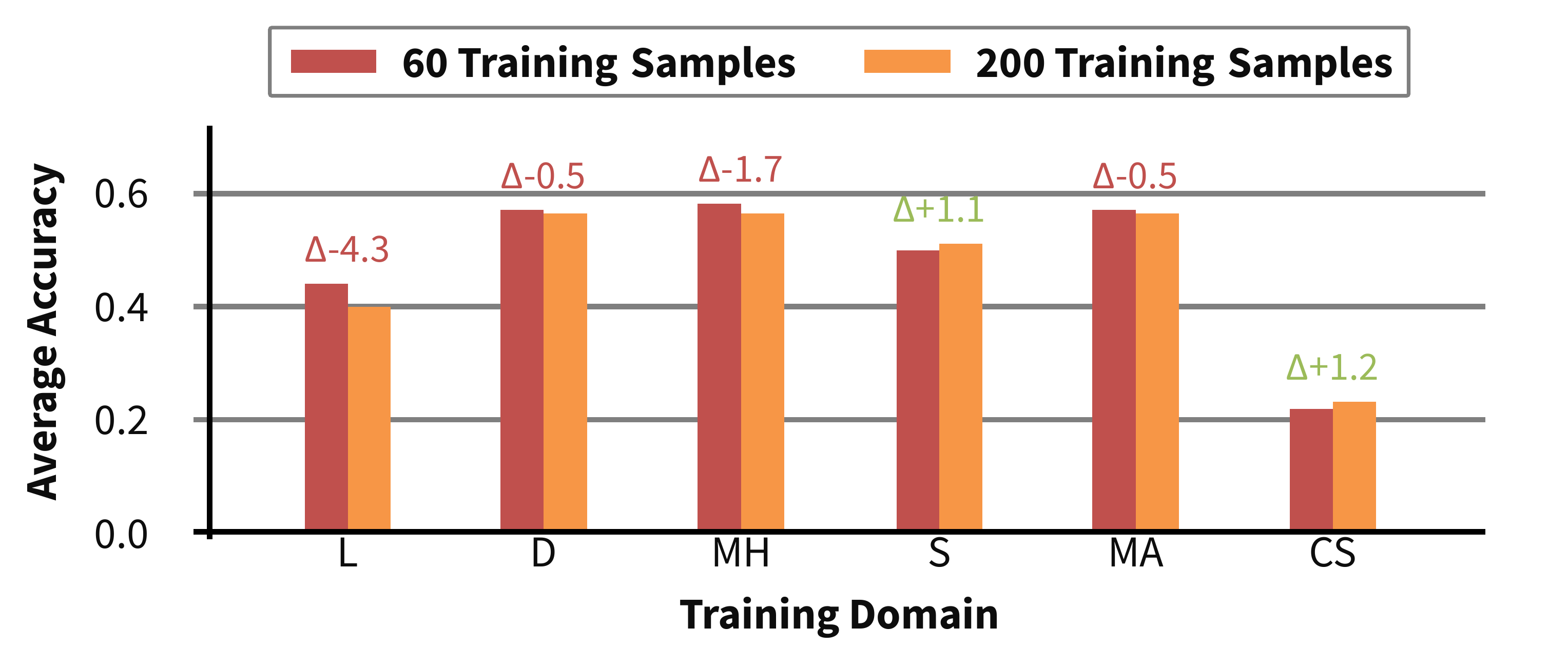}
    \caption{
    AgentDropout generalization performance under different training-set sizes.
    Here, generalization performance denotes the mean score across test domains for a fixed training domain, and $\Delta$ denotes the change in this mean score when the training set increases from 60 to 200 examples.
    }
    \label{fig:training_size}
\end{figure}
Figure~\ref{fig:training_size} provides a complementary view of training-set size for AgentDropout with \texttt{GPT-oss-20B}. 
Increasing the single-domain training budget from 60 to 200 examples yields only modest gains in average held-out accuracy for SciBench (+1.1) and StrategyQA (+1.2). 
However, the pattern is not consistent: performance declines for COM$^2$ (-4.3), MuSiQue (-1.7), and both CaseHOLD and TheoremQA (-0.5 each). 
These mixed results suggest that simply adding more in-domain examples does not reliably improve cross-domain generalization.

\subsection{Interchange Ablation}\label{sec:results_interchange}

The right block of Table~\ref{tab:rebuttal_analysis} reports an interchange ablation for \textbf{AgentDropout} with \textbf{GPT-oss-20B}, designed to isolate the relative contribution of roles and connections in the learned topology.
In \textit{Connection-OOD}, agent roles are fixed to the test domain while the inter-agent connections are swapped; in \textit{Role-OOD}, the connections are fixed while the agent roles are swapped.

Across most datasets, changing roles under \textit{Role-OOD} causes substantially larger degradation than changing connections under \textit{Connection-OOD}, suggesting that learned role assignments are generally a more performance-critical component of the topology.
MuSiQue is a notable exception: under \textit{Connection-OOD}, performance drops by 5.36 points, compared with an average drop of 0.418 points across the other five datasets.
This pattern is consistent with the stronger \textbf{Acc--$\mathcal{O}$} trend observed for MuSiQue in Table~\ref{tab:rebuttal_analysis}, suggesting that connection structure matters more for communication-intensive multi-hop reasoning.
\subsection{Three-Run Results for Cross-Domain Generalization}
Adaptive MAS methods can converge to different collaboration topologies due to training data and stochastic optimization. 
To assess robustness, we therefore report results from three independent train--test runs for each setting. 
Detailed per-run performance for \texttt{GPT-oss-20B} is provided in Table~\ref{tab:performance_measure_appendix}, while the corresponding results for \texttt{Qwen3-30B-A3B} are provided in Table~\ref{tab:performance_measure_appendix_raw}.

\subsection{Three-Run Results for Diagnosing Illusory Coordination}
\label{sec:appendix_internal_failure}
For the same reason, we also report three-run outcomes of our structural diagnostics to verify that \emph{Illusory Coordination} is reproducible rather than seed-specific.
Table~\ref{tab:structural_failure_appendix} presents the per-run results for AgentDropout with \texttt{GPT-oss-20B} across all transfer settings, while Table~\ref{tab:new_matric_qwen_three_run} reports the corresponding results for AgentDropout with \texttt{Qwen3-30B-A3B}.

\section{Multi-Agent System Failure Taxonomy}\label{sec:mast_taxonomy}
Table~\ref{tab:mast_taxonomy} summarizes the MAST categories and failure-mode definitions used throughout our analysis.

To systematically describe error patterns observed in MAS execution traces, we adopt the \textbf{Multi-Agent System Failure Taxonomy (MAST)} introduced by Cemri et al., an empirically derived taxonomy built from large-scale analyses of MAS failure traces.
MAST organizes failures into \textbf{14 failure modes} under three broad categories: \textit{System Design Issues} (violations of task/role specifications or missing termination conditions), \textit{Inter-Agent Misalignment} (breakdowns in coordination such as conversation resets, task derailment, or ignoring other agents' inputs), and \textit{Task Verification} (insufficient or incorrect verification of intermediate/final results).
In addition, each failure mode is associated with when it typically emerges in the end-to-end MAS pipeline (pre-execution, execution, and post-execution), highlighting that some issues can span multiple stages and propagate through later interactions.

\begin{table*}[ht]
    \centering
    \small
    \begin{tabular}{lp{4.5cm}p{7.5cm}}
    \toprule
    \textbf{Broad Category} & \textbf{Failure Mode} & \textbf{Description} \\ \midrule
    
    \multirow{6}{*}{\makecell[l]{\textbf{System Design} \\ \textbf{Issues}}} 
     & \textbf{FM-1.1: Disobey task}\newline\hspace*{3.8em}\textbf{specification} & Failure to adhere to specified constraints or requirements. \\ \cmidrule{2-3}
     & \textbf{FM-1.2: Disobey role}\newline\hspace*{3.8em}\textbf{specification} & Failure to adhere to defined responsibilities of its role. \\ \cmidrule{2-3}
     & \textbf{FM-1.3: Step repetition} & Unnecessary reiteration of previously completed steps. \\ \cmidrule{2-3}
     & \textbf{FM-1.4: Loss of conversation}\newline\hspace*{3.8em}\textbf{history} & Unexpected context truncation, reverting to previous state. \\ \cmidrule{2-3}
     & \textbf{FM-1.5: Unaware of }\newline \hspace*{3.8em}\textbf{terminal condition} & Lack of recognition for criteria that triggers interaction end. \\ \midrule
    
    \multirow{7}{*}{\makecell[l]{\textbf{Inter-Agent} \\ \textbf{Misalignment}}} 
     & \textbf{FM-2.1: Conversation reset} & Unexpected restarting of a dialogue, losing progress. \\ \cmidrule{2-3}
     & \textbf{FM-2.2: Fail to ask clarification} & Inability to request info when faced with unclear data. \\ \cmidrule{2-3}
     & \textbf{FM-2.3: Task derailment} & Deviation from the intended objective of a given task. \\ \cmidrule{2-3}
     & \textbf{FM-2.4: Information}\newline \hspace*{3.8em}\textbf{withholding} & Failure to share important data impacting decision-making. \\ \cmidrule{2-3}
     & \textbf{FM-2.5: Ignored agent input} & Disregarding input provided by other agents in the system. \\ \cmidrule{2-3}
     & \makecell[l]{\textbf{FM-2.6: Reasoning-action} \\ \hspace*{3.5em} \textbf{mismatch}} & Discrepancy between reasoning and actual actions taken. \\ \midrule
    
    \multirow{4}{*}{\makecell[l]{\textbf{Task} \\ \textbf{Verification}}} 
     & \textbf{FM-3.1: Premature termination} & Ending interaction before objectives are met. \\ \cmidrule{2-3}
     & \textbf{FM-3.2: No/incomplete} \newline \hspace*{3.5em} \textbf{verification} & Omission of checking of task outcomes or system outputs. \\ \cmidrule{2-3}
     & \textbf{FM-3.3: Incorrect verification} & Failure to adequately validate information during iterations. \\ \bottomrule
     
    \end{tabular}
    \caption{Taxonomy of Multi-Agent System Failure Modes (MAST Framework)}
    \label{tab:mast_taxonomy}
\end{table*}
\begin{table*}[!t]
    \centering
    \scriptsize
    \renewcommand{\arraystretch}{1.3}
    \setlength{\tabcolsep}{2.7pt} 


    \vspace{2em}

    \begin{minipage}{0.49\textwidth}
        \centering
        \newcommand{\RoneAFOne}{0.648} \newcommand{\RoneAFTwo}{0.506} 
        \newcommand{\RoneAFThree}{0.585} \newcommand{\RoneAFFour}{0.442} 
        \newcommand{\RoneAFFive}{0.656} \newcommand{\RoneAFSix}{0.755}
        \begin{subtable}{\textwidth}
            \centering
            \begin{tabular}{l c c c c c c}
                \toprule
                \textbf{Training / Test (Domain)} & \textbf{C} & \textbf{D} & \textbf{MH} & \textbf{S} & \textbf{MA} & \textbf{C} \\
                \midrule
                CaseHOLD(\textbf{L}egal)         & \gradcell{0}{\RoneAFOne}{0.639}{0} & \gradcell{0}{\RoneAFTwo}{0.427}{0} & \gradcell{0}{\RoneAFThree}{0.585}{0} & \gradcell{0}{\RoneAFFour}{0.442}{0} & \gradcell{0}{\RoneAFFive}{0.656}{0} & \gradcell{0}{\RoneAFSix}{0.690}{0} \\
                COM$^2$(\textbf{D}etective)      & \gradcell{0}{\RoneAFOne}{0.513}{0} & \gradcell{0}{\RoneAFTwo}{0.480}{0} & \gradcell{0}{\RoneAFThree}{0.528}{0} & \gradcell{0}{\RoneAFFour}{0.343}{0} & \gradcell{0}{\RoneAFFive}{0.544}{0} & \gradcell{0}{\RoneAFSix}{0.170}{0} \\
                MuSiQue(\textbf{M}ulti-\textbf{H}op) & \gradcell{0}{\RoneAFOne}{0.648}{0} & \gradcell{0}{\RoneAFTwo}{0.506}{0} & \gradcell{0}{\RoneAFThree}{0.580}{0} & \gradcell{0}{\RoneAFFour}{0.403}{0} & \gradcell{0}{\RoneAFFive}{0.652}{0} & \gradcell{0}{\RoneAFSix}{0.742}{0} \\
                SciBench(\textbf{S}cience)       & \gradcell{0}{\RoneAFOne}{0.612}{0} & \gradcell{0}{\RoneAFTwo}{0.345}{0} & \gradcell{0}{\RoneAFThree}{0.543}{0} & \gradcell{0}{\RoneAFFour}{0.378}{0} & \gradcell{0}{\RoneAFFive}{0.631}{0} & \gradcell{0}{\RoneAFSix}{0.485}{0} \\
                TheoremQA(\textbf{Ma}th)         & \gradcell{0}{\RoneAFOne}{0.622}{0} & \gradcell{0}{\RoneAFTwo}{0.471}{0} & \gradcell{0}{\RoneAFThree}{0.558}{0} & \gradcell{0}{\RoneAFFour}{0.361}{0} & \gradcell{0}{\RoneAFFive}{0.640}{0} & \gradcell{0}{\RoneAFSix}{0.738}{0} \\
                StrategyQA(\textbf{C}ommonsense) & \gradcell{0}{\RoneAFOne}{0.005}{0} & \gradcell{0}{\RoneAFTwo}{0.004}{0} & \gradcell{0}{\RoneAFThree}{0.417}{0} & \gradcell{0}{\RoneAFFour}{0.002}{0} & \gradcell{0}{\RoneAFFive}{0.153}{0} & \gradcell{0}{\RoneAFSix}{0.707}{0} \\
                \midrule
                Multi-Domain Training & \gradcell{0}{\RoneAFOne}{0.607}{0} & \gradcell{0}{\RoneAFTwo}{0.459}{0} & \gradcell{0}{\RoneAFThree}{0.541}{0} & \gradcell{0}{\RoneAFFour}{0.411}{0} & \gradcell{0}{\RoneAFFive}{0.643}{0} & \gradcell{0}{\RoneAFSix}{0.755}{0} \\
                \bottomrule
            \end{tabular}
            \caption{Run 1: AgentDropout Results}
        \end{subtable}
    \end{minipage}
    \hfill
    \begin{minipage}{0.49\textwidth}
        \centering
        \newcommand{\RoneDOOne}{0.635} \newcommand{\RoneDOTwo}{0.429} 
        \newcommand{\RoneDOThree}{0.588} \newcommand{\RoneDOFour}{0.438} 
        \newcommand{\RoneDOFive}{0.669} \newcommand{\RoneDOSix}{0.773}

        \begin{subtable}{\textwidth}
            \centering
            \begin{tabular}{l c c c c c c}
                \toprule
                \textbf{Training / Test (Domain)} & \textbf{C} & \textbf{D} & \textbf{MH} & \textbf{S} & \textbf{MA} & \textbf{C} \\
                \midrule
                CaseHOLD(\textbf{L}egal)  & \gradcell{0}{\RoneDOOne}{0.612}{0} & \gradcell{0}{\RoneDOTwo}{0.397}{0} & \gradcell{0}{\RoneDOThree}{0.452}{0} & \gradcell{0}{\RoneDOFour}{0.417}{0} & \gradcell{0}{\RoneDOFive}{0.651}{0} & \gradcell{0}{\RoneDOSix}{0.699}{0} \\
                COM$^2$(\textbf{D}etective)    & \gradcell{0}{\RoneDOOne}{0.635}{0} & \gradcell{0}{\RoneDOTwo}{0.429}{0} & \gradcell{0}{\RoneDOThree}{0.519}{0} & \gradcell{0}{\RoneDOFour}{0.399}{0} & \gradcell{0}{\RoneDOFive}{0.649}{0} & \gradcell{0}{\RoneDOSix}{0.773}{0} \\
                MuSiQue(\textbf{M}ulti-\textbf{H}op)    & \gradcell{0}{\RoneDOOne}{0.446}{0} & \gradcell{0}{\RoneDOTwo}{0.419}{0} & \gradcell{0}{\RoneDOThree}{0.482}{0} & \gradcell{0}{\RoneDOFour}{0.388}{0} & \gradcell{0}{\RoneDOFive}{0.644}{0} & \gradcell{0}{\RoneDOSix}{0.751}{0} \\
                SciBench(\textbf{S}cience)   & \gradcell{0}{\RoneDOOne}{0.433}{0} & \gradcell{0}{\RoneDOTwo}{0.418}{0} & \gradcell{0}{\RoneDOThree}{0.469}{0} & \gradcell{0}{\RoneDOFour}{0.391}{0} & \gradcell{0}{\RoneDOFive}{0.530}{0} & \gradcell{0}{\RoneDOSix}{0.769}{0} \\
                TheoremQA(\textbf{Ma}th)  & \gradcell{0}{\RoneDOOne}{0.392}{0} & \gradcell{0}{\RoneDOTwo}{0.397}{0} & \gradcell{0}{\RoneDOThree}{0.444}{0} & \gradcell{0}{\RoneDOFour}{0.297}{0} & \gradcell{0}{\RoneDOFive}{0.555}{0} & \gradcell{0}{\RoneDOSix}{0.699}{0} \\
                StrategyQA(\textbf{C}ommonsense) & \gradcell{0}{\RoneDOOne}{0.503}{0} & \gradcell{0}{\RoneDOTwo}{0.340}{0} & \gradcell{0}{\RoneDOThree}{0.388}{0} & \gradcell{0}{\RoneDOFour}{0.357}{0} & \gradcell{0}{\RoneDOFive}{0.616}{0} & \gradcell{0}{\RoneDOSix}{0.764}{0} \\
                \midrule
                Multi-Domain Training & \gradcell{0}{\RoneDOOne}{0.614}{0} & \gradcell{0}{\RoneDOTwo}{0.422}{0} & \gradcell{0}{\RoneDOThree}{0.588}{0} & \gradcell{0}{\RoneDOFour}{0.438}{0} & \gradcell{0}{\RoneDOFive}{0.669}{0} & \gradcell{0}{\RoneDOSix}{0.725}{0} \\
                \bottomrule
            \end{tabular}
            \caption{Run 1: AFlow Results}
        \end{subtable}
    \end{minipage}


    \vspace{3em}

    \begin{minipage}{0.49\textwidth}
        \centering
        \newcommand{\RtwoAFOne}{0.643} \newcommand{\RtwoAFTwo}{0.483} 
        \newcommand{\RtwoAFThree}{0.598} \newcommand{\RtwoAFFour}{0.411} 
        \newcommand{\RtwoAFFive}{0.663} \newcommand{\RtwoAFSix}{0.760}

        \begin{subtable}{\textwidth}
            \centering
            \begin{tabular}{l c c c c c c}
                \toprule
                \textbf{Training / Test (Domain)} & \textbf{C} & \textbf{D} & \textbf{MH} & \textbf{S} & \textbf{MA} & \textbf{C} \\
                \midrule
                CaseHOLD(\textbf{L}egal)   & \gradcell{0}{\RtwoAFOne}{0.643}{0} & \gradcell{0}{\RtwoAFTwo}{0.483}{0} & \gradcell{0}{\RtwoAFThree}{0.560}{0} & \gradcell{0}{\RtwoAFFour}{0.411}{0} & \gradcell{0}{\RtwoAFFive}{0.663}{0} & \gradcell{0}{\RtwoAFSix}{0.703}{0} \\
                COM$^2$(\textbf{D}etective)    & \gradcell{0}{\RtwoAFOne}{0.558}{0} & \gradcell{0}{\RtwoAFTwo}{0.483}{0} & \gradcell{0}{\RtwoAFThree}{0.546}{0} & \gradcell{0}{\RtwoAFFour}{0.361}{0} & \gradcell{0}{\RtwoAFFive}{0.528}{0} & \gradcell{0}{\RtwoAFSix}{0.170}{0} \\
                MuSiQue(\textbf{M}ulti-\textbf{H}op)  & \gradcell{0}{\RtwoAFOne}{0.619}{0} & \gradcell{0}{\RtwoAFTwo}{0.473}{0} & \gradcell{0}{\RtwoAFThree}{0.583}{0} & \gradcell{0}{\RtwoAFFour}{0.389}{0} & \gradcell{0}{\RtwoAFFive}{0.636}{0} & \gradcell{0}{\RtwoAFSix}{0.747}{0} \\
                SciBench(\textbf{S}cience)  & \gradcell{0}{\RtwoAFOne}{0.624}{0} & \gradcell{0}{\RtwoAFTwo}{0.335}{0} & \gradcell{0}{\RtwoAFThree}{0.560}{0} & \gradcell{0}{\RtwoAFFour}{0.392}{0} & \gradcell{0}{\RtwoAFFive}{0.627}{0} & \gradcell{0}{\RtwoAFSix}{0.476}{0} \\
                TheoremQA(\textbf{Ma}th)  & \gradcell{0}{\RtwoAFOne}{0.625}{0} & \gradcell{0}{\RtwoAFTwo}{0.475}{0} & \gradcell{0}{\RtwoAFThree}{0.598}{0} & \gradcell{0}{\RtwoAFFour}{0.365}{0} & \gradcell{0}{\RtwoAFFive}{0.643}{0} & \gradcell{0}{\RtwoAFSix}{0.747}{0} \\
                StrategyQA(\textbf{C}ommonsense) & \gradcell{0}{\RtwoAFOne}{0.007}{0} & \gradcell{0}{\RtwoAFTwo}{0.005}{0} & \gradcell{0}{\RtwoAFThree}{0.420}{0} & \gradcell{0}{\RtwoAFFour}{0.002}{0} & \gradcell{0}{\RtwoAFFive}{0.162}{0} & \gradcell{0}{\RtwoAFSix}{0.760}{0} \\
                \midrule
                Multi-Domain Training & \gradcell{0}{\RtwoAFOne}{0.592}{0} & \gradcell{0}{\RtwoAFTwo}{0.465}{0} & \gradcell{0}{\RtwoAFThree}{0.514}{0} & \gradcell{0}{\RtwoAFFour}{0.409}{0} & \gradcell{0}{\RtwoAFFive}{0.645}{0} & \gradcell{0}{\RtwoAFSix}{0.751}{0} \\
                \bottomrule
            \end{tabular}
            \caption{Run 2: AgentDropout Results}
        \end{subtable}
    \end{minipage}
    \hfill
    \begin{minipage}{0.49\textwidth}
        \centering
        \newcommand{\RtwoDOOne}{0.630} \newcommand{\RtwoDOTwo}{0.424} 
        \newcommand{\RtwoDOThree}{0.491} \newcommand{\RtwoDOFour}{0.402} 
        \newcommand{\RtwoDOFive}{0.670} \newcommand{\RtwoDOSix}{0.769}

        \begin{subtable}{\textwidth}
            \centering
            \begin{tabular}{l c c c c c c}
                \toprule
                \textbf{Training / Test (Domain)} & \textbf{C} & \textbf{D} & \textbf{MH} & \textbf{S} & \textbf{MA} & \textbf{C} \\
                \midrule
                CaseHOLD(\textbf{L}egal)   & \gradcell{0}{\RtwoDOOne}{0.630}{0} & \gradcell{0}{\RtwoDOTwo}{0.423}{0} & \gradcell{0}{\RtwoDOThree}{0.491}{0} & \gradcell{0}{\RtwoDOFour}{0.397}{0} & \gradcell{0}{\RtwoDOFive}{0.661}{0} & \gradcell{0}{\RtwoDOSix}{0.681}{0} \\
                COM$^2$(\textbf{D}etective)    & \gradcell{0}{\RtwoDOOne}{0.629}{0} & \gradcell{0}{\RtwoDOTwo}{0.423}{0} & \gradcell{0}{\RtwoDOThree}{0.474}{0} & \gradcell{0}{\RtwoDOFour}{0.379}{0} & \gradcell{0}{\RtwoDOFive}{0.568}{0} & \gradcell{0}{\RtwoDOSix}{0.218}{0} \\
                MuSiQue(\textbf{M}ulti-\textbf{H}op)  & \gradcell{0}{\RtwoDOOne}{0.624}{0} & \gradcell{0}{\RtwoDOTwo}{0.414}{0} & \gradcell{0}{\RtwoDOThree}{0.491}{0} & \gradcell{0}{\RtwoDOFour}{0.400}{0} & \gradcell{0}{\RtwoDOFive}{0.647}{0} & \gradcell{0}{\RtwoDOSix}{0.659}{0} \\
                SciBench(\textbf{S}cience) & \gradcell{0}{\RtwoDOOne}{0.620}{0} & \gradcell{0}{\RtwoDOTwo}{0.370}{0} & \gradcell{0}{\RtwoDOThree}{0.489}{0} & \gradcell{0}{\RtwoDOFour}{0.328}{0} & \gradcell{0}{\RtwoDOFive}{0.618}{0} & \gradcell{0}{\RtwoDOSix}{0.712}{0} \\
                TheoremQA(\textbf{Ma}th)  & \gradcell{0}{\RtwoDOOne}{0.449}{0} & \gradcell{0}{\RtwoDOTwo}{0.328}{0} & \gradcell{0}{\RtwoDOThree}{0.462}{0} & \gradcell{0}{\RtwoDOFour}{0.384}{0} & \gradcell{0}{\RtwoDOFive}{0.573}{0} & \gradcell{0}{\RtwoDOSix}{0.367}{0} \\
                StrategyQA(\textbf{C}ommonsense) & \gradcell{0}{\RtwoDOOne}{0.344}{0} & \gradcell{0}{\RtwoDOTwo}{0.417}{0} & \gradcell{0}{\RtwoDOThree}{0.412}{0} & \gradcell{0}{\RtwoDOFour}{0.402}{0} & \gradcell{0}{\RtwoDOFive}{0.670}{0} & \gradcell{0}{\RtwoDOSix}{0.769}{0} \\
                \midrule
                Multi-Domain Training & \gradcell{0}{\RtwoDOOne}{0.628}{0} & \gradcell{0}{\RtwoDOTwo}{0.424}{0} & \gradcell{0}{\RtwoDOThree}{0.469}{0} & \gradcell{0}{\RtwoDOFour}{0.395}{0} & \gradcell{0}{\RtwoDOFive}{0.658}{0} & \gradcell{0}{\RtwoDOSix}{0.747}{0} \\
                \bottomrule
            \end{tabular}
            \caption{Run 2: AFlow Results}
        \end{subtable}
    \end{minipage}

    \vspace{3.0em}
    

    \begin{minipage}{0.49\textwidth}
        \centering
        \newcommand{\RthreeAFOne}{0.629} \newcommand{\RthreeAFTwo}{0.492} 
        \newcommand{\RthreeAFThree}{0.588} \newcommand{\RthreeAFFour}{0.412} 
        \newcommand{\RthreeAFFive}{0.674} \newcommand{\RthreeAFSix}{0.769}

        \begin{subtable}{\textwidth}
            \centering
            \begin{tabular}{l c c c c c c}
                \toprule
                \textbf{Training / Test (Domain)} & \textbf{C} & \textbf{D} & \textbf{MH} & \textbf{S} & \textbf{MA} & \textbf{C} \\
                \midrule
                CaseHOLD(\textbf{L}egal)   & \gradcell{0}{\RthreeAFOne}{0.624}{0} & \gradcell{0}{\RthreeAFTwo}{0.473}{0} & \gradcell{0}{\RthreeAFThree}{0.578}{0} & \gradcell{0}{\RthreeAFFour}{0.403}{0} & \gradcell{0}{\RthreeAFFive}{0.645}{0} & \gradcell{0}{\RthreeAFSix}{0.707}{0} \\
                COM$^2$(\textbf{D}etective)    & \gradcell{0}{\RthreeAFOne}{0.624}{0} & \gradcell{0}{\RthreeAFTwo}{0.473}{0} & \gradcell{0}{\RthreeAFThree}{0.578}{0} & \gradcell{0}{\RthreeAFFour}{0.403}{0} & \gradcell{0}{\RthreeAFFive}{0.645}{0} & \gradcell{0}{\RthreeAFSix}{0.707}{0} \\
                MuSiQue(\textbf{M}ulti-\textbf{H}op)  & \gradcell{0}{\RthreeAFOne}{0.629}{0} & \gradcell{0}{\RthreeAFTwo}{0.492}{0} & \gradcell{0}{\RthreeAFThree}{0.588}{0} & \gradcell{0}{\RthreeAFFour}{0.412}{0} & \gradcell{0}{\RthreeAFFive}{0.674}{0} & \gradcell{0}{\RthreeAFSix}{0.725}{0} \\
                SciBench(\textbf{S}cience) & \gradcell{0}{\RthreeAFOne}{0.617}{0} & \gradcell{0}{\RthreeAFTwo}{0.346}{0} & \gradcell{0}{\RthreeAFThree}{0.543}{0} & \gradcell{0}{\RthreeAFFour}{0.396}{0} & \gradcell{0}{\RthreeAFFive}{0.625}{0} & \gradcell{0}{\RthreeAFSix}{0.463}{0} \\
                TheoremQA(\textbf{Ma}th)  & \gradcell{0}{\RthreeAFOne}{0.619}{0} & \gradcell{0}{\RthreeAFTwo}{0.470}{0} & \gradcell{0}{\RthreeAFThree}{0.570}{0} & \gradcell{0}{\RthreeAFFour}{0.380}{0} & \gradcell{0}{\RthreeAFFive}{0.632}{0} & \gradcell{0}{\RthreeAFSix}{0.769}{0} \\
                StrategyQA(\textbf{C}ommonsense) & \gradcell{0}{\RthreeAFOne}{0.006}{0} & \gradcell{0}{\RthreeAFTwo}{0.007}{0} & \gradcell{0}{\RthreeAFThree}{0.407}{0} & \gradcell{0}{\RthreeAFFour}{0.000}{0} & \gradcell{0}{\RthreeAFFive}{0.157}{0} & \gradcell{0}{\RthreeAFSix}{0.707}{0} \\
                \midrule
                Multi-Domain Training & \gradcell{0}{\RthreeAFOne}{0.607}{0} & \gradcell{0}{\RthreeAFTwo}{0.476}{0} & \gradcell{0}{\RthreeAFThree}{0.533}{0} & \gradcell{0}{\RthreeAFFour}{0.412}{0} & \gradcell{0}{\RthreeAFFive}{0.645}{0} & \gradcell{0}{\RthreeAFSix}{0.751}{0} \\
                \bottomrule
            \end{tabular}
            \caption{Run 3: AgentDropout Results}
        \end{subtable}
    \end{minipage}
    \hfill
    \begin{minipage}{0.49\textwidth}
        \centering
        \newcommand{\RthreeDOOne}{0.635} \newcommand{\RthreeDOTwo}{0.422} 
        \newcommand{\RthreeDOThree}{0.506} \newcommand{\RthreeDOFour}{0.409} 
        \newcommand{\RthreeDOFive}{0.679} \newcommand{\RthreeDOSix}{0.734}

        \begin{subtable}{\textwidth}
            \centering
            \begin{tabular}{l c c c c c c}
                \toprule
                \textbf{Training / Test (Domain)} & \textbf{C} & \textbf{D} & \textbf{MH} & \textbf{S} & \textbf{MA} & \textbf{C} \\
                \midrule
                CaseHOLD(\textbf{L}egal)   & \gradcell{0}{\RthreeDOOne}{0.609}{0} & \gradcell{0}{\RthreeDOTwo}{0.422}{0} & \gradcell{0}{\RthreeDOThree}{0.494}{0} & \gradcell{0}{\RthreeDOFour}{0.388}{0} & \gradcell{0}{\RthreeDOFive}{0.679}{0} & \gradcell{0}{\RthreeDOSix}{0.686}{0} \\
                COM$^2$(\textbf{D}etective)    & \gradcell{0}{\RthreeDOOne}{0.635}{0} & \gradcell{0}{\RthreeDOTwo}{0.422}{0} & \gradcell{0}{\RthreeDOThree}{0.494}{0} & \gradcell{0}{\RthreeDOFour}{0.388}{0} & \gradcell{0}{\RthreeDOFive}{0.654}{0} & \gradcell{0}{\RthreeDOSix}{0.664}{0} \\
                MuSiQue(\textbf{M}ulti-\textbf{H}op)  & \gradcell{0}{\RthreeDOOne}{0.481}{0} & \gradcell{0}{\RthreeDOTwo}{0.409}{0} & \gradcell{0}{\RthreeDOThree}{0.469}{0} & \gradcell{0}{\RthreeDOFour}{0.391}{0} & \gradcell{0}{\RthreeDOFive}{0.569}{0} & \gradcell{0}{\RthreeDOSix}{0.057}{0} \\
                SciBench(\textbf{S}cience)  & \gradcell{0}{\RthreeDOOne}{0.628}{0} & \gradcell{0}{\RthreeDOTwo}{0.388}{0} & \gradcell{0}{\RthreeDOThree}{0.469}{0} & \gradcell{0}{\RthreeDOFour}{0.272}{0} & \gradcell{0}{\RthreeDOFive}{0.568}{0} & \gradcell{0}{\RthreeDOSix}{0.664}{0} \\
                TheoremQA(\textbf{Ma}th)  & \gradcell{0}{\RthreeDOOne}{0.615}{0} & \gradcell{0}{\RthreeDOTwo}{0.407}{0} & \gradcell{0}{\RthreeDOThree}{0.496}{0} & \gradcell{0}{\RthreeDOFour}{0.397}{0} & \gradcell{0}{\RthreeDOFive}{0.667}{0} & \gradcell{0}{\RthreeDOSix}{0.677}{0} \\
                StrategyQA(\textbf{C}ommonsense) & \gradcell{0}{\RthreeDOOne}{0.299}{0} & \gradcell{0}{\RthreeDOTwo}{0.274}{0} & \gradcell{0}{\RthreeDOThree}{0.215}{0} & \gradcell{0}{\RthreeDOFour}{0.266}{0} & \gradcell{0}{\RthreeDOFive}{0.393}{0} & \gradcell{0}{\RthreeDOSix}{0.681}{0} \\
                \midrule
                Multi-Domain Training & \gradcell{0}{\RthreeDOOne}{0.624}{0} & \gradcell{0}{\RthreeDOTwo}{0.373}{0} & \gradcell{0}{\RthreeDOThree}{0.506}{0} & \gradcell{0}{\RthreeDOFour}{0.337}{0} & \gradcell{0}{\RthreeDOFive}{0.652}{0} & \gradcell{0}{\RthreeDOSix}{0.734}{0} \\
                \bottomrule
            \end{tabular}
            \caption{Run 3: AFlow Results}
        \end{subtable}
    \end{minipage}

    \vspace{0.8em}
    \caption{
    Full raw domain-transfer performance results on \texttt{GPT-oss-20B} across three independent runs.
    Left sub-tables show \textbf{AgentDropout} results and right sub-tables show \textbf{AFlow} results.
    Rows indicate the training domain and columns indicate the test domain, ordered as Legal (L), Detective (D), Multi-Hop (MH), Science (S), Math (MA), and Commonsense (C).
    Multi-Domain Training corresponds to multitask training.
    }    
    \label{tab:performance_measure_appendix}
\end{table*}
\begin{table*}[!t]
    \centering
    \scriptsize
    \renewcommand{\arraystretch}{1.3}
    \setlength{\tabcolsep}{2.7pt} 


    \vspace{2em}

    \begin{minipage}{0.49\textwidth}
        \centering
        \newcommand{\RoneADOne}{0.7010} \newcommand{\RoneADTwo}{0.4213} 
        \newcommand{\RoneADThree}{0.3556} \newcommand{\RoneADFour}{0.4599} 
        \newcommand{\RoneADFive}{0.6829} \newcommand{\RoneADSix}{0.6450}
        \begin{subtable}{\textwidth}
            \centering
            \begin{tabular}{l c c c c c c}
                \toprule
                \textbf{Training / Test (Domain)} & \textbf{L} & \textbf{D} & \textbf{MH} & \textbf{S} & \textbf{MA} & \textbf{C} \\
                \midrule
                CaseHOLD(\textbf{L}egal)      & \gradcell{0}{\RoneADOne}{0.7000}{0} & \gradcell{0}{\RoneADTwo}{0.3327}{0} & \gradcell{0}{\RoneADThree}{0.3556}{0} & \gradcell{0}{\RoneADFour}{0.4599}{0} & \gradcell{0}{\RoneADFive}{0.6829}{0} & \gradcell{0}{\RoneADSix}{0.6376}{0} \\
                COM$^2$(\textbf{D}etective)    & \gradcell{0}{\RoneADOne}{0.6940}{0} & \gradcell{0}{\RoneADTwo}{0.4213}{0} & \gradcell{0}{\RoneADThree}{0.2599}{0} & \gradcell{0}{\RoneADFour}{0.1679}{0} & \gradcell{0}{\RoneADFive}{0.2829}{0} & \gradcell{0}{\RoneADSix}{0.0742}{0} \\
                MuSiQue(\textbf{M}ulti-\textbf{H}op) & \gradcell{0}{\RoneADOne}{0.6790}{0} & \gradcell{0}{\RoneADTwo}{0.3127}{0} & \gradcell{0}{\RoneADThree}{0.2815}{0} & \gradcell{0}{\RoneADFour}{0.3668}{0} & \gradcell{0}{\RoneADFive}{0.6054}{0} & \gradcell{0}{\RoneADSix}{0.1135}{0} \\
                SciBench(\textbf{S}cience)     & \gradcell{0}{\RoneADOne}{0.7000}{0} & \gradcell{0}{\RoneADTwo}{0.2888}{0} & \gradcell{0}{\RoneADThree}{0.2074}{0} & \gradcell{0}{\RoneADFour}{0.4234}{0} & \gradcell{0}{\RoneADFive}{0.6234}{0} & \gradcell{0}{\RoneADSix}{0.3755}{0} \\
                TheoremQA(\textbf{Ma}th)      & \gradcell{0}{\RoneADOne}{0.6840}{0} & \gradcell{0}{\RoneADTwo}{0.3386}{0} & \gradcell{0}{\RoneADThree}{0.2642}{0} & \gradcell{0}{\RoneADFour}{0.4069}{0} & \gradcell{0}{\RoneADFive}{0.5982}{0} & \gradcell{0}{\RoneADSix}{0.4672}{0} \\
                StrategyQA(\textbf{C}ommonsense) & \gradcell{0}{\RoneADOne}{0.1080}{0} & \gradcell{0}{\RoneADTwo}{0.0996}{0} & \gradcell{0}{\RoneADThree}{0.0173}{0} & \gradcell{0}{\RoneADFour}{0.0018}{0} & \gradcell{0}{\RoneADFive}{0.1550}{0} & \gradcell{0}{\RoneADSix}{0.6450}{0} \\
                \midrule
                Multi-Domain Training         & \gradcell{0}{\RoneADOne}{0.7010}{0} & \gradcell{0}{\RoneADTwo}{0.2652}{0} & \gradcell{0}{\RoneADThree}{0.0815}{0} & \gradcell{0}{\RoneADFour}{0.0547}{0} & \gradcell{0}{\RoneADFive}{0.1351}{0} & \gradcell{0}{\RoneADSix}{0.0917}{0} \\
                \bottomrule
            \end{tabular}
            \caption{Run 1: AgentDropout Results}
        \end{subtable}
    \end{minipage}
    \hfill
    \begin{minipage}{0.49\textwidth}
        \centering
        \newcommand{\RoneAFOne}{0.7070} \newcommand{\RoneAFTwo}{0.4094} 
        \newcommand{\RoneAFThree}{0.5407} \newcommand{\RoneAFFour}{0.4384} 
        \newcommand{\RoneAFFive}{0.6486} \newcommand{\RoneAFSix}{0.8777}
        \begin{subtable}{\textwidth}
            \centering
            \begin{tabular}{l c c c c c c}
                \toprule
                \textbf{Training / Test (Domain)} & \textbf{L} & \textbf{D} & \textbf{MH} & \textbf{S} & \textbf{MA} & \textbf{C} \\
                \midrule
                CaseHOLD(\textbf{L}egal)      & \gradcell{0}{\RoneAFOne}{0.7070}{0} & \gradcell{0}{\RoneAFTwo}{0.3536}{0} & \gradcell{0}{\RoneAFThree}{0.4074}{0} & \gradcell{0}{\RoneAFFour}{0.4312}{0} & \gradcell{0}{\RoneAFFive}{0.6198}{0} & \gradcell{0}{\RoneAFSix}{0.7205}{0} \\
                COM$^2$(\textbf{D}etective)    & \gradcell{0}{\RoneAFOne}{0.6981}{0} & \gradcell{0}{\RoneAFTwo}{0.3984}{0} & \gradcell{0}{\RoneAFThree}{0.4716}{0} & \gradcell{0}{\RoneAFFour}{0.4384}{0} & \gradcell{0}{\RoneAFFive}{0.6342}{0} & \gradcell{0}{\RoneAFSix}{0.8777}{0} \\
                MuSiQue(\textbf{M}ulti-\textbf{H}op) & \gradcell{0}{\RoneAFOne}{0.7030}{0} & \gradcell{0}{\RoneAFTwo}{0.3546}{0} & \gradcell{0}{\RoneAFThree}{0.3926}{0} & \gradcell{0}{\RoneAFFour}{0.4257}{0} & \gradcell{0}{\RoneAFFive}{0.6234}{0} & \gradcell{0}{\RoneAFSix}{0.7249}{0} \\
                SciBench(\textbf{S}cience)     & \gradcell{0}{\RoneAFOne}{0.7000}{0} & \gradcell{0}{\RoneAFTwo}{0.3506}{0} & \gradcell{0}{\RoneAFThree}{0.4222}{0} & \gradcell{0}{\RoneAFFour}{0.4312}{0} & \gradcell{0}{\RoneAFFive}{0.6306}{0} & \gradcell{0}{\RoneAFSix}{0.7293}{0} \\
                TheoremQA(\textbf{Ma}th)      & \gradcell{0}{\RoneAFOne}{0.6820}{0} & \gradcell{0}{\RoneAFTwo}{0.3108}{0} & \gradcell{0}{\RoneAFThree}{0.3432}{0} & \gradcell{0}{\RoneAFFour}{0.3460}{0} & \gradcell{0}{\RoneAFFive}{0.6342}{0} & \gradcell{0}{\RoneAFSix}{0.7293}{0} \\
                StrategyQA(\textbf{C}ommonsense) & \gradcell{0}{\RoneAFOne}{0.6830}{0} & \gradcell{0}{\RoneAFTwo}{0.4094}{0} & \gradcell{0}{\RoneAFThree}{0.5407}{0} & \gradcell{0}{\RoneAFFour}{0.3659}{0} & \gradcell{0}{\RoneAFFive}{0.5189}{0} & \gradcell{0}{\RoneAFSix}{0.7031}{0} \\
                \midrule
                Multi-Domain Training         & \gradcell{0}{\RoneAFOne}{0.4680}{0} & \gradcell{0}{\RoneAFTwo}{0.3506}{0} & \gradcell{0}{\RoneAFThree}{0.4173}{0} & \gradcell{0}{\RoneAFFour}{0.3659}{0} & \gradcell{0}{\RoneAFFive}{0.6486}{0} & \gradcell{0}{\RoneAFSix}{0.7467}{0} \\
                \bottomrule
            \end{tabular}
            \caption{Run 1: AFlow Results}
        \end{subtable}
    \end{minipage}


    \vspace{3em}

    \begin{minipage}{0.49\textwidth}
        \centering
        \newcommand{\RtwoADOne}{0.7040} \newcommand{\RtwoADTwo}{0.3805} 
        \newcommand{\RtwoADThree}{0.3802} \newcommand{\RtwoADFour}{0.4854} 
        \newcommand{\RtwoADFive}{0.6523} \newcommand{\RtwoADSix}{0.6550}
        \begin{subtable}{\textwidth}
            \centering
            \begin{tabular}{l c c c c c c}
                \toprule
                \textbf{Training / Test (Domain)} & \textbf{L} & \textbf{D} & \textbf{MH} & \textbf{S} & \textbf{MA} & \textbf{C} \\
                \midrule
                CaseHOLD(\textbf{L}egal)      & \gradcell{0}{\RtwoADOne}{0.6970}{0} & \gradcell{0}{\RtwoADTwo}{0.3357}{0} & \gradcell{0}{\RtwoADThree}{0.3802}{0} & \gradcell{0}{\RtwoADFour}{0.4854}{0} & \gradcell{0}{\RtwoADFive}{0.6523}{0} & \gradcell{0}{\RtwoADSix}{0.6550}{0} \\
                COM$^2$(\textbf{D}etective)    & \gradcell{0}{\RtwoADOne}{0.6900}{0} & \gradcell{0}{\RtwoADTwo}{0.3805}{0} & \gradcell{0}{\RtwoADThree}{0.2420}{0} & \gradcell{0}{\RtwoADFour}{0.1734}{0} & \gradcell{0}{\RtwoADFive}{0.2919}{0} & \gradcell{0}{\RtwoADSix}{0.0568}{0} \\
                MuSiQue(\textbf{M}ulti-\textbf{H}op) & \gradcell{0}{\RtwoADOne}{0.6710}{0} & \gradcell{0}{\RtwoADTwo}{0.3127}{0} & \gradcell{0}{\RtwoADThree}{0.2840}{0} & \gradcell{0}{\RtwoADFour}{0.3631}{0} & \gradcell{0}{\RtwoADFive}{0.6090}{0} & \gradcell{0}{\RtwoADSix}{0.1310}{0} \\
                SciBench(\textbf{S}cience)     & \gradcell{0}{\RtwoADOne}{0.6870}{0} & \gradcell{0}{\RtwoADTwo}{0.2888}{0} & \gradcell{0}{\RtwoADThree}{0.1877}{0} & \gradcell{0}{\RtwoADFour}{0.4234}{0} & \gradcell{0}{\RtwoADFive}{0.6198}{0} & \gradcell{0}{\RtwoADSix}{0.3275}{0} \\
                TheoremQA(\textbf{Ma}th)      & \gradcell{0}{\RtwoADOne}{0.7040}{0} & \gradcell{0}{\RtwoADTwo}{0.3386}{0} & \gradcell{0}{\RtwoADThree}{0.2667}{0} & \gradcell{0}{\RtwoADFour}{0.3869}{0} & \gradcell{0}{\RtwoADFive}{0.6180}{0} & \gradcell{0}{\RtwoADSix}{0.5852}{0} \\
                StrategyQA(\textbf{C}ommonsense) & \gradcell{0}{\RtwoADOne}{0.0870}{0} & \gradcell{0}{\RtwoADTwo}{0.0876}{0} & \gradcell{0}{\RtwoADThree}{0.0247}{0} & \gradcell{0}{\RtwoADFour}{0.0000}{0} & \gradcell{0}{\RtwoADFive}{0.1622}{0} & \gradcell{0}{\RtwoADSix}{0.6157}{0} \\
                \midrule
                Multi-Domain Training         & \gradcell{0}{\RtwoADOne}{0.7000}{0} & \gradcell{0}{\RtwoADTwo}{0.2732}{0} & \gradcell{0}{\RtwoADThree}{0.0842}{0} & \gradcell{0}{\RtwoADFour}{0.0566}{0} & \gradcell{0}{\RtwoADFive}{0.1189}{0} & \gradcell{0}{\RtwoADSix}{0.0524}{0} \\
                \bottomrule
            \end{tabular}
            \caption{Run 2: AgentDropout Results}
        \end{subtable}
    \end{minipage}
    \hfill
    \begin{minipage}{0.49\textwidth}
        \centering
        \newcommand{\RtwoAFOne}{0.7020} \newcommand{\RtwoAFTwo}{0.4054} 
        \newcommand{\RtwoAFThree}{0.5358} \newcommand{\RtwoAFFour}{0.4275} 
        \newcommand{\RtwoAFFive}{0.6324} \newcommand{\RtwoAFSix}{0.7511}
        \begin{subtable}{\textwidth}
            \centering
            \begin{tabular}{l c c c c c c}
                \toprule
                \textbf{Training / Test (Domain)} & \textbf{L} & \textbf{D} & \textbf{MH} & \textbf{S} & \textbf{MA} & \textbf{C} \\
                \midrule
                CaseHOLD(\textbf{L}egal)      & \gradcell{0}{\RtwoAFOne}{0.6830}{0} & \gradcell{0}{\RtwoAFTwo}{0.4054}{0} & \gradcell{0}{\RtwoAFThree}{0.5284}{0} & \gradcell{0}{\RtwoAFFour}{0.2192}{0} & \gradcell{0}{\RtwoAFFive}{0.3894}{0} & \gradcell{0}{\RtwoAFSix}{0.6463}{0} \\
                COM$^2$(\textbf{D}etective)    & \gradcell{0}{\RtwoAFOne}{0.6800}{0} & \gradcell{0}{\RtwoAFTwo}{0.3775}{0} & \gradcell{0}{\RtwoAFThree}{0.5284}{0} & \gradcell{0}{\RtwoAFFour}{0.4149}{0} & \gradcell{0}{\RtwoAFFive}{0.6324}{0} & \gradcell{0}{\RtwoAFSix}{0.7511}{0} \\
                MuSiQue(\textbf{M}ulti-\textbf{H}op) & \gradcell{0}{\RtwoAFOne}{0.6940}{0} & \gradcell{0}{\RtwoAFTwo}{0.3785}{0} & \gradcell{0}{\RtwoAFThree}{0.5358}{0} & \gradcell{0}{\RtwoAFFour}{0.4058}{0} & \gradcell{0}{\RtwoAFFive}{0.6180}{0} & \gradcell{0}{\RtwoAFSix}{0.7380}{0} \\
                SciBench(\textbf{S}cience)     & \gradcell{0}{\RtwoAFOne}{0.7020}{0} & \gradcell{0}{\RtwoAFTwo}{0.3396}{0} & \gradcell{0}{\RtwoAFThree}{0.4099}{0} & \gradcell{0}{\RtwoAFFour}{0.4275}{0} & \gradcell{0}{\RtwoAFFive}{0.6288}{0} & \gradcell{0}{\RtwoAFSix}{0.7074}{0} \\
                TheoremQA(\textbf{Ma}th)      & \gradcell{0}{\RtwoAFOne}{0.6520}{0} & \gradcell{0}{\RtwoAFTwo}{0.3875}{0} & \gradcell{0}{\RtwoAFThree}{0.2395}{0} & \gradcell{0}{\RtwoAFFour}{0.1105}{0} & \gradcell{0}{\RtwoAFFive}{0.2703}{0} & \gradcell{0}{\RtwoAFSix}{0.6114}{0} \\
                StrategyQA(\textbf{C}ommonsense) & \gradcell{0}{\RtwoAFOne}{0.6850}{0} & \gradcell{0}{\RtwoAFTwo}{0.3725}{0} & \gradcell{0}{\RtwoAFThree}{0.2296}{0} & \gradcell{0}{\RtwoAFFour}{0.1196}{0} & \gradcell{0}{\RtwoAFFive}{0.3297}{0} & \gradcell{0}{\RtwoAFSix}{0.6725}{0} \\
                \midrule
                Multi-Domain Training         & \gradcell{0}{\RtwoAFOne}{0.6800}{0} & \gradcell{0}{\RtwoAFTwo}{0.4044}{0} & \gradcell{0}{\RtwoAFThree}{0.3086}{0} & \gradcell{0}{\RtwoAFFour}{0.1395}{0} & \gradcell{0}{\RtwoAFFive}{0.3387}{0} & \gradcell{0}{\RtwoAFSix}{0.2052}{0} \\
                \bottomrule
            \end{tabular}
            \caption{Run 2: AFlow Results}
        \end{subtable}
    \end{minipage}


    \vspace{3.0em}

    \begin{minipage}{0.49\textwidth}
        \centering
        \newcommand{\RthreeADOne}{0.6990} \newcommand{\RthreeADTwo}{0.4044} 
        \newcommand{\RthreeADThree}{0.3679} \newcommand{\RthreeADFour}{0.4872} 
        \newcommand{\RthreeADFive}{0.6486} \newcommand{\RthreeADSix}{0.6288}
        \begin{subtable}{\textwidth}
            \centering
            \begin{tabular}{l c c c c c c}
                \toprule
                \textbf{Training / Test (Domain)} & \textbf{L} & \textbf{D} & \textbf{MH} & \textbf{S} & \textbf{MA} & \textbf{C} \\
                \midrule
                CaseHOLD(\textbf{L}egal)      & \gradcell{0}{\RthreeADOne}{0.6970}{0} & \gradcell{0}{\RthreeADTwo}{0.3476}{0} & \gradcell{0}{\RthreeADThree}{0.3679}{0} & \gradcell{0}{\RthreeADFour}{0.4872}{0} & \gradcell{0}{\RthreeADFive}{0.6486}{0} & \gradcell{0}{\RthreeADSix}{0.6114}{0} \\
                COM$^2$(\textbf{D}etective)    & \gradcell{0}{\RthreeADOne}{0.6950}{0} & \gradcell{0}{\RthreeADTwo}{0.4044}{0} & \gradcell{0}{\RthreeADThree}{0.2272}{0} & \gradcell{0}{\RthreeADFour}{0.1788}{0} & \gradcell{0}{\RthreeADFive}{0.3081}{0} & \gradcell{0}{\RthreeADSix}{0.0655}{0} \\
                MuSiQue(\textbf{M}ulti-\textbf{H}op) & \gradcell{0}{\RthreeADOne}{0.6850}{0} & \gradcell{0}{\RthreeADTwo}{0.3277}{0} & \gradcell{0}{\RthreeADThree}{0.2741}{0} & \gradcell{0}{\RthreeADFour}{0.3850}{0} & \gradcell{0}{\RthreeADFive}{0.6018}{0} & \gradcell{0}{\RthreeADSix}{0.1528}{0} \\
                SciBench(\textbf{S}cience)     & \gradcell{0}{\RthreeADOne}{0.6890}{0} & \gradcell{0}{\RthreeADTwo}{0.3108}{0} & \gradcell{0}{\RthreeADThree}{0.1951}{0} & \gradcell{0}{\RthreeADFour}{0.4270}{0} & \gradcell{0}{\RthreeADFive}{0.6054}{0} & \gradcell{0}{\RthreeADSix}{0.3231}{0} \\
                TheoremQA(\textbf{Ma}th)      & \gradcell{0}{\RthreeADOne}{0.6980}{0} & \gradcell{0}{\RthreeADTwo}{0.3327}{0} & \gradcell{0}{\RthreeADThree}{0.2642}{0} & \gradcell{0}{\RthreeADFour}{0.3923}{0} & \gradcell{0}{\RthreeADFive}{0.6036}{0} & \gradcell{0}{\RthreeADSix}{0.5371}{0} \\
                StrategyQA(\textbf{C}ommonsense) & \gradcell{0}{\RthreeADOne}{0.0750}{0} & \gradcell{0}{\RthreeADTwo}{0.0926}{0} & \gradcell{0}{\RthreeADThree}{0.0148}{0} & \gradcell{0}{\RthreeADFour}{0.0018}{0} & \gradcell{0}{\RthreeADFive}{0.1514}{0} & \gradcell{0}{\RthreeADSix}{0.6288}{0} \\
                \midrule
                Multi-Domain Training         & \gradcell{0}{\RthreeADOne}{0.6990}{0} & \gradcell{0}{\RthreeADTwo}{0.2829}{0} & \gradcell{0}{\RthreeADThree}{0.1062}{0} & \gradcell{0}{\RthreeADFour}{0.0566}{0} & \gradcell{0}{\RthreeADFive}{0.1261}{0} & \gradcell{0}{\RthreeADSix}{0.0655}{0} \\
                \bottomrule
            \end{tabular}
            \caption{Run 3: AgentDropout Results}
        \end{subtable}
    \end{minipage}
    \hfill
    \begin{minipage}{0.49\textwidth}
        \centering
        \newcommand{\RthreeAFOne}{0.7070} \newcommand{\RthreeAFTwo}{0.3865} 
        \newcommand{\RthreeAFThree}{0.4716} \newcommand{\RthreeAFFour}{0.4348} 
        \newcommand{\RthreeAFFive}{0.6414} \newcommand{\RthreeAFSix}{0.7598}
        \begin{subtable}{\textwidth}
            \centering
            \begin{tabular}{l c c c c c c}
                \toprule
                \textbf{Training / Test (Domain)} & \textbf{L} & \textbf{D} & \textbf{MH} & \textbf{S} & \textbf{MA} & \textbf{C} \\
                \midrule
                CaseHOLD(\textbf{L}egal)      & \gradcell{0}{\RthreeAFOne}{0.7010}{0} & \gradcell{0}{\RthreeAFTwo}{0.3496}{0} & \gradcell{0}{\RthreeAFThree}{0.4247}{0} & \gradcell{0}{\RthreeAFFour}{0.4293}{0} & \gradcell{0}{\RthreeAFFive}{0.6342}{0} & \gradcell{0}{\RthreeAFSix}{0.6987}{0} \\
                COM$^2$(\textbf{D}etective)    & \gradcell{0}{\RthreeAFOne}{0.6240}{0} & \gradcell{0}{\RthreeAFTwo}{0.3865}{0} & \gradcell{0}{\RthreeAFThree}{0.4716}{0} & \gradcell{0}{\RthreeAFFour}{0.4275}{0} & \gradcell{0}{\RthreeAFFive}{0.6414}{0} & \gradcell{0}{\RthreeAFSix}{0.7511}{0} \\
                MuSiQue(\textbf{M}ulti-\textbf{H}op) & \gradcell{0}{\RthreeAFOne}{0.6620}{0} & \gradcell{0}{\RthreeAFTwo}{0.3695}{0} & \gradcell{0}{\RthreeAFThree}{0.3802}{0} & \gradcell{0}{\RthreeAFFour}{0.3460}{0} & \gradcell{0}{\RthreeAFFive}{0.5982}{0} & \gradcell{0}{\RthreeAFSix}{0.6201}{0} \\
                SciBench(\textbf{S}cience)     & \gradcell{0}{\RthreeAFOne}{0.6880}{0} & \gradcell{0}{\RthreeAFTwo}{0.2273}{0} & \gradcell{0}{\RthreeAFThree}{0.3155}{0} & \gradcell{0}{\RthreeAFFour}{0.3134}{0} & \gradcell{0}{\RthreeAFFive}{0.5153}{0} & \gradcell{0}{\RthreeAFSix}{0.3917}{0} \\
                TheoremQA(\textbf{Ma}th)      & \gradcell{0}{\RthreeAFOne}{0.6990}{0} & \gradcell{0}{\RthreeAFTwo}{0.3375}{0} & \gradcell{0}{\RthreeAFThree}{0.4272}{0} & \gradcell{0}{\RthreeAFFour}{0.4348}{0} & \gradcell{0}{\RthreeAFFive}{0.6216}{0} & \gradcell{0}{\RthreeAFSix}{0.7598}{0} \\
                StrategyQA(\textbf{C}ommonsense) & \gradcell{0}{\RthreeAFOne}{0.7070}{0} & \gradcell{0}{\RthreeAFTwo}{0.3466}{0} & \gradcell{0}{\RthreeAFThree}{0.4123}{0} & \gradcell{0}{\RthreeAFFour}{0.4239}{0} & \gradcell{0}{\RthreeAFFive}{0.6306}{0} & \gradcell{0}{\RthreeAFSix}{0.7031}{0} \\
                \midrule
                Multi-Domain Training         & \gradcell{0}{\RthreeAFOne}{0.6950}{0} & \gradcell{0}{\RthreeAFTwo}{0.3466}{0} & \gradcell{0}{\RthreeAFThree}{0.4296}{0} & \gradcell{0}{\RthreeAFFour}{0.4149}{0} & \gradcell{0}{\RthreeAFFive}{0.6234}{0} & \gradcell{0}{\RthreeAFSix}{0.7118}{0} \\
                \bottomrule
            \end{tabular}
            \caption{Run 3: AFlow Results}
        \end{subtable}
    \end{minipage}

    \vspace{0.8em}
    \caption{
    Full raw domain-transfer performance results on \texttt{Qwen3-30B-A3B} across three independent runs.
    Left sub-tables show \textbf{AgentDropout} results and right sub-tables show \textbf{AFlow} results.
    Rows indicate the training domain and columns indicate the test domain, ordered as Legal (L), Detective (D), Multi-Hop (MH), Science (S), Math (MA), and Commonsense (C).
    Multi-Domain Training corresponds to multitask training.
    }    
    \label{tab:performance_measure_appendix_raw}
\end{table*}
\begin{table*}[!t]
    \centering
    \scriptsize
    \renewcommand{\arraystretch}{1.2}
    \setlength{\tabcolsep}{3.5pt} 
    { 
        \begin{minipage}{0.48\textwidth}
\centering
\newcommand{\cOneMaxAF}{1} 
\newcommand{\cTwoMaxAF}{1} 
\newcommand{\cThreeMaxAF}{1}
\newcommand{\cFourMaxAF}{1} 
\newcommand{\cFiveMaxAF}{1} 
\newcommand{\cSixMaxAF}{1}
\newcommand{\cSevenMaxAF}{1}
\begin{subtable}{\textwidth}
        \centering
        \setlength{\tabcolsep}{2pt}
        \begin{tabular}{l c c c c c c}
            \toprule
            \textbf{Training / Test (Domain)} & \textbf{L} & \textbf{D} & \textbf{MH} & \textbf{S} & \textbf{MA} & \textbf{C}\\
            \midrule
            CaseHOLD(\textbf{L}egal)   & 
            \gradcellStructural{1}{\cOneMaxAF}{1}{0} &
            \gradcellStructural{0}{\cOneMaxAF}{0.573839316}{0} & 
            \gradcellStructural{0}{\cOneMaxAF}{0.040577254}{0} &
            \gradcellStructural{0}{\cOneMaxAF}{0.21705212}{0} & 
            \gradcellStructural{0}{\cOneMaxAF}{0.256143353}{0} & 
            \gradcellStructural{0}{\cOneMaxAF}{0.565298908}{0}\\
            COM$^2$(\textbf{D}etective)    & 
            \gradcellStructural{0}{\cTwoMaxAF}{0.802496966}{0} & 
            \gradcellStructural{1}{\cTwoMaxAF}{1}{0} & 
            \gradcellStructural{0}{\cTwoMaxAF}{0.048350259}{0} &
            \gradcellStructural{0}{\cTwoMaxAF}{0.433760732}{0} & 
            \gradcellStructural{0}{\cTwoMaxAF}{0.468436035}{0} &
            \gradcellStructural{0}{\cTwoMaxAF}{0.834385904}{0} \\
            MuSiQue(\textbf{M}ulti-\textbf{H}op)    &
            \gradcellStructural{0}{\cThreeMaxAF}{0.660862785}{0} & 
            \gradcellStructural{0}{\cThreeMaxAF}{1}{0} & 
            \gradcellStructural{1}{\cThreeMaxAF}{0.394676626}{0} & 
            \gradcellStructural{0}{\cThreeMaxAF}{0.449974733}{0} & 
            \gradcellStructural{0}{\cThreeMaxAF}{0.536879528}{0} & 
            \gradcellStructural{0}{\cThreeMaxAF}{0.595971743}{0}\\
            SciBench(\textbf{S}cience)   & 
            \gradcellStructural{0}{\cFourMaxAF}{0.486879289}{0} & 
            \gradcellStructural{0}{\cFourMaxAF}{0.518275105}{0} &
            \gradcellStructural{0}{\cFourMaxAF}{0.046425926}{0} & 
            \gradcellStructural{1}{\cFourMaxAF}{1}{0} & 
            \gradcellStructural{0}{\cFourMaxAF}{0.644901007}{0} & 
            \gradcellStructural{0}{\cFourMaxAF}{0.432702807}{0} \\
            TheoremQA(\textbf{Ma}th)  & 
            \gradcellStructural{0}{\cFiveMaxAF}{0.38460137}{0} &
            \gradcellStructural{0}{\cFiveMaxAF}{0.366589174}{0} &
            \gradcellStructural{0}{\cFiveMaxAF}{0.045866639}{0} &
            \gradcellStructural{0}{\cFiveMaxAF}{0.643020248}{0} &
            \gradcellStructural{1}{\cFiveMaxAF}{1}{0} &
            \gradcellStructural{0}{\cFiveMaxAF}{0.356247337}{0} \\
            StrategyQA(\textbf{C}ommonsense) & 
            \gradcellStructural{0}{\cSixMaxAF}{1}{0} &
            \gradcellStructural{0}{\cSixMaxAF}{0.979035983}{0} & 
            \gradcellStructural{0}{\cSixMaxAF}{0.069430788}{0} &
            \gradcellStructural{0}{\cSixMaxAF}{0.335716056}{0} &
            \gradcellStructural{0}{\cSixMaxAF}{0.468291617}{0} & 
            \gradcellStructural{1}{\cSixMaxAF}{0.867386909}{0} \\
            \midrule
            Multi-Domain Training    & 
            \gradcellStructural{0}{\cSevenMaxAF}{0.938465674}{0} &
            \gradcellStructural{0}{\cSevenMaxAF}{1}{0} & 
            \gradcellStructural{0}{\cSevenMaxAF}{0.280946226}{0} & 
            \gradcellStructural{0}{\cSevenMaxAF}{0.627870245}{0} &
            \gradcellStructural{0}{\cSevenMaxAF}{0.678068142}{0} &
            \gradcellStructural{0}{\cSevenMaxAF}{0.974538325}{0} \\
            \bottomrule
        \end{tabular}
        \caption{Run 1: Role Alignment ($\mathcal{R}$) Heatmap}
    \end{subtable}
\end{minipage}

        \hfill
        \begin{minipage}{0.48\textwidth}
\centering
\newcommand{\cOneMaxAF}{1} 
\newcommand{\cTwoMaxAF}{1} 
\newcommand{\cThreeMaxAF}{1}
\newcommand{\cFourMaxAF}{1} 
\newcommand{\cFiveMaxAF}{1} 
\newcommand{\cSixMaxAF}{1}
\newcommand{\cSevenMaxAF}{1}
\begin{subtable}{\textwidth}
        \centering
        {\setlength{\tabcolsep}{1.2pt}
        \begin{tabular}{l c c c c c c}
            \toprule
            \textbf{Training / Test (Domain)} & \textbf{L} & \textbf{D} & \textbf{MH} & \textbf{S} & \textbf{MA} & \textbf{C}\\
            \midrule
            CaseHOLD(\textbf{L}egal)   & 
            \gradcellStructural{1}{\cOneMaxAF}{1}{0} & 
            \gradcellStructural{0}{\cOneMaxAF}{-0.073499819}{0} & 
            \gradcellStructural{0}{\cOneMaxAF}{-1.944008217}{0} & 
            \gradcellStructural{0}{\cOneMaxAF}{-2.323982252}{0} & 
            \gradcellStructural{0}{\cOneMaxAF}{-2.125454396}{0} & 
            \gradcellStructural{0}{\cOneMaxAF}{-1.738938885}{0}\\
            COM$^2$(\textbf{D}etective)    & 
            \gradcellStructural{0}{\cTwoMaxAF}{0.836610336}{0} & 
            \gradcellStructural{1}{\cTwoMaxAF}{1}{0} & 
            \gradcellStructural{0}{\cTwoMaxAF}{0.130065867}{0} & 
            \gradcellStructural{0}{\cTwoMaxAF}{0.458865854}{0} &
            \gradcellStructural{0}{\cTwoMaxAF}{0.410255407}{0} & 
            \gradcellStructural{0}{\cTwoMaxAF}{0.207075225}{0}\\
            MuSiQue(\textbf{M}ulti-\textbf{H}op)    & 
            \gradcellStructural{0}{\cThreeMaxAF}{1}{0} &
            \gradcellStructural{0}{\cThreeMaxAF}{0.954025868}{0} & 
            \gradcellStructural{1}{\cThreeMaxAF}{0.156357384}{0} & 
            \gradcellStructural{0}{\cThreeMaxAF}{0.961693613}{0} & 
            \gradcellStructural{0}{\cThreeMaxAF}{0.737156686}{0} & 
            \gradcellStructural{0}{\cThreeMaxAF}{0.955170744}{0} \\
            SciBench(\textbf{S}cience)   & 
            \gradcellStructural{0}{\cFourMaxAF}{-0.793577931}{0} &
            \gradcellStructural{0}{\cFourMaxAF}{-0.532619907}{0} &
            \gradcellStructural{0}{\cFourMaxAF}{-0.571052452}{0} &
            \gradcellStructural{1}{\cFourMaxAF}{1}{0} &
            \gradcellStructural{0}{\cFourMaxAF}{0.856036059}{0} &
            \gradcellStructural{0}{\cFourMaxAF}{-0.014136576}{0} \\
            TheoremQA(\textbf{Ma}th)  &
            \gradcellStructural{0}{\cSixMaxAF}{0.051930468}{0} &
            \gradcellStructural{0}{\cSixMaxAF}{0.414851013}{0} &
            \gradcellStructural{0}{\cSixMaxAF}{0.007468762}{0} &
            \gradcellStructural{0}{\cSixMaxAF}{1}{0} &
            \gradcellStructural{1}{\cSixMaxAF}{0.920790799}{0} &
            \gradcellStructural{0}{\cSixMaxAF}{0.515924917}{0}\\
            StrategyQA(\textbf{C}ommonsense) & 
            \gradcellStructural{0}{\cFiveMaxAF}{0.905753968}{0} & 
            \gradcellStructural{0}{\cFiveMaxAF}{0.935029314}{0} & 
            \gradcellStructural{0}{\cFiveMaxAF}{0.13642007}{0} & 
            \gradcellStructural{0}{\cFiveMaxAF}{1}{0} & 
            \gradcellStructural{0}{\cFiveMaxAF}{0.886047646}{0} & 
            \gradcellStructural{1}{\cFiveMaxAF}{0.815450689}{0} \\
            \midrule
            Multi-Domain Training    & 
            \gradcellStructural{0}{\cSevenMaxAF}{0.805784833}{0} &
            \gradcellStructural{0}{\cSevenMaxAF}{0.976807246}{0} &
            \gradcellStructural{0}{\cSevenMaxAF}{-0.05334082}{0} &
            \gradcellStructural{0}{\cSevenMaxAF}{1}{0} &
            \gradcellStructural{0}{\cSevenMaxAF}{0.956645885}{0} &
            \gradcellStructural{0}{\cSevenMaxAF}{0.957668669}{0} \\
            \bottomrule
        \end{tabular}
        }
        \caption{Run 1: Connection Significance ($\mathcal{O}$) Heatmap}
    \end{subtable}
\end{minipage}

    } 
    { 
        \begin{minipage}{0.48\textwidth}
\centering
\newcommand{\cOneMaxAF}{1} 
\newcommand{\cTwoMaxAF}{1} 
\newcommand{\cThreeMaxAF}{1}
\newcommand{\cFourMaxAF}{1} 
\newcommand{\cFiveMaxAF}{1} 
\newcommand{\cSixMaxAF}{1}
\newcommand{\cSevenMaxAF}{1}
\begin{subtable}{\textwidth}
        \centering
        \setlength{\tabcolsep}{2pt}
        \begin{tabular}{l c c c c c c}
            \toprule
            \textbf{Training / Test (Domain)} & \textbf{L} & \textbf{D} & \textbf{MH} & \textbf{S} & \textbf{MA} & \textbf{C}\\
            \midrule
            CaseHOLD(\textbf{L}egal)   & 
            \gradcellStructural{1}{\cOneMaxAF}{1}{0} &
            \gradcellStructural{0}{\cOneMaxAF}{0.555176318}{0} & 
            \gradcellStructural{0}{\cOneMaxAF}{0.042448544}{0} &
            \gradcellStructural{0}{\cOneMaxAF}{0.217219522}{0} & 
            \gradcellStructural{0}{\cOneMaxAF}{0.255005182}{0} & 
            \gradcellStructural{0}{\cOneMaxAF}{0.534649299}{0}\\
            COM$^2$(\textbf{D}etective)    & 
            \gradcellStructural{0}{\cTwoMaxAF}{0.780178133}{0} & 
            \gradcellStructural{1}{\cTwoMaxAF}{1}{0} & 
            \gradcellStructural{0}{\cTwoMaxAF}{0.040236558}{0} &
            \gradcellStructural{0}{\cTwoMaxAF}{0.451676252}{0} & 
            \gradcellStructural{0}{\cTwoMaxAF}{0.48863783}{0} &
            \gradcellStructural{0}{\cTwoMaxAF}{0.828505065}{0} \\
            MuSiQue(\textbf{M}ulti-\textbf{H}op)    &
            \gradcellStructural{0}{\cThreeMaxAF}{0.703010364}{0} & 
            \gradcellStructural{0}{\cThreeMaxAF}{1}{0} & 
            \gradcellStructural{1}{\cThreeMaxAF}{0.391233615}{0} & 
            \gradcellStructural{0}{\cThreeMaxAF}{0.523291577}{0} & 
            \gradcellStructural{0}{\cThreeMaxAF}{0.623289619}{0} & 
            \gradcellStructural{0}{\cThreeMaxAF}{0.60950277}{0}\\
            SciBench(\textbf{S}cience)   & 
            \gradcellStructural{0}{\cFourMaxAF}{0.451736522}{0} & 
            \gradcellStructural{0}{\cFourMaxAF}{0.520153617}{0} &
            \gradcellStructural{0}{\cFourMaxAF}{0.042920016}{0} & 
            \gradcellStructural{1}{\cFourMaxAF}{1}{0} & 
            \gradcellStructural{0}{\cFourMaxAF}{0.63185569}{0} & 
            \gradcellStructural{0}{\cFourMaxAF}{0.515064262}{0} \\
            TheoremQA(\textbf{Ma}th)  & 
            \gradcellStructural{0}{\cFiveMaxAF}{0.386232705}{0} &
            \gradcellStructural{0}{\cFiveMaxAF}{0.368376004}{0} &
            \gradcellStructural{0}{\cFiveMaxAF}{0.043621087}{0} &
            \gradcellStructural{0}{\cFiveMaxAF}{0.639157337}{0} &
            \gradcellStructural{1}{\cFiveMaxAF}{1}{0} &
            \gradcellStructural{0}{\cFiveMaxAF}{0.326711224}{0} \\
            StrategyQA(\textbf{C}ommonsense) & 
            \gradcellStructural{0}{\cSixMaxAF}{0.901821789}{0} &
            \gradcellStructural{0}{\cSixMaxAF}{0.828433416}{0} & 
            \gradcellStructural{0}{\cSixMaxAF}{0.061296529}{0} &
            \gradcellStructural{0}{\cSixMaxAF}{0.267751061}{0} &
            \gradcellStructural{0}{\cSixMaxAF}{0.370691335}{0} & 
            \gradcellStructural{1}{\cSixMaxAF}{1}{0} \\
            \midrule
            Multi-Domain Training    & 
            \gradcellStructural{0}{\cSevenMaxAF}{0.865855301}{0} &
            \gradcellStructural{0}{\cSevenMaxAF}{1}{0} & 
            \gradcellStructural{0}{\cSevenMaxAF}{0.325132675}{0} & 
            \gradcellStructural{0}{\cSevenMaxAF}{0.580326536}{0} &
            \gradcellStructural{0}{\cSevenMaxAF}{0.609573703}{0} &
            \gradcellStructural{0}{\cSevenMaxAF}{0.861925213}{0} \\
            \bottomrule
        \end{tabular}
        \caption{Run 2: Role Alignment ($\mathcal{R}$) Heatmap}
    \end{subtable}
\end{minipage}

        \hfill
        \begin{minipage}{0.48\textwidth}
\centering
\newcommand{\cOneMaxAF}{1} 
\newcommand{\cTwoMaxAF}{1} 
\newcommand{\cThreeMaxAF}{1}
\newcommand{\cFourMaxAF}{1} 
\newcommand{\cFiveMaxAF}{1} 
\newcommand{\cSixMaxAF}{1}
\newcommand{\cSevenMaxAF}{1}
\begin{subtable}{\textwidth}
        \centering
        {\setlength{\tabcolsep}{1.2pt}
        \begin{tabular}{l c c c c c c}
            \toprule
            \textbf{Training / Test (Domain)} & \textbf{L} & \textbf{D} & \textbf{MH} & \textbf{S} & \textbf{MA} & \textbf{C}\\
            \midrule
            CaseHOLD(\textbf{L}egal)   & 
            \gradcellStructural{1}{\cOneMaxAF}{1}{0} & 
            \gradcellStructural{0}{\cOneMaxAF}{0.280792275}{0} & 
            \gradcellStructural{0}{\cOneMaxAF}{-1.666719551}{0} & 
            \gradcellStructural{0}{\cOneMaxAF}{-1.946485482}{0} & 
            \gradcellStructural{0}{\cOneMaxAF}{-1.764066854}{0} & 
            \gradcellStructural{0}{\cOneMaxAF}{-1.50121318}{0}\\
            COM$^2$(\textbf{D}etective)    & 
            \gradcellStructural{0}{\cTwoMaxAF}{0.975332964}{0} & 
            \gradcellStructural{1}{\cTwoMaxAF}{0.994542354}{0} & 
            \gradcellStructural{0}{\cTwoMaxAF}{0.268041624}{0} & 
            \gradcellStructural{0}{\cTwoMaxAF}{1}{0} &
            \gradcellStructural{0}{\cTwoMaxAF}{0.916279837}{0} & 
            \gradcellStructural{0}{\cTwoMaxAF}{0.805368086}{0}\\
            MuSiQue(\textbf{M}ulti-\textbf{H}op)    & 
            \gradcellStructural{0}{\cThreeMaxAF}{1}{0} &
            \gradcellStructural{0}{\cThreeMaxAF}{0.990035641}{0} & 
            \gradcellStructural{1}{\cThreeMaxAF}{0.146115111}{0} & 
            \gradcellStructural{0}{\cThreeMaxAF}{0.916740808}{0} & 
            \gradcellStructural{0}{\cThreeMaxAF}{0.806245798}{0} & 
            \gradcellStructural{0}{\cThreeMaxAF}{0.712020633}{0} \\
            SciBench(\textbf{S}cience)   & 
            \gradcellStructural{0}{\cFourMaxAF}{-0.799359211}{0} &
            \gradcellStructural{0}{\cFourMaxAF}{-0.454358799}{0} &
            \gradcellStructural{0}{\cFourMaxAF}{-0.478000193}{0} &
            \gradcellStructural{1}{\cFourMaxAF}{1}{0} &
            \gradcellStructural{0}{\cFourMaxAF}{0.730266327}{0} &
            \gradcellStructural{0}{\cFourMaxAF}{-0.112358003}{0} \\
            TheoremQA(\textbf{Ma}th)  &
            \gradcellStructural{0}{\cSixMaxAF}{-0.492397868}{0} &
            \gradcellStructural{0}{\cSixMaxAF}{-0.067168665}{0} &
            \gradcellStructural{0}{\cSixMaxAF}{-0.186777737}{0} &
            \gradcellStructural{0}{\cSixMaxAF}{1}{0} &
            \gradcellStructural{1}{\cSixMaxAF}{0.878993004}{0} &
            \gradcellStructural{0}{\cSixMaxAF}{0.284768763}{0}\\
            StrategyQA(\textbf{C}ommonsense) & 
            \gradcellStructural{0}{\cFiveMaxAF}{0.953198959}{0} & 
            \gradcellStructural{0}{\cFiveMaxAF}{0.98912615}{0} & 
            \gradcellStructural{0}{\cFiveMaxAF}{0.195566451}{0} & 
            \gradcellStructural{0}{\cFiveMaxAF}{1}{0} & 
            \gradcellStructural{0}{\cFiveMaxAF}{0.921767845}{0} & 
            \gradcellStructural{1}{\cFiveMaxAF}{0.802994687}{0} \\
            \midrule
            Multi-Domain Training    & 
            \gradcellStructural{0}{\cSevenMaxAF}{0.775014393}{0} &
            \gradcellStructural{0}{\cSevenMaxAF}{0.972886007}{0} &
            \gradcellStructural{0}{\cSevenMaxAF}{-0.03384998}{0} &
            \gradcellStructural{0}{\cSevenMaxAF}{1}{0} &
            \gradcellStructural{0}{\cSevenMaxAF}{0.990146414}{0} &
            \gradcellStructural{0}{\cSevenMaxAF}{0.922183148}{0} \\
            \bottomrule
        \end{tabular}
        }
        \caption{Run 2: Connection Significance ($\mathcal{O}$) Heatmap}
    \end{subtable}
\end{minipage}
    } 
    { 
        \begin{minipage}{0.48\textwidth}
\centering
\newcommand{\cOneMaxAF}{1} 
\newcommand{\cTwoMaxAF}{1} 
\newcommand{\cThreeMaxAF}{1}
\newcommand{\cFourMaxAF}{1} 
\newcommand{\cFiveMaxAF}{1} 
\newcommand{\cSixMaxAF}{1}
\newcommand{\cSevenMaxAF}{1}
\begin{subtable}{\textwidth}
        \centering
        \setlength{\tabcolsep}{2pt}
        \begin{tabular}{l c c c c c c}
            \toprule
            \textbf{Training / Test (Domain)} & \textbf{L} & \textbf{D} & \textbf{MH} & \textbf{S} & \textbf{MA} & \textbf{C}\\
            \midrule
            CaseHOLD(\textbf{L}egal)   & 
            \gradcellStructural{1}{\cOneMaxAF}{1}{0} &
            \gradcellStructural{0}{\cOneMaxAF}{0.556704367}{0} & 
            \gradcellStructural{0}{\cOneMaxAF}{0.043491919}{0} &
            \gradcellStructural{0}{\cOneMaxAF}{0.214976523}{0} & 
            \gradcellStructural{0}{\cOneMaxAF}{0.246230369}{0} & 
            \gradcellStructural{0}{\cOneMaxAF}{0.512591186}{0}\\
            COM$^2$(\textbf{D}etective)    & 
            \gradcellStructural{0}{\cTwoMaxAF}{0.778178932}{0} & 
            \gradcellStructural{1}{\cTwoMaxAF}{1}{0} & 
            \gradcellStructural{0}{\cTwoMaxAF}{0.045242296}{0} &
            \gradcellStructural{0}{\cTwoMaxAF}{0.403660721}{0} & 
            \gradcellStructural{0}{\cTwoMaxAF}{0.45375823}{0} &
            \gradcellStructural{0}{\cTwoMaxAF}{0.804531872}{0} \\
            MuSiQue(\textbf{M}ulti-\textbf{H}op)    &
            \gradcellStructural{0}{\cThreeMaxAF}{0.690771059}{0} & 
            \gradcellStructural{0}{\cThreeMaxAF}{1}{0} & 
            \gradcellStructural{1}{\cThreeMaxAF}{0.361515335}{0} & 
            \gradcellStructural{0}{\cThreeMaxAF}{0.528750193}{0} & 
            \gradcellStructural{0}{\cThreeMaxAF}{0.582242567}{0} & 
            \gradcellStructural{0}{\cThreeMaxAF}{0.535329456}{0}\\
            SciBench(\textbf{S}cience)   & 
            \gradcellStructural{0}{\cFourMaxAF}{0.397949877}{0} & 
            \gradcellStructural{0}{\cFourMaxAF}{0.428265032}{0} &
            \gradcellStructural{0}{\cFourMaxAF}{0.041107951}{0} & 
            \gradcellStructural{1}{\cFourMaxAF}{1}{0} & 
            \gradcellStructural{0}{\cFourMaxAF}{0.587365705}{0} & 
            \gradcellStructural{0}{\cFourMaxAF}{0.430620502}{0} \\
            TheoremQA(\textbf{Ma}th)  & 
            \gradcellStructural{0}{\cFiveMaxAF}{0.365332479}{0} &
            \gradcellStructural{0}{\cFiveMaxAF}{0.336617837}{0} &
            \gradcellStructural{0}{\cFiveMaxAF}{0.033651429}{0} &
            \gradcellStructural{0}{\cFiveMaxAF}{0.523328068}{0} &
            \gradcellStructural{1}{\cFiveMaxAF}{1}{0} &
            \gradcellStructural{0}{\cFiveMaxAF}{0.274173917}{0} \\
            StrategyQA(\textbf{C}ommonsense) & 
            \gradcellStructural{0}{\cSixMaxAF}{1}{0} &
            \gradcellStructural{0}{\cSixMaxAF}{0.99355841}{0} & 
            \gradcellStructural{0}{\cSixMaxAF}{0.07144266}{0} &
            \gradcellStructural{0}{\cSixMaxAF}{0.323042469}{0} &
            \gradcellStructural{0}{\cSixMaxAF}{0.470005084}{0} & 
            \gradcellStructural{1}{\cSixMaxAF}{0.875488021}{0} \\
            \midrule
            Multi-Domain Training    & 
            \gradcellStructural{0}{\cSevenMaxAF}{0.791968269}{0} &
            \gradcellStructural{0}{\cSevenMaxAF}{0.910432246}{0} & 
            \gradcellStructural{0}{\cSevenMaxAF}{0.296051997}{0} & 
            \gradcellStructural{0}{\cSevenMaxAF}{0.475229944}{0} &
            \gradcellStructural{0}{\cSevenMaxAF}{0.517515557}{0} &
            \gradcellStructural{0}{\cSevenMaxAF}{1}{0} \\
            \bottomrule
        \end{tabular}
        \caption{Run 3: Role Alignment ($\mathcal{R}$) Heatmap}
    \end{subtable}
\end{minipage}

        \hfill
        \begin{minipage}{0.48\textwidth}
\centering
\newcommand{\cOneMaxAF}{1} 
\newcommand{\cTwoMaxAF}{1} 
\newcommand{\cThreeMaxAF}{1}
\newcommand{\cFourMaxAF}{1} 
\newcommand{\cFiveMaxAF}{1} 
\newcommand{\cSixMaxAF}{1}
\newcommand{\cSevenMaxAF}{1}
\begin{subtable}{\textwidth}
        \centering
        {\setlength{\tabcolsep}{1.2pt}
        \begin{tabular}{l c c c c c c}
            \toprule
            \textbf{Training / Test (Domain)} & \textbf{L} & \textbf{D} & \textbf{MH} & \textbf{S} & \textbf{MA} & \textbf{C}\\
            \midrule
            CaseHOLD(\textbf{L}egal)   & 
            \gradcellStructural{1}{\cOneMaxAF}{1}{0} & 
            \gradcellStructural{0}{\cOneMaxAF}{-0.040934773}{0} & 
            \gradcellStructural{0}{\cOneMaxAF}{-1.799517794}{0} & 
            \gradcellStructural{0}{\cOneMaxAF}{-1.993788922}{0} & 
            \gradcellStructural{0}{\cOneMaxAF}{-1.84511969}{0} & 
            \gradcellStructural{0}{\cOneMaxAF}{-1.480042612}{0}\\
            COM$^2$(\textbf{D}etective)    & 
            \gradcellStructural{0}{\cTwoMaxAF}{0.869134853}{0} & 
            \gradcellStructural{1}{\cTwoMaxAF}{1}{0} & 
            \gradcellStructural{0}{\cTwoMaxAF}{0.106722289}{0} & 
            \gradcellStructural{0}{\cTwoMaxAF}{0.420422167}{0} &
            \gradcellStructural{0}{\cTwoMaxAF}{0.360136434}{0} & 
            \gradcellStructural{0}{\cTwoMaxAF}{0.294236226}{0}\\
            MuSiQue(\textbf{M}ulti-\textbf{H}op)    & 
            \gradcellStructural{0}{\cThreeMaxAF}{1}{0} &
            \gradcellStructural{0}{\cThreeMaxAF}{0.946141832}{0} & 
            \gradcellStructural{1}{\cThreeMaxAF}{0.15130774}{0} & 
            \gradcellStructural{0}{\cThreeMaxAF}{0.971386934}{0} & 
            \gradcellStructural{0}{\cThreeMaxAF}{0.847364416}{0} & 
            \gradcellStructural{0}{\cThreeMaxAF}{0.915314385}{0} \\
            SciBench(\textbf{S}cience)   & 
            \gradcellStructural{0}{\cFourMaxAF}{-0.669796831}{0} &
            \gradcellStructural{0}{\cFourMaxAF}{-0.516750036}{0} &
            \gradcellStructural{0}{\cFourMaxAF}{-0.646172333}{0} &
            \gradcellStructural{1}{\cFourMaxAF}{1}{0} &
            \gradcellStructural{0}{\cFourMaxAF}{0.719209415}{0} &
            \gradcellStructural{0}{\cFourMaxAF}{-0.079651962}{0} \\
            TheoremQA(\textbf{Ma}th)  &
            \gradcellStructural{0}{\cSixMaxAF}{0.087407314}{0} &
            \gradcellStructural{0}{\cSixMaxAF}{0.28335378}{0} &
            \gradcellStructural{0}{\cSixMaxAF}{-0.034784749}{0} &
            \gradcellStructural{0}{\cSixMaxAF}{1}{0} &
            \gradcellStructural{1}{\cSixMaxAF}{0.850477962}{0} &
            \gradcellStructural{0}{\cSixMaxAF}{0.621957443}{0}\\
            StrategyQA(\textbf{C}ommonsense) & 
            \gradcellStructural{0}{\cFiveMaxAF}{0.92117}{0} & 
            \gradcellStructural{0}{\cFiveMaxAF}{0.916145758}{0} & 
            \gradcellStructural{0}{\cFiveMaxAF}{0.187939895}{0} & 
            \gradcellStructural{0}{\cFiveMaxAF}{1}{0} & 
            \gradcellStructural{0}{\cFiveMaxAF}{0.885327857}{0} & 
            \gradcellStructural{1}{\cFiveMaxAF}{0.821435819}{0} \\
            \midrule
            Multi-Domain Training    & 
            \gradcellStructural{0}{\cSevenMaxAF}{0.257149042}{0} &
            \gradcellStructural{0}{\cSevenMaxAF}{0.857789978}{0} &
            \gradcellStructural{0}{\cSevenMaxAF}{-0.783007531}{0} &
            \gradcellStructural{0}{\cSevenMaxAF}{0.327537298}{0} &
            \gradcellStructural{0}{\cSevenMaxAF}{0.280614728}{0} &
            \gradcellStructural{0}{\cSevenMaxAF}{1}{0} \\
            \bottomrule
        \end{tabular}
        }
        \caption{Run 3: Connection Significance ($\mathcal{O}$) Heatmap}
    \end{subtable}
\end{minipage}

    } 
    \caption{
    Per-run (3 independent runs) \textbf{Role Alignment} (left) and \textbf{Connection Significance} (right) heatmaps for \textbf{AgentDropout} with \texttt{GPT-oss-20B} under domain transfer across six domains (L, D, MH, S, MA, C; see Table~\ref{tab:main_table}).
    Rows indicate the \emph{training} domain and columns the \emph{test} domain.
    All entries are \emph{row-wise normalized} by the maximum value in each row (each cell reports $v/\max(\text{row})$, so the row maximum is $1.00$).
    Cells are shaded \textcolor{blue}{blue} for normalized values $\ge 0.70$ (successful transfer) and \textcolor{red}{red} for values $< 0.70$ (failed transfer); intermediate values are left unshaded.
    Boldface indicates \textbf{in-domain} (diagonal) results.
    }
    \label{tab:structural_failure_appendix}
\end{table*}
\begin{table*}[!t]
    \centering
    \scriptsize
    \renewcommand{\arraystretch}{1.2}
    \setlength{\tabcolsep}{3.5pt}

    {%
    \begin{minipage}{0.48\textwidth}
        \centering
        \newcommand{\cOneMaxAF}{1}
        \newcommand{\cTwoMaxAF}{1}
        \newcommand{\cThreeMaxAF}{1}
        \newcommand{\cFourMaxAF}{1}
        \newcommand{\cFiveMaxAF}{1}
        \newcommand{\cSixMaxAF}{1}
        \newcommand{\cSevenMaxAF}{1}
        \begin{subtable}{\textwidth}
            \centering
            \setlength{\tabcolsep}{2pt}
            \begin{tabular}{l c c c c c c}
                \toprule
                \textbf{Training / Test (Domain)} & \textbf{L} & \textbf{D} & \textbf{MH} & \textbf{S} & \textbf{MA} & \textbf{C}\\
                \midrule
                CaseHOLD(\textbf{L}egal) &
                \gradcellStructural{1}{\cOneMaxAF}{1.000000}{0} &
                \gradcellStructural{0}{\cOneMaxAF}{0.431909}{0} &
                \gradcellStructural{0}{\cOneMaxAF}{0.160277}{0} &
                \gradcellStructural{0}{\cOneMaxAF}{0.346112}{0} &
                \gradcellStructural{0}{\cOneMaxAF}{0.241329}{0} &
                \gradcellStructural{0}{\cOneMaxAF}{0.664476}{0}\\
                COM$^2$(\textbf{D}etective) &
                \gradcellStructural{0}{\cTwoMaxAF}{0.225559}{0} &
                \gradcellStructural{1}{\cTwoMaxAF}{1.000000}{0} &
                \gradcellStructural{0}{\cTwoMaxAF}{0.232921}{0} &
                \gradcellStructural{0}{\cTwoMaxAF}{0.321260}{0} &
                \gradcellStructural{0}{\cTwoMaxAF}{0.360130}{0} &
                \gradcellStructural{0}{\cTwoMaxAF}{0.325088}{0}\\
                MuSiQue(\textbf{M}ulti-\textbf{H}op) &
                \gradcellStructural{0}{\cThreeMaxAF}{0.816881}{0} &
                \gradcellStructural{0}{\cThreeMaxAF}{0.962070}{0} &
                \gradcellStructural{1}{\cThreeMaxAF}{0.910505}{0} &
                \gradcellStructural{0}{\cThreeMaxAF}{1.000000}{0} &
                \gradcellStructural{0}{\cThreeMaxAF}{0.961110}{0} &
                \gradcellStructural{0}{\cThreeMaxAF}{0.834838}{0}\\
                SciBench(\textbf{S}cience) &
                \gradcellStructural{0}{\cFourMaxAF}{0.813357}{0} &
                \gradcellStructural{0}{\cFourMaxAF}{0.833481}{0} &
                \gradcellStructural{0}{\cFourMaxAF}{0.801415}{0} &
                \gradcellStructural{1}{\cFourMaxAF}{0.901371}{0} &
                \gradcellStructural{0}{\cFourMaxAF}{1.000000}{0} &
                \gradcellStructural{0}{\cFourMaxAF}{0.938302}{0}\\
                TheoremQA(\textbf{Ma}th) &
                \gradcellStructural{0}{\cFiveMaxAF}{0.741794}{0} &
                \gradcellStructural{0}{\cFiveMaxAF}{0.747386}{0} &
                \gradcellStructural{0}{\cFiveMaxAF}{0.536348}{0} &
                \gradcellStructural{0}{\cFiveMaxAF}{0.758814}{0} &
                \gradcellStructural{1}{\cFiveMaxAF}{1.000000}{0} &
                \gradcellStructural{0}{\cFiveMaxAF}{0.892050}{0}\\
                StrategyQA(\textbf{C}ommonsense) &
                \gradcellStructural{0}{\cSixMaxAF}{0.969729}{0} &
                \gradcellStructural{0}{\cSixMaxAF}{0.968058}{0} &
                \gradcellStructural{0}{\cSixMaxAF}{0.619624}{0} &
                \gradcellStructural{0}{\cSixMaxAF}{0.945511}{0} &
                \gradcellStructural{0}{\cSixMaxAF}{1.000000}{0} &
                \gradcellStructural{1}{\cSixMaxAF}{0.902923}{0}\\
                \midrule
                Multi-Domain Training &
                \gradcellStructural{0}{\cSevenMaxAF}{0.934093}{0} &
                \gradcellStructural{0}{\cSevenMaxAF}{0.934653}{0} &
                \gradcellStructural{0}{\cSevenMaxAF}{0.505041}{0} &
                \gradcellStructural{0}{\cSevenMaxAF}{0.604742}{0} &
                \gradcellStructural{0}{\cSevenMaxAF}{0.669156}{0} &
                \gradcellStructural{0}{\cSevenMaxAF}{1.000000}{0}\\
                \bottomrule
            \end{tabular}
            \caption{Run 1: Role Alignment ($\mathcal{R}$) Heatmap}
        \end{subtable}
    \end{minipage}
    }%
    \hfill
    {%
    \begin{minipage}{0.48\textwidth}
        \centering
        \newcommand{\cOneMaxBF}{1}
        \newcommand{\cTwoMaxBF}{1}
        \newcommand{\cThreeMaxBF}{1}
        \newcommand{\cFourMaxBF}{1}
        \newcommand{\cFiveMaxBF}{1}
        \newcommand{\cSixMaxBF}{1}
        \newcommand{\cSevenMaxBF}{1}
        \begin{subtable}{\textwidth}
            \centering
            \setlength{\tabcolsep}{2pt}
            \begin{tabular}{l c c c c c c}
                \toprule
                \textbf{Training / Test (Domain)} & \textbf{L} & \textbf{D} & \textbf{MH} & \textbf{S} & \textbf{MA} & \textbf{C}\\
                \midrule
                CaseHOLD(\textbf{L}egal) &
                \gradcellStructural{1}{\cOneMaxBF}{1.000000}{0} &
                \gradcellStructural{0}{\cOneMaxBF}{0.085067}{0} &
                \gradcellStructural{0}{\cOneMaxBF}{-0.922679}{0} &
                \gradcellStructural{0}{\cOneMaxBF}{-0.776329}{0} &
                \gradcellStructural{0}{\cOneMaxBF}{-0.852973}{0} &
                \gradcellStructural{0}{\cOneMaxBF}{-0.672825}{0}\\
                COM$^2$(\textbf{D}etective) &
                \gradcellStructural{0}{\cTwoMaxBF}{-0.132135}{0} &
                \gradcellStructural{1}{\cTwoMaxBF}{1.000000}{0} &
                \gradcellStructural{0}{\cTwoMaxBF}{-1.052864}{0} &
                \gradcellStructural{0}{\cTwoMaxBF}{-0.264680}{0} &
                \gradcellStructural{0}{\cTwoMaxBF}{-0.244531}{0} &
                \gradcellStructural{0}{\cTwoMaxBF}{-0.013676}{0}\\
                MuSiQue(\textbf{M}ulti-\textbf{H}op) &
                \gradcellStructural{0}{\cThreeMaxBF}{-0.832197}{0} &
                \gradcellStructural{0}{\cThreeMaxBF}{-1.000000}{0} &
                \gradcellStructural{1}{\cThreeMaxBF}{-0.749930}{0} &
                \gradcellStructural{0}{\cThreeMaxBF}{-0.895680}{0} &
                \gradcellStructural{0}{\cThreeMaxBF}{-0.855316}{0} &
                \gradcellStructural{0}{\cThreeMaxBF}{-0.928812}{0}\\
                SciBench(\textbf{S}cience) &
                \gradcellStructural{0}{\cFourMaxBF}{-0.026304}{0} &
                \gradcellStructural{0}{\cFourMaxBF}{0.486369}{0} &
                \gradcellStructural{0}{\cFourMaxBF}{-0.011617}{0} &
                \gradcellStructural{1}{\cFourMaxBF}{1.000000}{0} &
                \gradcellStructural{0}{\cFourMaxBF}{0.919974}{0} &
                \gradcellStructural{0}{\cFourMaxBF}{0.449566}{0}\\
                TheoremQA(\textbf{Ma}th) &
                \gradcellStructural{0}{\cFiveMaxBF}{-0.145618}{0} &
                \gradcellStructural{0}{\cFiveMaxBF}{0.479731}{0} &
                \gradcellStructural{0}{\cFiveMaxBF}{0.275943}{0} &
                \gradcellStructural{0}{\cFiveMaxBF}{1.000000}{0} &
                \gradcellStructural{1}{\cFiveMaxBF}{0.992074}{0} &
                \gradcellStructural{0}{\cFiveMaxBF}{0.457617}{0}\\
                StrategyQA(\textbf{C}ommonsense) &
                \gradcellStructural{0}{\cSixMaxBF}{0.547854}{0} &
                \gradcellStructural{0}{\cSixMaxBF}{0.758469}{0} &
                \gradcellStructural{0}{\cSixMaxBF}{0.912069}{0} &
                \gradcellStructural{0}{\cSixMaxBF}{0.085043}{0} &
                \gradcellStructural{0}{\cSixMaxBF}{0.004729}{0} &
                \gradcellStructural{1}{\cSixMaxBF}{1.000000}{0}\\
                \midrule
                Multi-Domain Training &
                \gradcellStructural{0}{\cSevenMaxBF}{1.000000}{0} &
                \gradcellStructural{0}{\cSevenMaxBF}{0.990738}{0} &
                \gradcellStructural{0}{\cSevenMaxBF}{-0.935856}{0} &
                \gradcellStructural{0}{\cSevenMaxBF}{0.786501}{0} &
                \gradcellStructural{0}{\cSevenMaxBF}{0.878051}{0} &
                \gradcellStructural{0}{\cSevenMaxBF}{0.429559}{0}\\
                \bottomrule
            \end{tabular}
            \caption{Run 1: Connection Significance ($\mathcal{O}$) Heatmap}
        \end{subtable}
    \end{minipage}
    }%

    \par\vspace{1.2em}

    {%
    \begin{minipage}{0.48\textwidth}
        \centering
        \newcommand{\cOneMaxCF}{1}
        \newcommand{\cTwoMaxCF}{1}
        \newcommand{\cThreeMaxCF}{1}
        \newcommand{\cFourMaxCF}{1}
        \newcommand{\cFiveMaxCF}{1}
        \newcommand{\cSixMaxCF}{1}
        \newcommand{\cSevenMaxCF}{1}
        \begin{subtable}{\textwidth}
            \centering
            \setlength{\tabcolsep}{2pt}
            \begin{tabular}{l c c c c c c}
                \toprule
                \textbf{Training / Test (Domain)} & \textbf{L} & \textbf{D} & \textbf{MH} & \textbf{S} & \textbf{MA} & \textbf{C}\\
                \midrule
                CaseHOLD(\textbf{L}egal) &
                \gradcellStructural{1}{\cOneMaxCF}{1.000000}{0} &
                \gradcellStructural{0}{\cOneMaxCF}{0.301304}{0} &
                \gradcellStructural{0}{\cOneMaxCF}{0.102265}{0} &
                \gradcellStructural{0}{\cOneMaxCF}{0.153054}{0} &
                \gradcellStructural{0}{\cOneMaxCF}{0.117364}{0} &
                \gradcellStructural{0}{\cOneMaxCF}{0.330130}{0}\\
                COM$^2$(\textbf{D}etective) &
                \gradcellStructural{0}{\cTwoMaxCF}{0.190255}{0} &
                \gradcellStructural{1}{\cTwoMaxCF}{1.000000}{0} &
                \gradcellStructural{0}{\cTwoMaxCF}{0.203145}{0} &
                \gradcellStructural{0}{\cTwoMaxCF}{0.117040}{0} &
                \gradcellStructural{0}{\cTwoMaxCF}{0.156999}{0} &
                \gradcellStructural{0}{\cTwoMaxCF}{0.398814}{0}\\
                MuSiQue(\textbf{M}ulti-\textbf{H}op) &
                \gradcellStructural{0}{\cThreeMaxCF}{0.746224}{0} &
                \gradcellStructural{0}{\cThreeMaxCF}{0.935881}{0} &
                \gradcellStructural{1}{\cThreeMaxCF}{0.857257}{0} &
                \gradcellStructural{0}{\cThreeMaxCF}{1.000000}{0} &
                \gradcellStructural{0}{\cThreeMaxCF}{0.958688}{0} &
                \gradcellStructural{0}{\cThreeMaxCF}{0.788236}{0}\\
                SciBench(\textbf{S}cience) &
                \gradcellStructural{0}{\cFourMaxCF}{0.816329}{0} &
                \gradcellStructural{0}{\cFourMaxCF}{0.821918}{0} &
                \gradcellStructural{0}{\cFourMaxCF}{0.804932}{0} &
                \gradcellStructural{1}{\cFourMaxCF}{1.000000}{0} &
                \gradcellStructural{0}{\cFourMaxCF}{0.996384}{0} &
                \gradcellStructural{0}{\cFourMaxCF}{0.922849}{0}\\
                TheoremQA(\textbf{Ma}th) &
                \gradcellStructural{0}{\cFiveMaxCF}{0.717686}{0} &
                \gradcellStructural{0}{\cFiveMaxCF}{0.706736}{0} &
                \gradcellStructural{0}{\cFiveMaxCF}{0.550107}{0} &
                \gradcellStructural{0}{\cFiveMaxCF}{0.724589}{0} &
                \gradcellStructural{1}{\cFiveMaxCF}{1.000000}{0} &
                \gradcellStructural{0}{\cFiveMaxCF}{0.859795}{0}\\
                StrategyQA(\textbf{C}ommonsense) &
                \gradcellStructural{0}{\cSixMaxCF}{0.702645}{0} &
                \gradcellStructural{0}{\cSixMaxCF}{0.747224}{0} &
                \gradcellStructural{0}{\cSixMaxCF}{0.488896}{0} &
                \gradcellStructural{0}{\cSixMaxCF}{0.702482}{0} &
                \gradcellStructural{0}{\cSixMaxCF}{0.765349}{0} &
                \gradcellStructural{1}{\cSixMaxCF}{1.000000}{0}\\
                \midrule
                Multi-Domain Training &
                \gradcellStructural{0}{\cSevenMaxCF}{0.961428}{0} &
                \gradcellStructural{0}{\cSevenMaxCF}{0.954135}{0} &
                \gradcellStructural{0}{\cSevenMaxCF}{0.600461}{0} &
                \gradcellStructural{0}{\cSevenMaxCF}{0.619267}{0} &
                \gradcellStructural{0}{\cSevenMaxCF}{0.678181}{0} &
                \gradcellStructural{0}{\cSevenMaxCF}{1.000000}{0}\\
                \bottomrule
            \end{tabular}
            \caption{Run 2: Role Alignment ($\mathcal{R}$) Heatmap}
        \end{subtable}
    \end{minipage}
    }%
    \hfill
    {%
    \begin{minipage}{0.48\textwidth}
        \centering
        \newcommand{\cOneMaxDF}{1}
        \newcommand{\cTwoMaxDF}{1}
        \newcommand{\cThreeMaxDF}{1}
        \newcommand{\cFourMaxDF}{1}
        \newcommand{\cFiveMaxDF}{1}
        \newcommand{\cSixMaxDF}{1}
        \newcommand{\cSevenMaxDF}{1}
        \begin{subtable}{\textwidth}
            \centering
            \setlength{\tabcolsep}{2pt}
            \begin{tabular}{l c c c c c c}
                \toprule
                \textbf{Training / Test (Domain)} & \textbf{L} & \textbf{D} & \textbf{MH} & \textbf{S} & \textbf{MA} & \textbf{C}\\
                \midrule
                CaseHOLD(\textbf{L}egal) &
                \gradcellStructural{1}{\cOneMaxDF}{1.000000}{0} &
                \gradcellStructural{0}{\cOneMaxDF}{-0.290030}{0} &
                \gradcellStructural{0}{\cOneMaxDF}{-1.033017}{0} &
                \gradcellStructural{0}{\cOneMaxDF}{-0.887185}{0} &
                \gradcellStructural{0}{\cOneMaxDF}{-0.936366}{0} &
                \gradcellStructural{0}{\cOneMaxDF}{-0.806789}{0}\\
                COM$^2$(\textbf{D}etective) &
                \gradcellStructural{0}{\cTwoMaxDF}{-1.053227}{0} &
                \gradcellStructural{1}{\cTwoMaxDF}{1.000000}{0} &
                \gradcellStructural{0}{\cTwoMaxDF}{-1.449968}{0} &
                \gradcellStructural{0}{\cTwoMaxDF}{-1.179275}{0} &
                \gradcellStructural{0}{\cTwoMaxDF}{-1.198698}{0} &
                \gradcellStructural{0}{\cTwoMaxDF}{-0.880594}{0}\\
                MuSiQue(\textbf{M}ulti-\textbf{H}op) &
                \gradcellStructural{0}{\cThreeMaxDF}{-0.853528}{0} &
                \gradcellStructural{0}{\cThreeMaxDF}{-1.000000}{0} &
                \gradcellStructural{1}{\cThreeMaxDF}{-0.745273}{0} &
                \gradcellStructural{0}{\cThreeMaxDF}{-0.914729}{0} &
                \gradcellStructural{0}{\cThreeMaxDF}{-0.872625}{0} &
                \gradcellStructural{0}{\cThreeMaxDF}{-0.952065}{0}\\
                SciBench(\textbf{S}cience) &
                \gradcellStructural{0}{\cFourMaxDF}{-0.177412}{0} &
                \gradcellStructural{0}{\cFourMaxDF}{0.339033}{0} &
                \gradcellStructural{0}{\cFourMaxDF}{0.148452}{0} &
                \gradcellStructural{1}{\cFourMaxDF}{1.000000}{0} &
                \gradcellStructural{0}{\cFourMaxDF}{0.946648}{0} &
                \gradcellStructural{0}{\cFourMaxDF}{0.361461}{0}\\
                TheoremQA(\textbf{Ma}th) &
                \gradcellStructural{0}{\cFiveMaxDF}{0.040358}{0} &
                \gradcellStructural{0}{\cFiveMaxDF}{0.645235}{0} &
                \gradcellStructural{0}{\cFiveMaxDF}{0.266567}{0} &
                \gradcellStructural{0}{\cFiveMaxDF}{1.000000}{0} &
                \gradcellStructural{1}{\cFiveMaxDF}{0.997420}{0} &
                \gradcellStructural{0}{\cFiveMaxDF}{0.718297}{0}\\
                StrategyQA(\textbf{C}ommonsense) &
                \gradcellStructural{0}{\cSixMaxDF}{0.344265}{0} &
                \gradcellStructural{0}{\cSixMaxDF}{0.789710}{0} &
                \gradcellStructural{0}{\cSixMaxDF}{0.977687}{0} &
                \gradcellStructural{0}{\cSixMaxDF}{0.103047}{0} &
                \gradcellStructural{0}{\cSixMaxDF}{-0.053357}{0} &
                \gradcellStructural{1}{\cSixMaxDF}{1.000000}{0}\\
                \midrule
                Multi-Domain Training &
                \gradcellStructural{0}{\cSevenMaxDF}{1.000000}{0} &
                \gradcellStructural{0}{\cSevenMaxDF}{0.997439}{0} &
                \gradcellStructural{0}{\cSevenMaxDF}{-0.525624}{0} &
                \gradcellStructural{0}{\cSevenMaxDF}{0.981352}{0} &
                \gradcellStructural{0}{\cSevenMaxDF}{0.979278}{0} &
                \gradcellStructural{0}{\cSevenMaxDF}{0.310181}{0}\\
                \bottomrule
            \end{tabular}
            \caption{Run 2: Connection Significance ($\mathcal{O}$) Heatmap}
        \end{subtable}
    \end{minipage}
    }%

    \par\vspace{1.2em}

    {%
    \begin{minipage}{0.48\textwidth}
        \centering
        \newcommand{\cOneMaxEF}{1}
        \newcommand{\cTwoMaxEF}{1}
        \newcommand{\cThreeMaxEF}{1}
        \newcommand{\cFourMaxEF}{1}
        \newcommand{\cFiveMaxEF}{1}
        \newcommand{\cSixMaxEF}{1}
        \newcommand{\cSevenMaxEF}{1}
        \begin{subtable}{\textwidth}
            \centering
            \setlength{\tabcolsep}{2pt}
            \begin{tabular}{l c c c c c c}
                \toprule
                \textbf{Training / Test (Domain)} & \textbf{L} & \textbf{D} & \textbf{MH} & \textbf{S} & \textbf{MA} & \textbf{C}\\
                \midrule
                CaseHOLD(\textbf{L}egal) &
                \gradcellStructural{1}{\cOneMaxEF}{1.000000}{0} &
                \gradcellStructural{0}{\cOneMaxEF}{0.303151}{0} &
                \gradcellStructural{0}{\cOneMaxEF}{0.107818}{0} &
                \gradcellStructural{0}{\cOneMaxEF}{0.128588}{0} &
                \gradcellStructural{0}{\cOneMaxEF}{0.105251}{0} &
                \gradcellStructural{0}{\cOneMaxEF}{0.379930}{0}\\
                COM$^2$(\textbf{D}etective) &
                \gradcellStructural{0}{\cTwoMaxEF}{0.254850}{0} &
                \gradcellStructural{1}{\cTwoMaxEF}{1.000000}{0} &
                \gradcellStructural{0}{\cTwoMaxEF}{0.253206}{0} &
                \gradcellStructural{0}{\cTwoMaxEF}{0.304505}{0} &
                \gradcellStructural{0}{\cTwoMaxEF}{0.353831}{0} &
                \gradcellStructural{0}{\cTwoMaxEF}{0.386386}{0}\\
                MuSiQue(\textbf{M}ulti-\textbf{H}op) &
                \gradcellStructural{0}{\cThreeMaxEF}{0.799882}{0} &
                \gradcellStructural{0}{\cThreeMaxEF}{0.981754}{0} &
                \gradcellStructural{1}{\cThreeMaxEF}{0.848931}{0} &
                \gradcellStructural{0}{\cThreeMaxEF}{1.000000}{0} &
                \gradcellStructural{0}{\cThreeMaxEF}{0.969884}{0} &
                \gradcellStructural{0}{\cThreeMaxEF}{0.848538}{0}\\
                SciBench(\textbf{S}cience) &
                \gradcellStructural{0}{\cFourMaxEF}{0.774773}{0} &
                \gradcellStructural{0}{\cFourMaxEF}{0.800459}{0} &
                \gradcellStructural{0}{\cFourMaxEF}{0.753785}{0} &
                \gradcellStructural{1}{\cFourMaxEF}{1.000000}{0} &
                \gradcellStructural{0}{\cFourMaxEF}{0.951968}{0} &
                \gradcellStructural{0}{\cFourMaxEF}{0.887439}{0}\\
                TheoremQA(\textbf{Ma}th) &
                \gradcellStructural{0}{\cFiveMaxEF}{0.757386}{0} &
                \gradcellStructural{0}{\cFiveMaxEF}{0.781172}{0} &
                \gradcellStructural{0}{\cFiveMaxEF}{0.568603}{0} &
                \gradcellStructural{0}{\cFiveMaxEF}{0.761392}{0} &
                \gradcellStructural{1}{\cFiveMaxEF}{1.000000}{0} &
                \gradcellStructural{0}{\cFiveMaxEF}{0.968453}{0}\\
                StrategyQA(\textbf{C}ommonsense) &
                \gradcellStructural{0}{\cSixMaxEF}{0.711335}{0} &
                \gradcellStructural{0}{\cSixMaxEF}{0.775521}{0} &
                \gradcellStructural{0}{\cSixMaxEF}{0.497269}{0} &
                \gradcellStructural{0}{\cSixMaxEF}{0.726357}{0} &
                \gradcellStructural{0}{\cSixMaxEF}{0.781837}{0} &
                \gradcellStructural{1}{\cSixMaxEF}{1.000000}{0}\\
                \midrule
                Multi-Domain Training &
                \gradcellStructural{0}{\cSevenMaxEF}{0.919015}{0} &
                \gradcellStructural{0}{\cSevenMaxEF}{0.916949}{0} &
                \gradcellStructural{0}{\cSevenMaxEF}{0.562570}{0} &
                \gradcellStructural{0}{\cSevenMaxEF}{0.609357}{0} &
                \gradcellStructural{0}{\cSevenMaxEF}{0.673243}{0} &
                \gradcellStructural{0}{\cSevenMaxEF}{1.000000}{0}\\
                \bottomrule
            \end{tabular}
            \caption{Run 3: Role Metric Heatmap}
        \end{subtable}
    \end{minipage}
    }%
    \hfill
    {%
    \begin{minipage}{0.48\textwidth}
        \centering
        \newcommand{\cOneMaxFF}{1}
        \newcommand{\cTwoMaxFF}{1}
        \newcommand{\cThreeMaxFF}{1}
        \newcommand{\cFourMaxFF}{1}
        \newcommand{\cFiveMaxFF}{1}
        \newcommand{\cSixMaxFF}{1}
        \newcommand{\cSevenMaxFF}{1}
        \begin{subtable}{\textwidth}
            \centering
            \setlength{\tabcolsep}{2pt}
            \begin{tabular}{l c c c c c c}
                \toprule
                \textbf{Training / Test (Domain)} & \textbf{L} & \textbf{D} & \textbf{MH} & \textbf{S} & \textbf{MA} & \textbf{C}\\
                \midrule
                CaseHOLD(\textbf{L}egal) &
                \gradcellStructural{1}{\cOneMaxFF}{1.000000}{0} &
                \gradcellStructural{0}{\cOneMaxFF}{0.160706}{0} &
                \gradcellStructural{0}{\cOneMaxFF}{-0.849712}{0} &
                \gradcellStructural{0}{\cOneMaxFF}{-0.431941}{0} &
                \gradcellStructural{0}{\cOneMaxFF}{-0.629899}{0} &
                \gradcellStructural{0}{\cOneMaxFF}{-0.460083}{0}\\
                COM$^2$(\textbf{D}etective) &
                \gradcellStructural{0}{\cTwoMaxFF}{-0.322906}{0} &
                \gradcellStructural{1}{\cTwoMaxFF}{1.000000}{0} &
                \gradcellStructural{0}{\cTwoMaxFF}{-1.147068}{0} &
                \gradcellStructural{0}{\cTwoMaxFF}{-0.438182}{0} &
                \gradcellStructural{0}{\cTwoMaxFF}{-0.418989}{0} &
                \gradcellStructural{0}{\cTwoMaxFF}{-0.212772}{0}\\
                MuSiQue(\textbf{M}ulti-\textbf{H}op) &
                \gradcellStructural{0}{\cThreeMaxFF}{-0.764491}{0} &
                \gradcellStructural{0}{\cThreeMaxFF}{-1.000000}{0} &
                \gradcellStructural{1}{\cThreeMaxFF}{-0.644220}{0} &
                \gradcellStructural{0}{\cThreeMaxFF}{-0.903660}{0} &
                \gradcellStructural{0}{\cThreeMaxFF}{-0.815933}{0} &
                \gradcellStructural{0}{\cThreeMaxFF}{-0.886287}{0}\\
                SciBench(\textbf{S}cience) &
                \gradcellStructural{0}{\cFourMaxFF}{-0.411373}{0} &
                \gradcellStructural{0}{\cFourMaxFF}{0.113423}{0} &
                \gradcellStructural{0}{\cFourMaxFF}{-0.151994}{0} &
                \gradcellStructural{1}{\cFourMaxFF}{1.000000}{0} &
                \gradcellStructural{0}{\cFourMaxFF}{0.960612}{0} &
                \gradcellStructural{0}{\cFourMaxFF}{0.316357}{0}\\
                TheoremQA(\textbf{Ma}th) &
                \gradcellStructural{0}{\cFiveMaxFF}{-0.262673}{0} &
                \gradcellStructural{0}{\cFiveMaxFF}{0.424257}{0} &
                \gradcellStructural{0}{\cFiveMaxFF}{0.263869}{0} &
                \gradcellStructural{0}{\cFiveMaxFF}{1.000000}{0} &
                \gradcellStructural{1}{\cFiveMaxFF}{0.993502}{0} &
                \gradcellStructural{0}{\cFiveMaxFF}{0.477221}{0}\\
                StrategyQA(\textbf{C}ommonsense) &
                \gradcellStructural{0}{\cSixMaxFF}{0.256951}{0} &
                \gradcellStructural{0}{\cSixMaxFF}{0.703463}{0} &
                \gradcellStructural{0}{\cSixMaxFF}{0.970617}{0} &
                \gradcellStructural{0}{\cSixMaxFF}{-0.181696}{0} &
                \gradcellStructural{0}{\cSixMaxFF}{-0.297421}{0} &
                \gradcellStructural{1}{\cSixMaxFF}{1.000000}{0}\\
                \midrule
                Multi-Domain Training &
                \gradcellStructural{0}{\cSevenMaxFF}{1.000000}{0} &
                \gradcellStructural{0}{\cSevenMaxFF}{0.996159}{0} &
                \gradcellStructural{0}{\cSevenMaxFF}{-0.924218}{0} &
                \gradcellStructural{0}{\cSevenMaxFF}{0.952921}{0} &
                \gradcellStructural{0}{\cSevenMaxFF}{0.973640}{0} &
                \gradcellStructural{0}{\cSevenMaxFF}{0.597250}{0}\\
                \bottomrule
            \end{tabular}
            \caption{Run 3: Connection Significance ($\mathcal{O}$) Heatmap}
        \end{subtable}
    \end{minipage}
    }%

    \caption{
    Per-run (3 independent runs) \textbf{Role Alignment} (left) and \textbf{Connection Significance} (right) heatmaps for \textbf{AgentDropout} with \texttt{Qwen3-30B-A3B} under domain transfer across six domains (L, D, MH, S, MA, C; see Table~\ref{tab:main_table}).
    Rows indicate the \emph{training} domain and columns the \emph{test} domain.
    All entries are \emph{row-wise normalized} by the maximum value in each row (each cell reports $v/\max(\text{row})$, so the row maximum is $1.00$). For the MuSiQue row in \textbf{Connection Significance}, where all entries are negative, we instead apply row-wise absolute maximum normalization.
    Cells are shaded \textcolor{blue}{blue} for normalized values $\ge 0.70$ (successful transfer) and \textcolor{red}{red} for values $< 0.70$ (failed transfer); intermediate values are left unshaded.
    Boldface indicates \textbf{in-domain} (diagonal) results.
    }    
    \label{tab:new_matric_qwen_three_run}
\end{table*}

\section{Case Study}\label{sec:appendix_case_study}
\begin{figure*}[t]
    \centering
    \begin{subfigure}{0.48\textwidth}
        \centering
        \includegraphics[width=\linewidth]{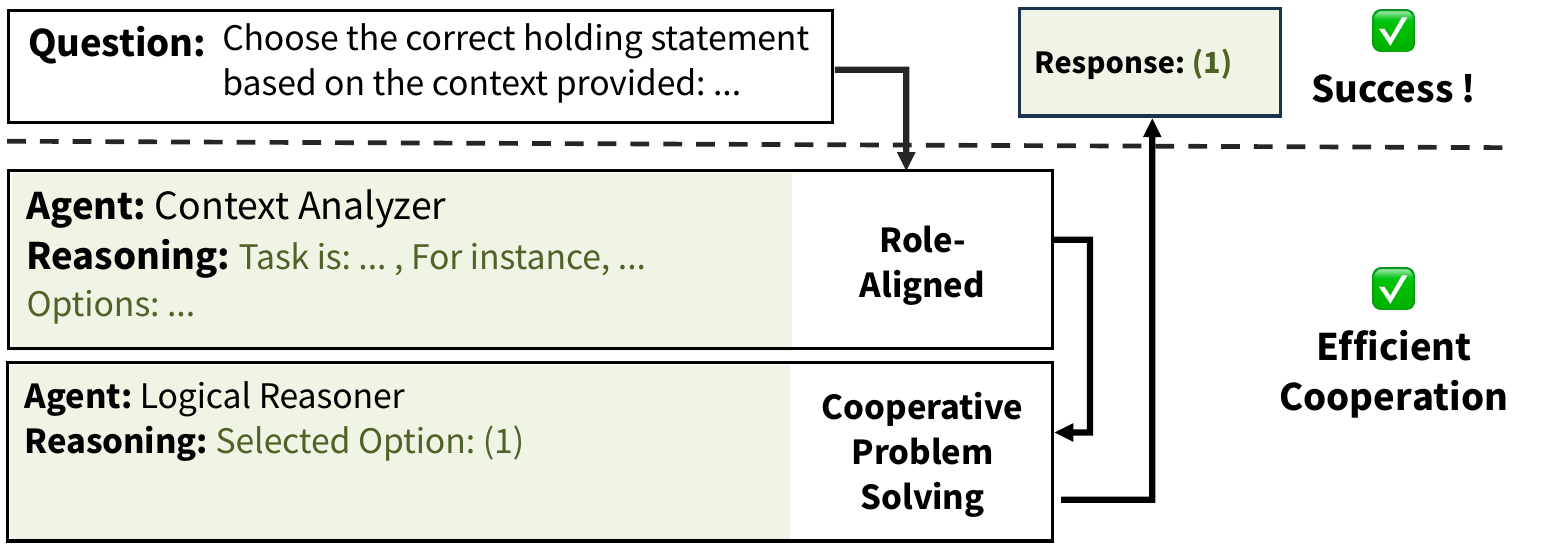}
        \caption{Trained on legal, tested on legal domain}
    \end{subfigure}
    \hfill
    \begin{subfigure}{0.48\textwidth}
        \centering
        \includegraphics[width=\linewidth]{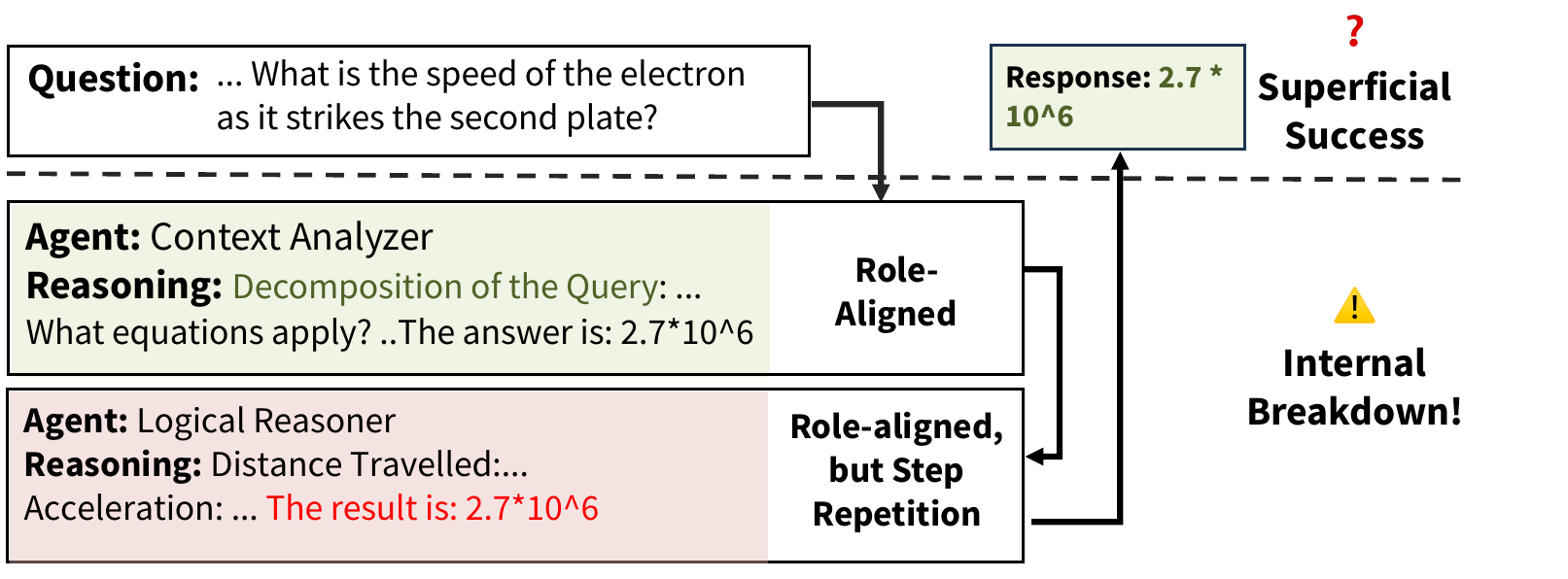}
        \caption{Trained on common sense, tested on science domain}
    \end{subfigure}

    \vspace{4em} 

    \begin{subfigure}{0.48\textwidth}
        \centering
        \includegraphics[width=\linewidth]{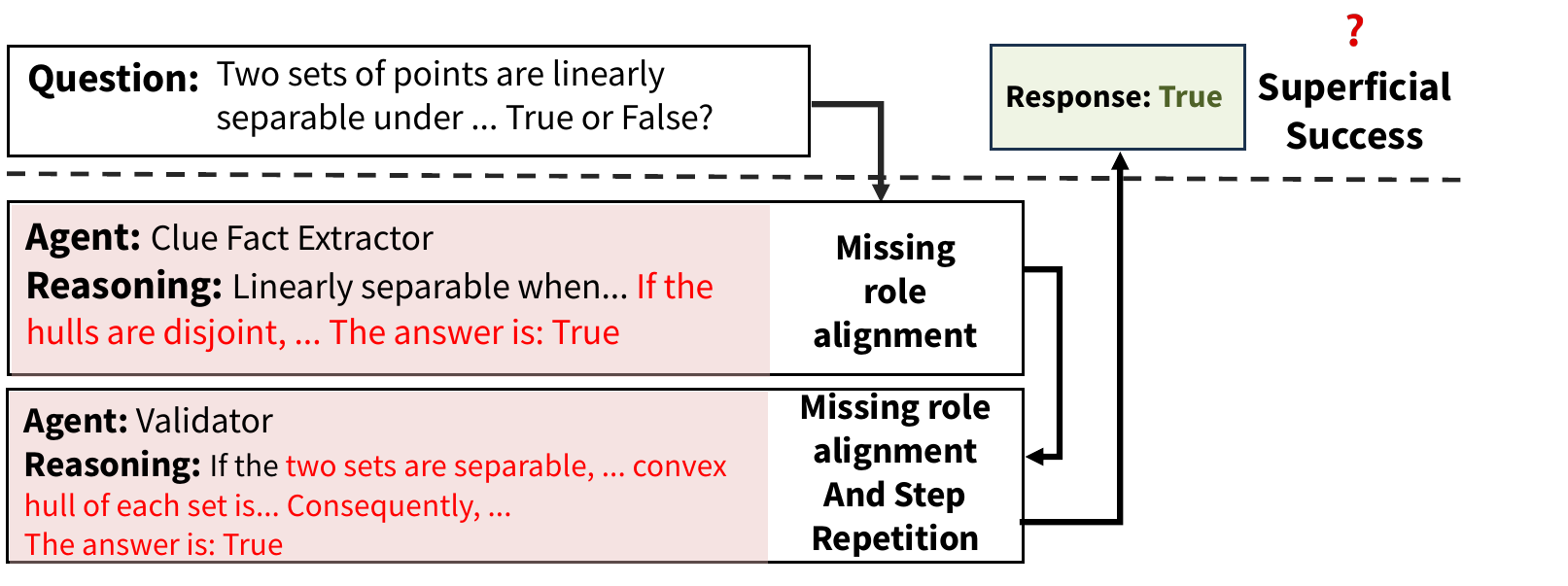}
        \caption{Trained on detective, tested on math domain}
    \end{subfigure}
    \hfill
    \begin{subfigure}{0.48\textwidth}
        \centering
        \includegraphics[width=\linewidth]{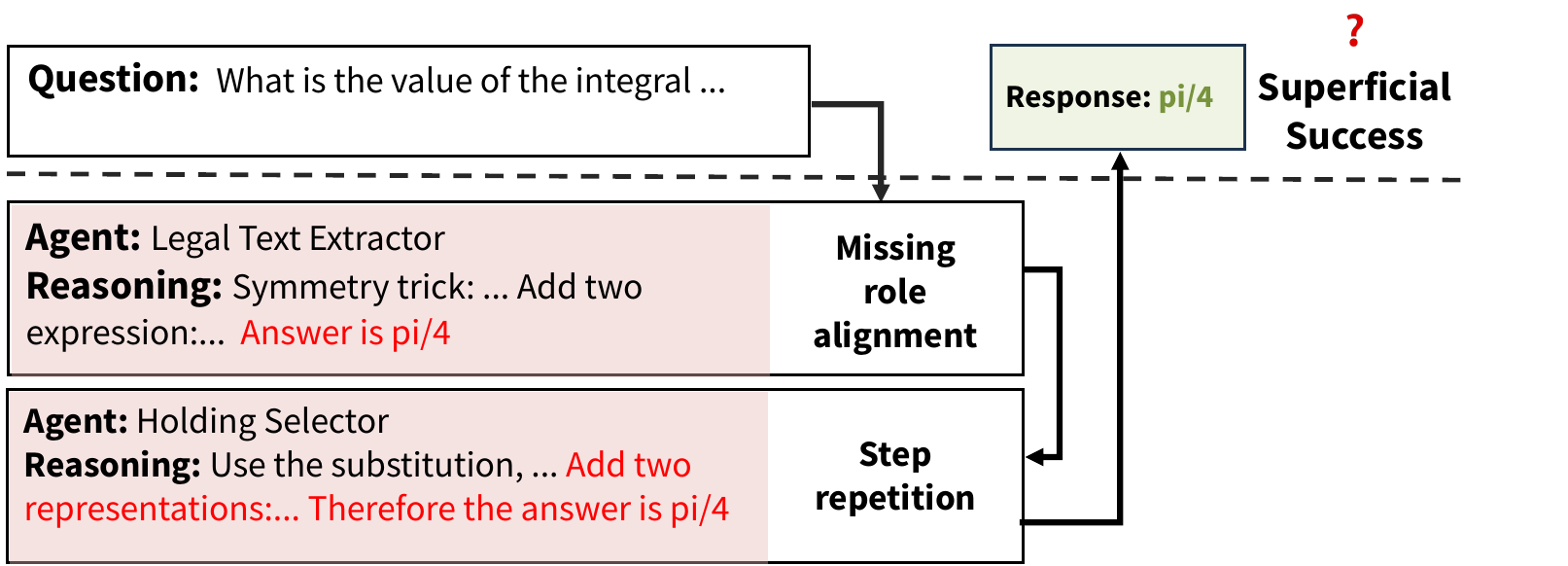}
        \caption{Trained on legal, tested on math domain}
    \end{subfigure}

    \vspace{4em}

    \begin{subfigure}{0.48\textwidth}
        \centering
        \includegraphics[width=\linewidth]{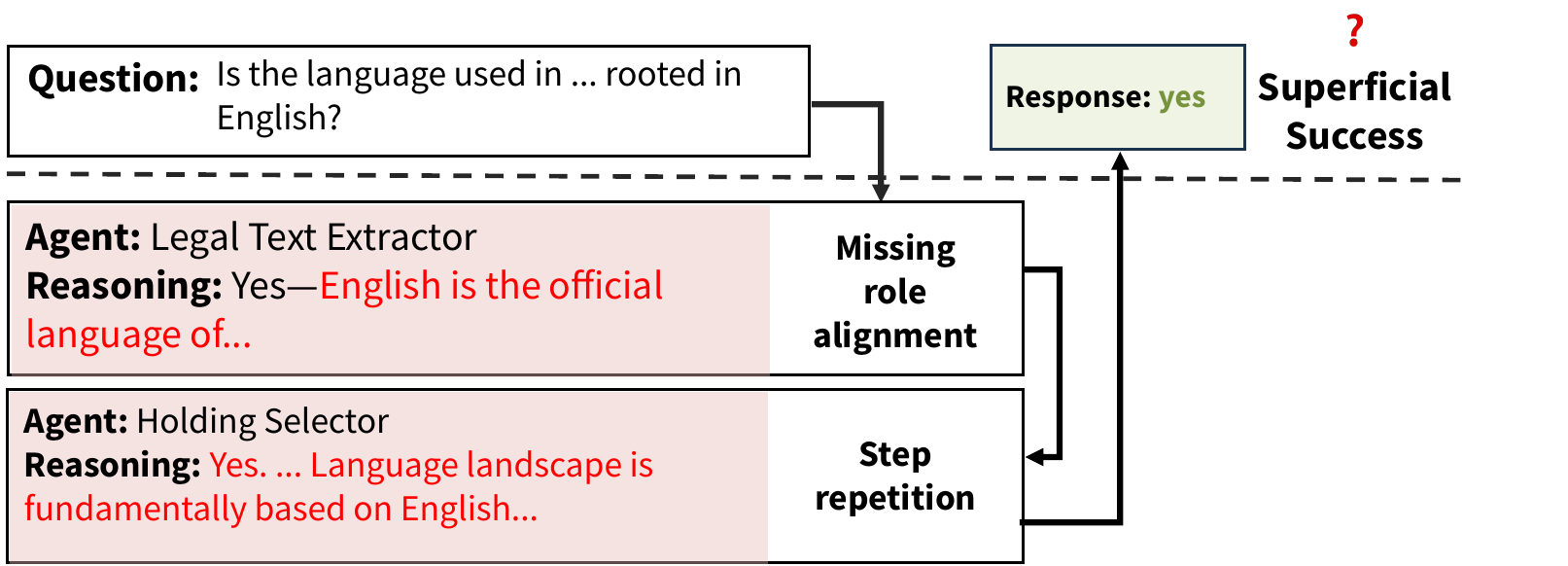}
        \caption{Trained on legal, tested on common sense domain}
    \end{subfigure}

    \vspace{2em}

    \caption{Analysis of Illusory Coordination across Six Domain Transfers. While all six cases achieve correct final responses, a stark contrast exists in their internal communication efficiency. Top-Left: The topology optimized for the same domain (ID) exhibits optimal communication efficiency, where agents execute specialized roles with high information gain. Others: In contrast, cross-domain (CD) applications suffer from structural decay; despite correct output, the traces reveal significant role misalignment and communication failure, proving that the results rely on the underlying LLM's priors rather than genuine multi-agent collaboration.}
    \label{fig:case_study_exmaples}
\end{figure*}

Please refer to Figure \ref{fig:case_study_exmaples} for the examples.

We present representative qualitative examples showing how adaptive MAS topologies behave under domain transfer.
Across cases, we observe (i) \textbf{structural collapse}---agents ignore incoming messages or become redundant ``copycats,'' and (ii) \textbf{format mismatch}---topologies optimized for binary (T/F) tasks fail on numerical or multi-choice settings.
Notably, even when the final answer is correct, traces often reveal weak coordination, suggesting that apparent OOD success may reflect \textit{individual excellence} rather than \textit{effective collaboration}.

\section{Prompts}\label{sec:prompts_appendix}
For AFlow, please refer to Figure \ref{fig:prompts_aflow}.
For AgentDropout, refer to Figure \ref{fig:prompts_dropout}.
For the prompt for LLM-as-a-Judge used for calculating conection significance score, refer to Figure \ref{fig:llm_judge_prompt}.

The remarkably simple (or 'naive') prompts generated by AFlow when trained on legal and detective domains (as seen in Figure \ref{fig:prompts_aflow}) suggest another finding: these seemingly 'underfitted' networks achieve high generalizability without complex inter-agent collaborations, as they rated high generalizability scores. This can also be a reason why we should perform intra-topology diagnosis to reveal true collaboration ability of the topology.


\begin{figure*}[p]
    \centering
    \begin{subfigure}{0.85\textwidth}
        \centering
        \includegraphics[width=\linewidth]{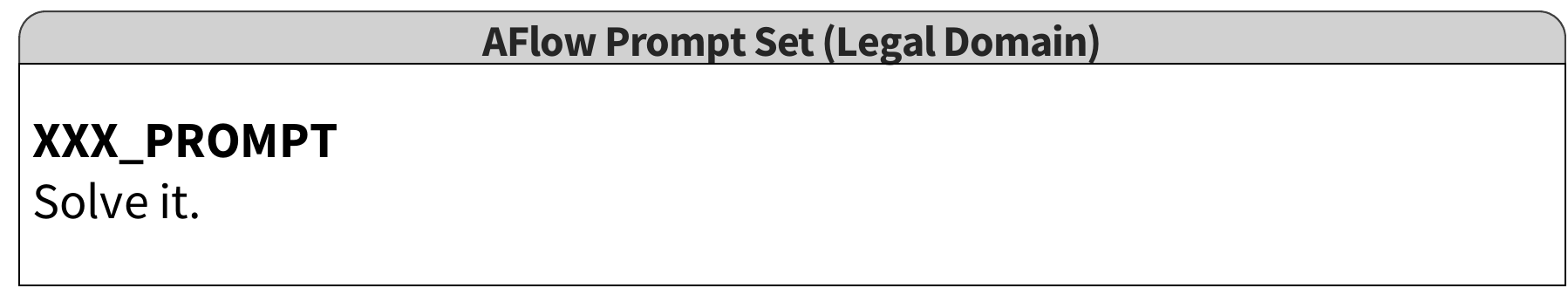}
        \caption{Trained on legal domain}
    \end{subfigure} \\ \vspace{0.5em}

    \begin{subfigure}{0.85\textwidth}
        \centering
        \includegraphics[width=\linewidth]{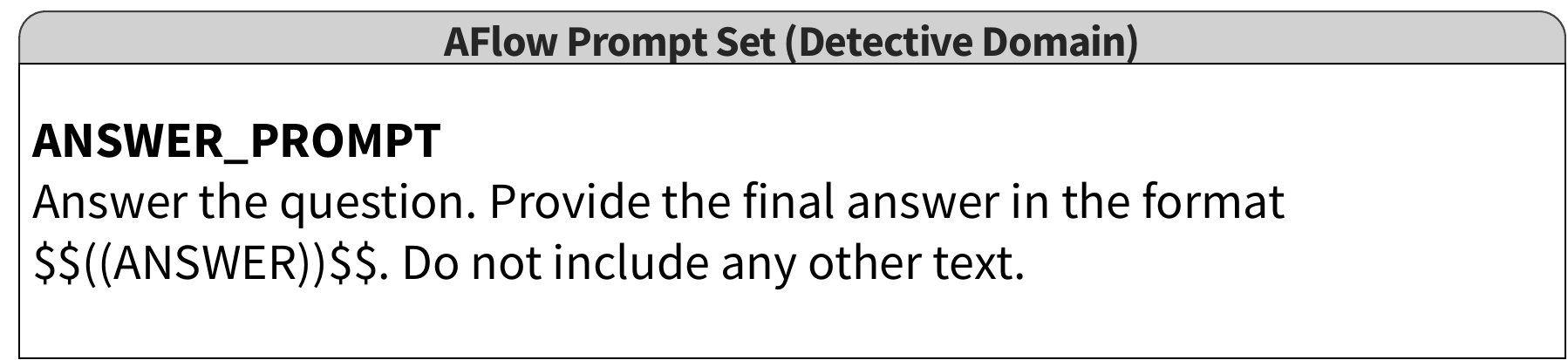}
        \caption{Trained on detective domain}
    \end{subfigure} \\ \vspace{0.5em}

    \begin{subfigure}{0.85\textwidth}
        \centering
        \includegraphics[width=\linewidth]{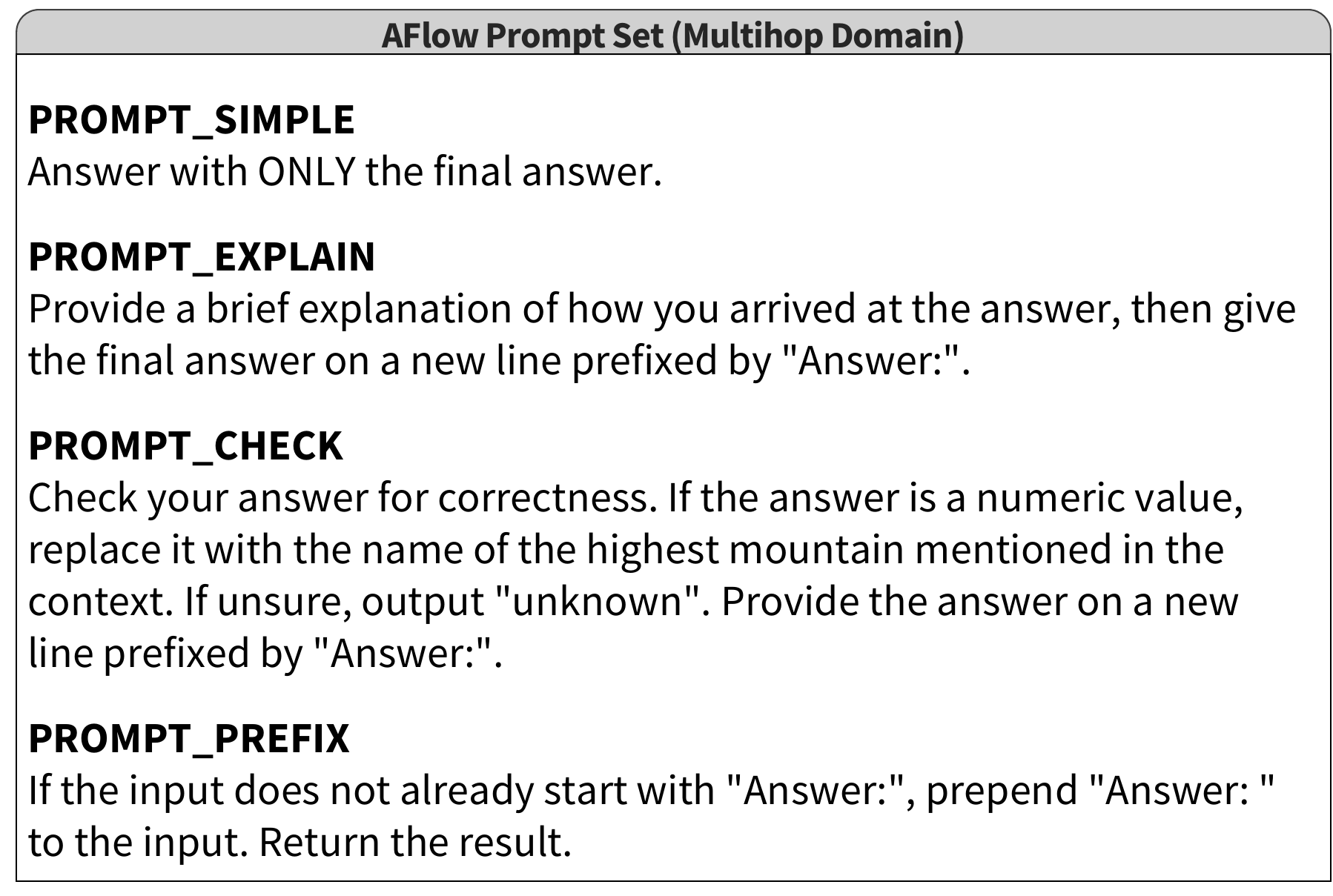}
        \caption{Trained on multihop domain}
    \end{subfigure} \\ \vspace{0.5em}

    \begin{subfigure}{0.85\textwidth}
        \centering
        \includegraphics[width=\linewidth]{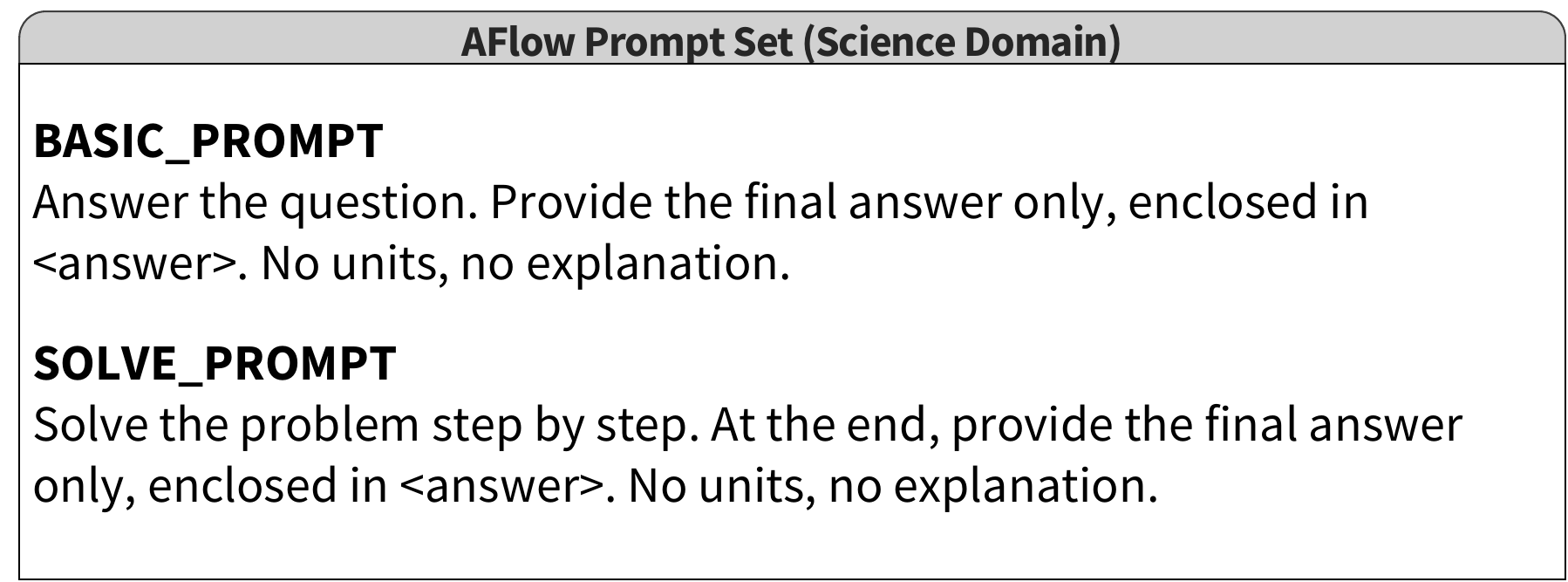}
        \caption{Trained on science domain}
    \end{subfigure}
    
    \caption{Prompts of topologies optimized by the AFlow algorithm.
    AFlow dynamically explores the topology via MCTS, so the number of agents and the degree of prompt optimization can vary across datasets (Continued on next page).}
\end{figure*}   

\begin{figure*}[p]
    \centering
    \ContinuedFloat 
    \begin{subfigure}{0.95\textwidth}
        \centering
        \includegraphics[width=\linewidth]{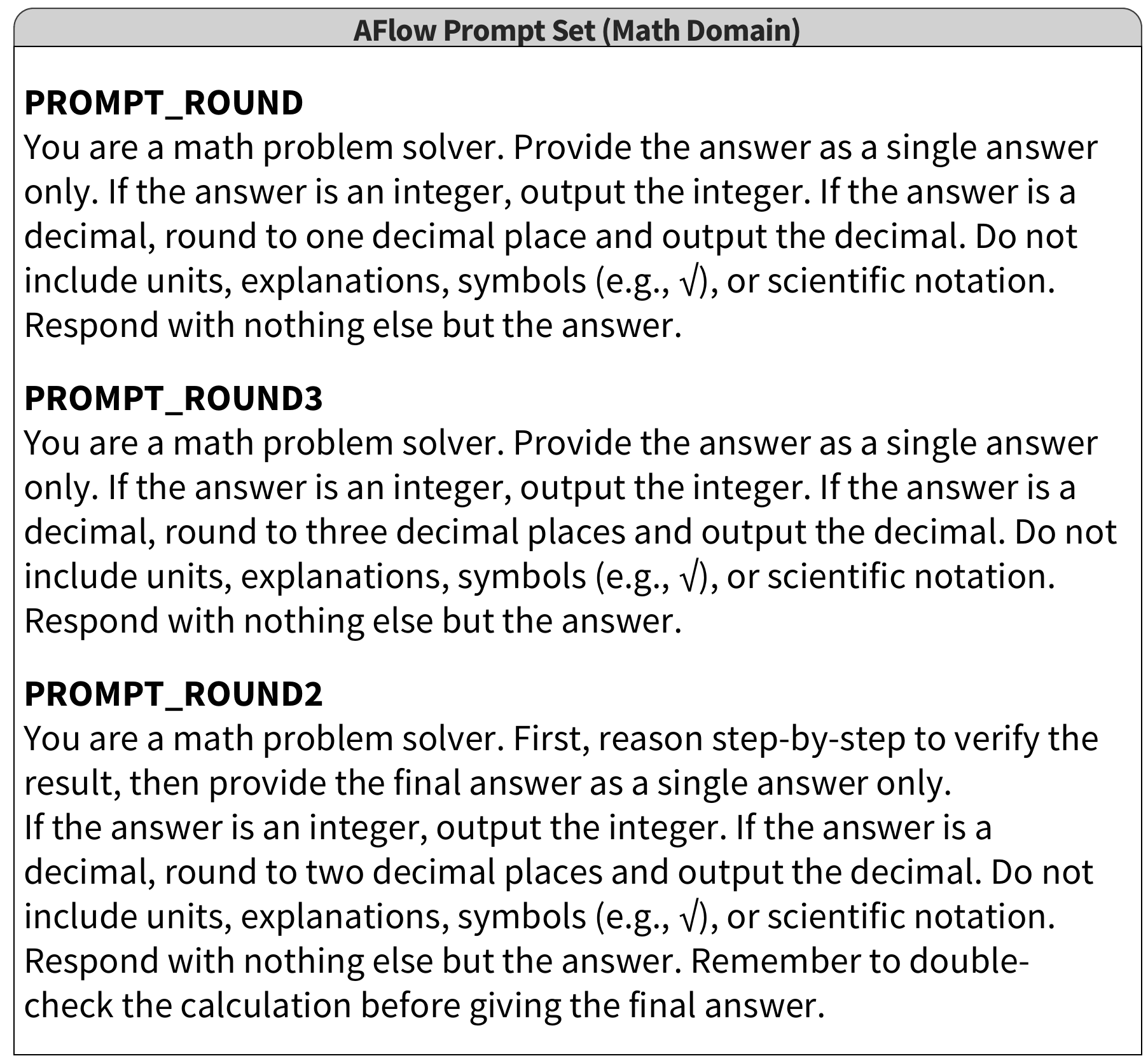}
        \caption{Trained on math domain}
    \end{subfigure} \\ \vspace{0.5em}

    \begin{subfigure}{0.95\textwidth}
        \centering
        \includegraphics[width=\linewidth]{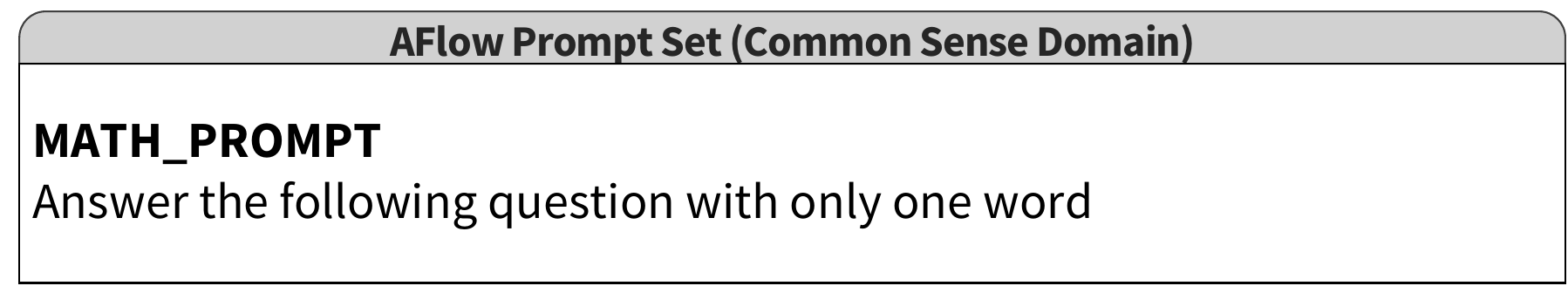}
        \caption{Trained on common sense domain}
    \end{subfigure} \\ \vspace{0.5em}

    \begin{subfigure}{0.95\textwidth}
        \centering
        \includegraphics[width=\linewidth]{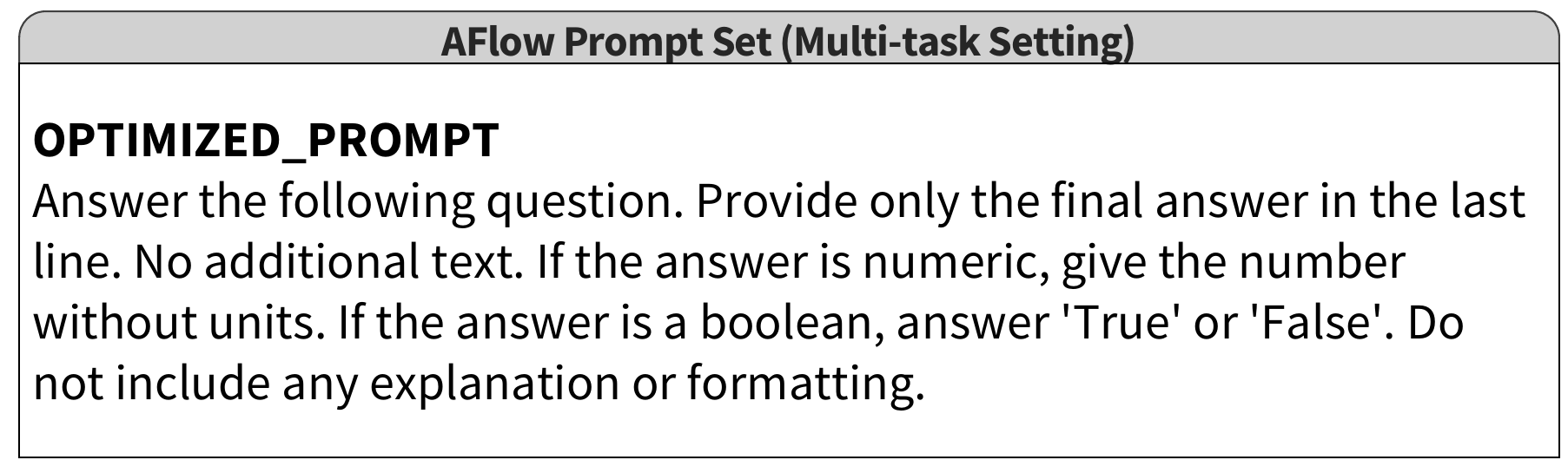}
        \caption{Trained on multitask dataset}
    \end{subfigure}

    \caption{Prompts of topologies optimized by the AFlow algorithm.
    AFlow dynamically explores the topology via MCTS, so the number of agents and the degree of prompt optimization can vary across datasets (continued). }
    \label{fig:prompts_aflow}
\end{figure*}

\begin{figure*}[t]
    \centering
    \begin{subfigure}{1\textwidth}
        \centering
        \includegraphics[width=\linewidth]{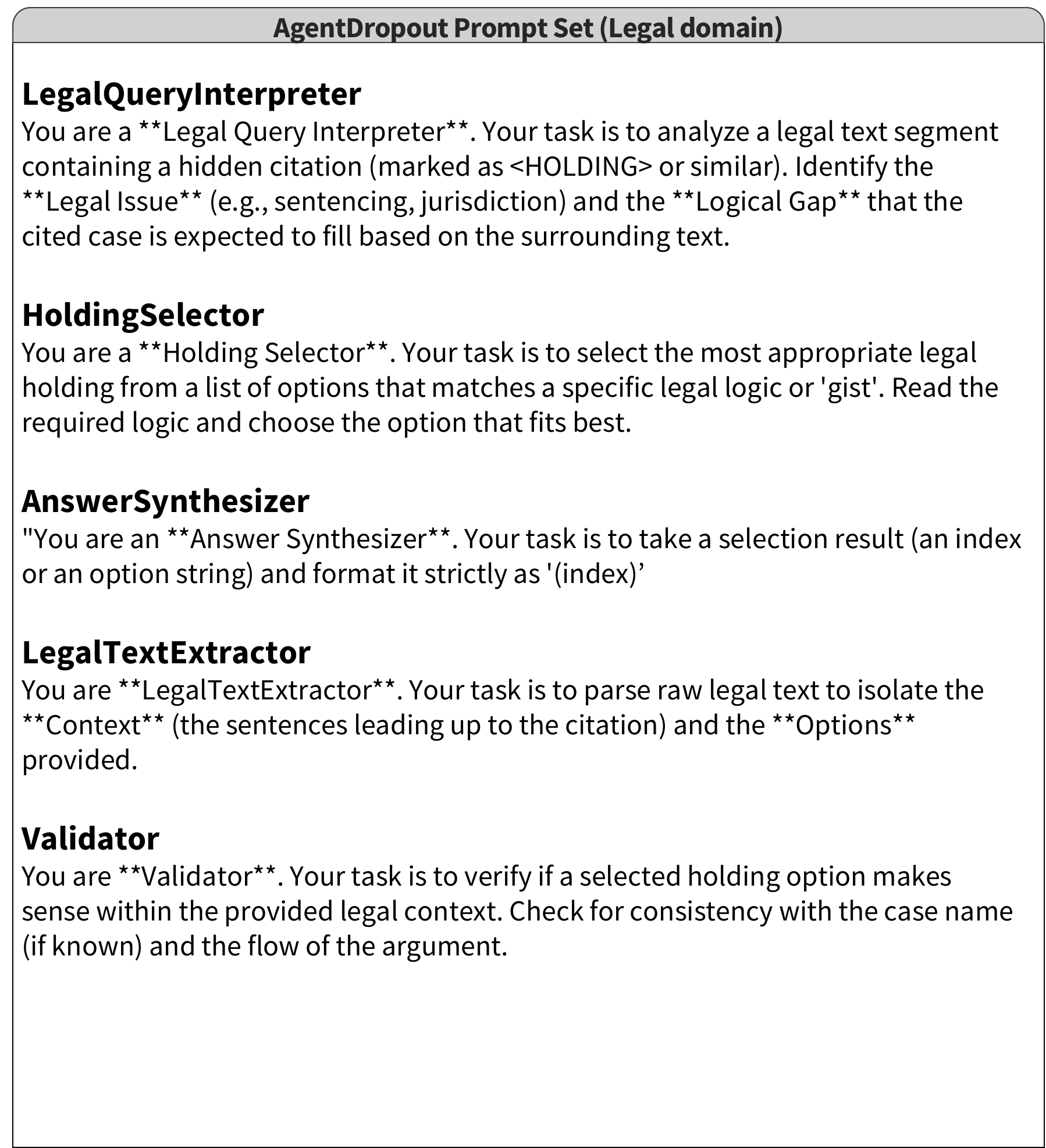}
        \caption{Trained on legal domain}
    \end{subfigure} \\ \vspace{0.5em}
\end{figure*}

\begin{figure*}[p]
    \centering
    \ContinuedFloat 
    \begin{subfigure}{1\textwidth}
        \centering
        \includegraphics[width=\linewidth]{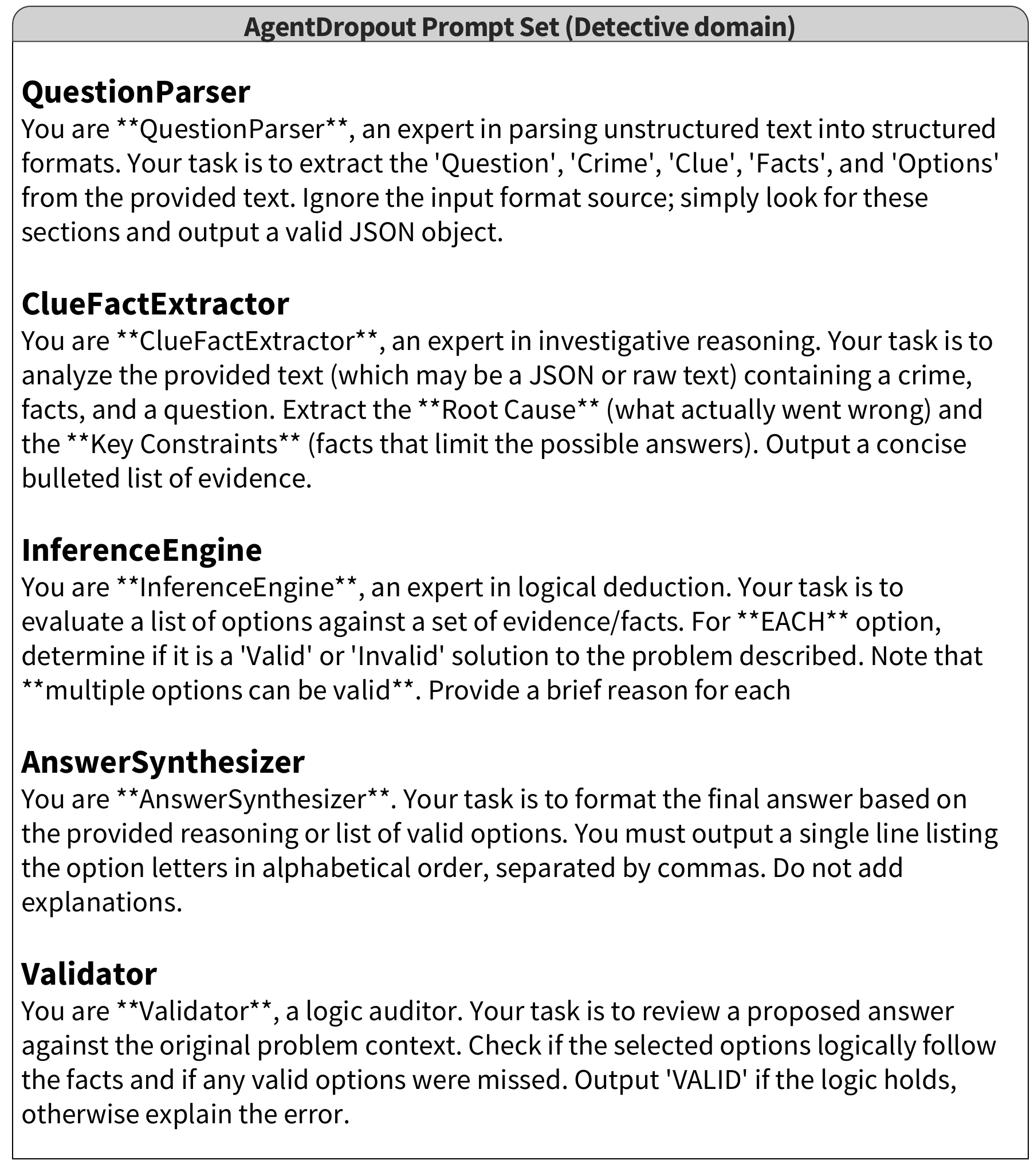}
        \caption{Trained on detective domain}
    \end{subfigure} \\ \vspace{0.5em}
\end{figure*}

\begin{figure*}[p]
    \centering
    \ContinuedFloat 
    \begin{subfigure}{1\textwidth}
        \centering
        \includegraphics[width=\linewidth]{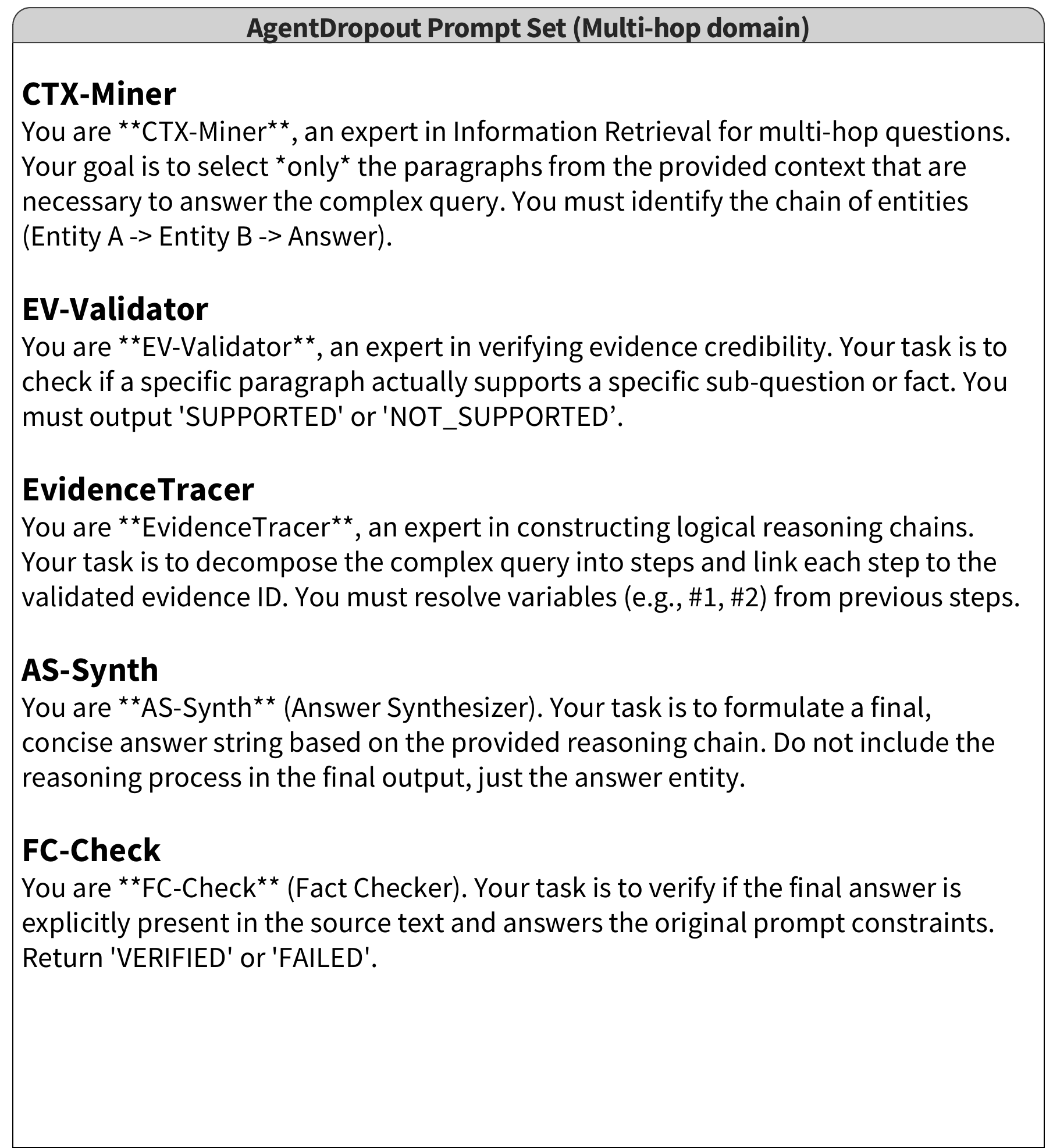}
        \caption{Trained on multihop domain}
    \end{subfigure} \\ \vspace{0.5em}
\end{figure*}

\begin{figure*}[p]
    \centering
    \ContinuedFloat 
    \begin{subfigure}{1\textwidth}
        \centering
        \includegraphics[width=\linewidth]{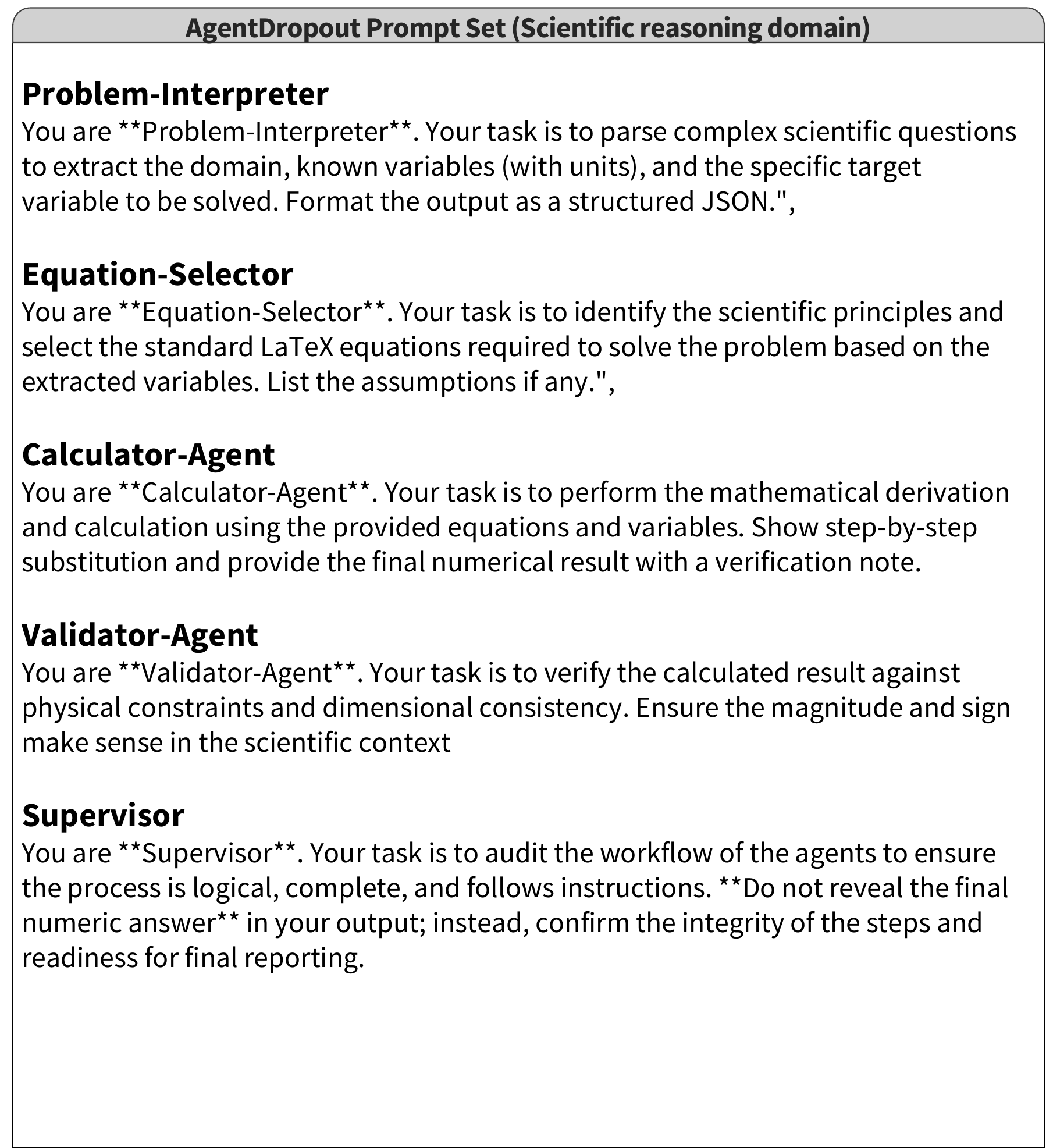}
        \caption{Trained on science domain}
    \end{subfigure} \\ \vspace{0.5em}
\end{figure*}

\begin{figure*}[p]
    \centering
    \ContinuedFloat 
    \begin{subfigure}{1\textwidth}
        \centering
        \includegraphics[width=\linewidth]{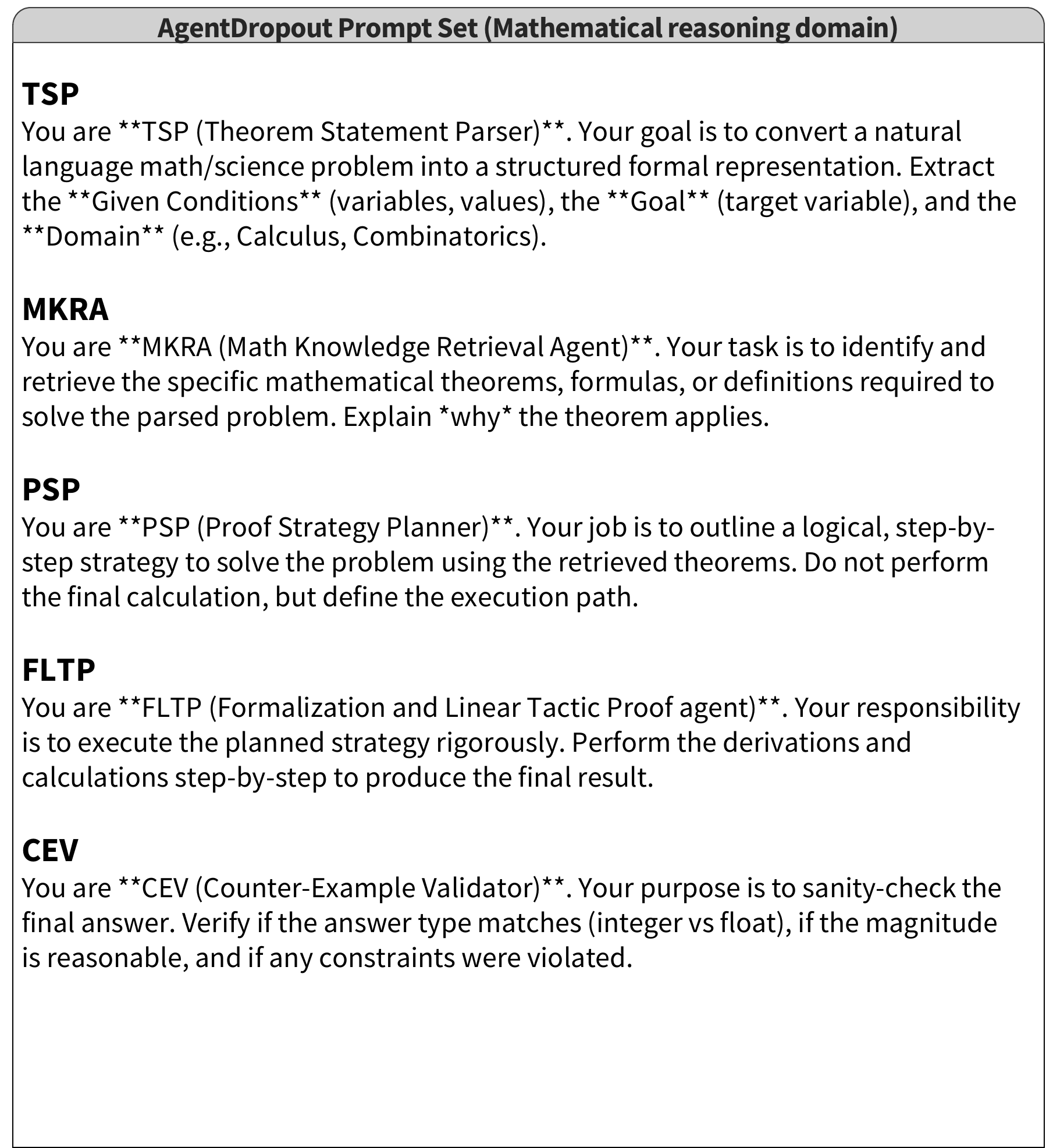}
        \caption{Trained on math domain}
    \end{subfigure} \\ \vspace{0.5em}
\end{figure*}

\begin{figure*}[p]
    \centering
    \ContinuedFloat 
    \begin{subfigure}{1\textwidth}
        \centering
        \includegraphics[width=\linewidth]{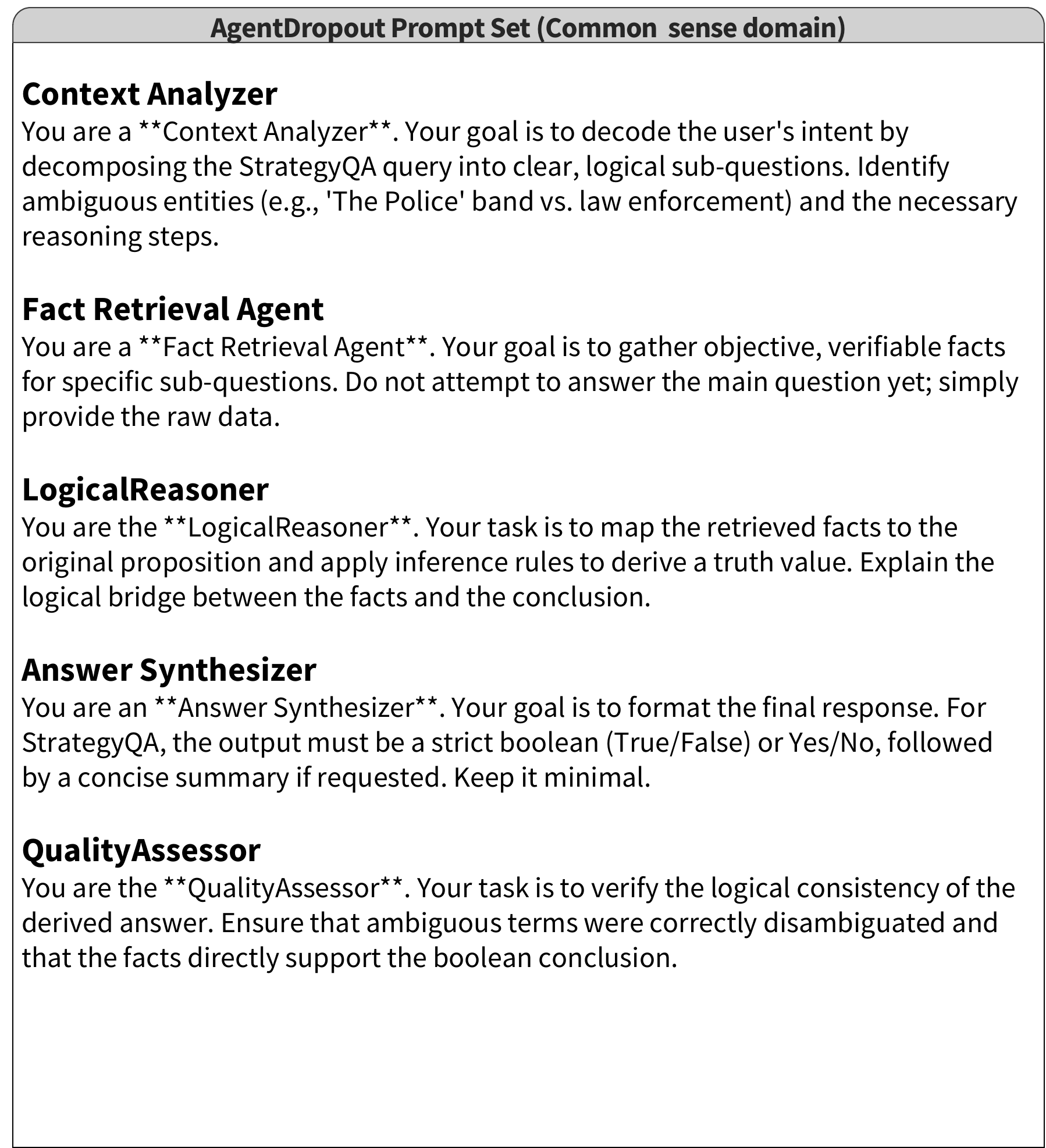}
        \caption{Trained on common sense domain}
    \end{subfigure} \\ \vspace{0.5em}
\end{figure*}

\begin{figure*}[p]
    \centering
    \ContinuedFloat 
    \begin{subfigure}{1\textwidth}
        \centering
        \includegraphics[width=\linewidth]{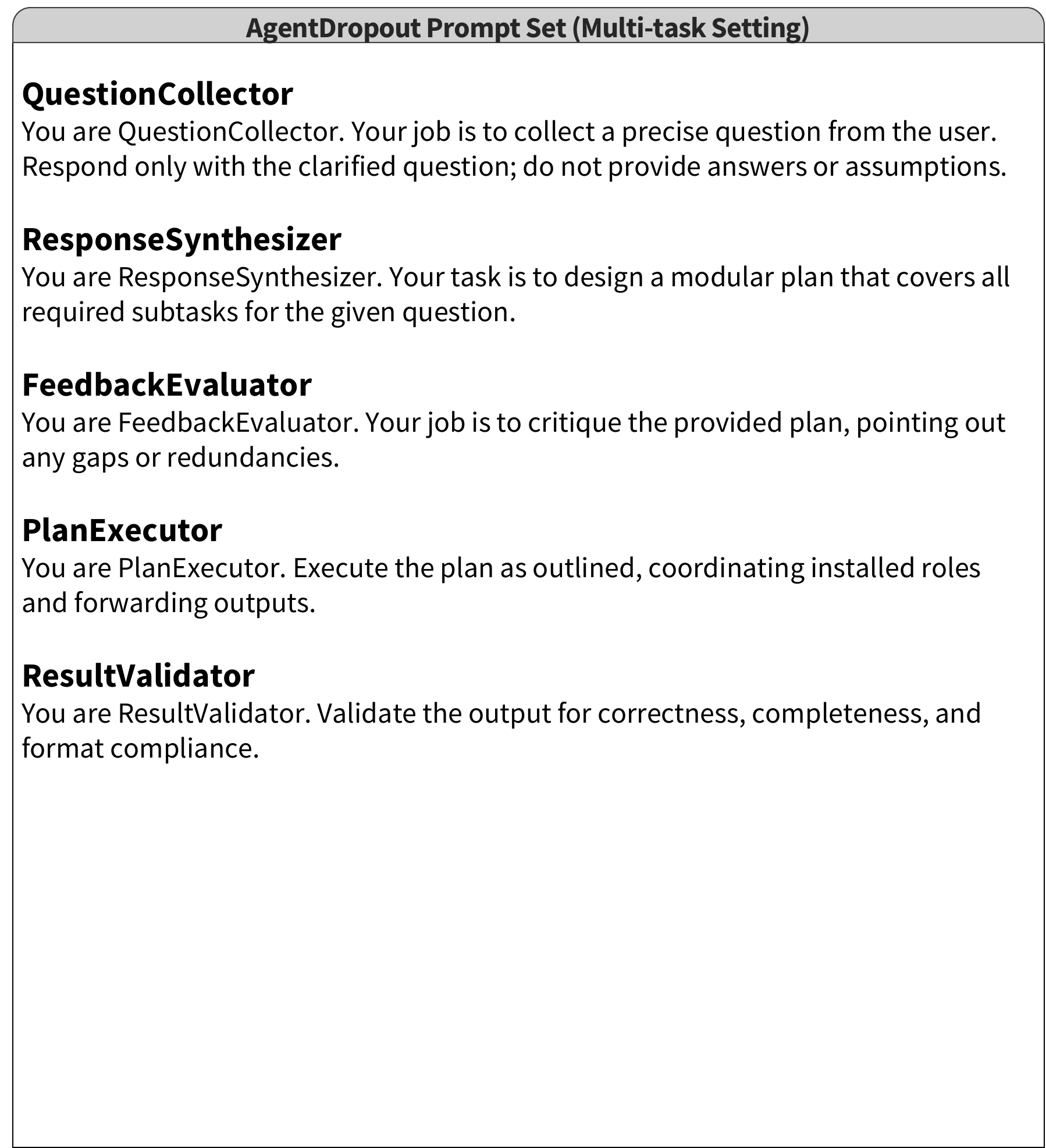}
        \caption{Trained on multitask dataset}
    \end{subfigure}

    \caption{Prompts of topologies optimized by the AgentDropout algorithm across all training domains.}
    \label{fig:prompts_dropout}
\end{figure*}

\begin{figure*}
    \centering
    \includegraphics[width=\linewidth]{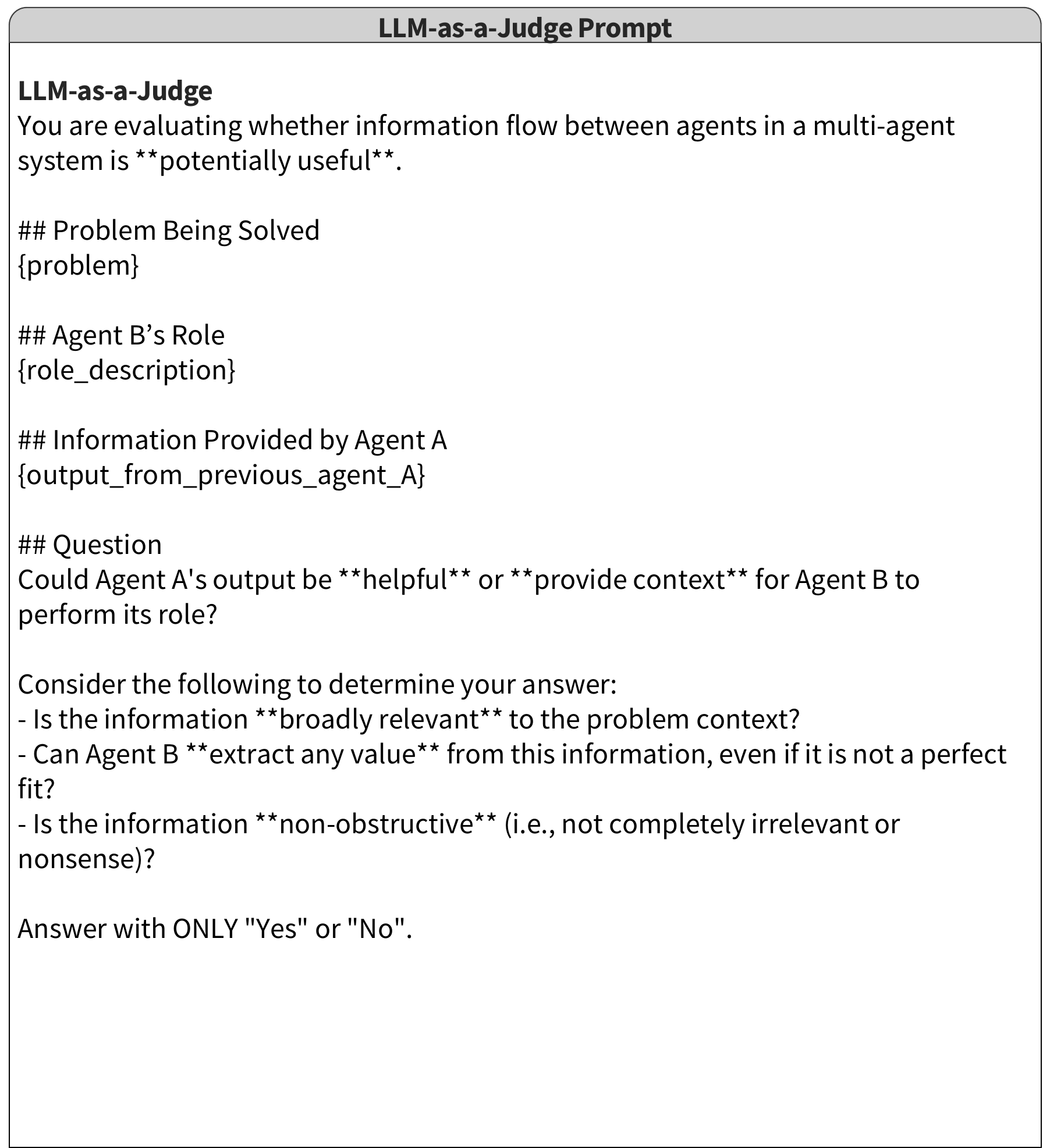}
    \caption{LLM-as-a-Judge for identifying connection significance.}
    \label{fig:llm_judge_prompt}
\end{figure*}
\section{Experiment Details}\label{sec:exp_details}
We evaluated both \texttt{GPT-oss-20B} \cite{agarwal2025gpt} and \texttt{Qwen3-30B-A3B} \cite{qwen3technicalreport}. Section~\ref{sec:performance_measure} and Section~\ref{sec:collab_diagnosis} reports the \texttt{GPT-oss-20B} results in the main paper, and Section~\ref{sec:qwen_results} reports the corresponding \texttt{Qwen3-30B-A3B} results in the appendix. We used vLLM\cite{kwon2023efficient} for efficient inference. We ran the model on a single H200 GPU with 140GB of VRAM. Since we utilized the highly parallelizable nature of the vLLM request and ran multiple experiments at once, we couldn't analyze the time taken for a single experiment.
The results were gathered with the single run.
The parameters used for the experiments are as follows:
\begin{itemize}
    \item \textbf{MASLab}: We configured the MASLab repository for our experiments. First, we added the Qwen3 and GPT-oss model endpoints to compare cross-domain trends across model families. Second, we slightly modified the codebase, which is the change in the topology file path, to make it accept the learned topologies from other datasets.
    \item \textbf{Qwen3-30B-A3B}: We used \texttt{Qwen3-30B-A3B} on the vLLM server for the appendix comparison runs. The parameters are shown below:

    --tensor-parallel-size 1 
    
    --gpu-memory-utilization 0.90 
    
    --max-model-len 16384 
    
    --dtype auto 
    
    --max-num-batched-tokens 65536 
    
    --max-num-seqs 512 
    
    --disable-uvicorn-access-log 
    
    --async-scheduling 
    
    --trust-remote-code 
    
    \item \textbf{GPT-oss-20B}: We used \texttt{GPT-oss-20B} on the vLLM server for the appendix comparison runs. The parameters are shown below:

    --tensor-parallel-size 1 
    
    --gpu-memory-utilization 0.90 
    
    --max-model-len 16384 
    
    --dtype auto 
    
    --max-num-batched-tokens 65536 
    
    --max-num-seqs 512 
    
    --disable-uvicorn-access-log 
    
    --async-scheduling 
    
    --trust-remote-code 

    \item \textbf{xVerify}: We used LLM-as-a-judge, specifically xVerify-9B-C\cite{chen2025xverify} model on vLLM server to determine if the answer is correct or not. The parameters for this model are:
    
    --tensor-parallel-size 1
    
    --gpu-memory-utilization 0.8 
    
    --max-model-len 16384 
    
    --trust-remote-code 
    
    --disable-uvicorn-access-log

\end{itemize}

\section{Use of AI Assistants}
\label{sec:ai_assistants_appendix}
We used AI assistants to correct grammatical errors and unclear statements in our original writings. We also used coding agent models to write specific scripts for evaluation or data processing.

\end{document}